\documentclass[%
    aip,
     amsmath,amssymb,
     reprint,%
]{revtex4-1}

\usepackage{graphicx}
\usepackage{dcolumn}
\usepackage{bm}

\usepackage[utf8]{inputenc}
\usepackage[T1]{fontenc}
\usepackage{mathptmx}
\usepackage{etoolbox}

\makeatletter
\def\@email#1#2{%
 \endgroup
 \patchcmd{\titleblock@produce}
  {\frontmatter@RRAPformat}
  {\frontmatter@RRAPformat{\produce@RRAP{*#1\href{mailto:#2}{#2}}}\frontmatter@RRAPformat}
  {}{}
}%
\makeatother

\usepackage[margin=2cm]{geometry}
\allowdisplaybreaks

\usepackage{subfig}

\usepackage{hyperref}
\usepackage[dvipsnames]{xcolor}
\usepackage{tikz}
\usetikzlibrary{quantikz2}

\usepackage{soul}  
\usepackage{pgfplots}
\pgfplotsset{compat=1.18}

\usepackage{mathbbol}
\usepackage{xifthen}

\usepackage[titletoc,title]{appendix}

\usepackage{multirow}

\colorlet{ygray}{gray!20}

\newcommand{\aeq}{\begin{equation}}
\newcommand{\eeq}{\end{equation}}
\newcommand{\aeqn}{\begin{eqnarray}}
\newcommand{\eeqn}{\end{eqnarray}}
\newcommand{\aeqns}{\begin{eqnarray*}}
\newcommand{\eeqns}{\end{eqnarray*}}

\newcommand{\yi}{\mathrm{i}}
\newcommand{\ydd}[2]{\frac{ \diff #1}{\diff #2}}

\newcommand{\oO}{\mathcal{O}}

\newcommand*\diff{\mathop{}\!\mathrm{d}}

\newcommand{\ylb}{\left(}
\newcommand{\yrb}{\right)}

\DeclarePairedDelimiter\floor{\lfloor}{\rfloor}
\DeclarePairedDelimiter{\yceil}{\lceil}{\rceil}

\newcommand{\yEx}[1][]{\ifthenelse{\isempty{#1}}{\tilde{E}_x}{\tilde{E}_{x,#1}}}
\newcommand{\yEy}[1][]{\ifthenelse{\isempty{#1}}{\tilde{E}_y}{\tilde{E}_{y,#1}}}
\newcommand{\yBz}[1][]{\ifthenelse{\isempty{#1}}{\tilde{B}_z}{\tilde{B}_{z,#1}}}

\newcommand{\yham}{\mathcal{H}}
\newcommand{\yhamu}{\yham^{\rm UW}}
\newcommand{\yhamc}{\yham^{\rm CD}}
\newcommand{\yhamd}{\yham^{\rm CDD}}

\newcommand{\yhamuA}{\yham^{\rm UW, Adv}}
\newcommand{\yhamuL}{\yham^{\rm UW, Liouville}}

\def\dim{{d_x}}
\def\dimN{{D}}

\def\tporder{{q_t}}
\def\cporder{p_x}
\def\cfl{c_t}
\def\nx{n_x}
\def\nt{n_t}
\def\Nx{N_x}
\def\Nt{N_t}
\def\Ns{N_s}
\newcommand{\poly}[1]{{{\rm poly\,}#1}}
\newcommand{\oOrder}[1]{ {\oO{ \left( #1 \right)} } }

\newcommand{\ypqc}{p_{\rm QC}}

\usepgfplotslibrary{fillbetween}

\begin{document}
\preprint{LLNL-JRNL-869590}

\title[]{Quantum algorithm for the advection-diffusion equation and the Koopman-von Neumann approach to nonlinear dynamical systems}
\author{I. Novikau}
\email{novikau1@llnl.gov}
\affiliation{Lawrence Livermore National Laboratory, Livermore, California 94550, USA}
\author{I. Joseph}
\affiliation{Lawrence Livermore National Laboratory, Livermore, California 94550, USA}

\date{\today}

\begin{abstract}
We propose an explicit algorithm based on the Linear Combination of Hamiltonian Simulations technique to simulate both the advection-diffusion equation and a nonunitary discretized version of the Koopman--von Neumann formulation of nonlinear dynamics.  
By including dissipation into the model, through an upwind discretization of the advection operator, we avoid spurious parasitic oscillations which usually accompany standard finite difference discretizations of the advection equation.
In contrast to prior works on quantum simulation of nonlinear problems, we explain in detail how different components of the algorithm can be implemented by using the Quantum Signal Processing (QSP) and Quantum Singular Value Transformation (QSVT) methods.
In addition, we discuss the general method for implementing the block-encoding (BE) required for QSP and QSVT circuits and provide explicit implementations of the BE oracles tailored to our specific test cases.
We simulate the resulting circuit on a digital emulator of quantum fault-tolerant computers and investigate its complexity and success probability.
The proposed algorithm is universal and can be used for modeling a broad class of linear and nonlinear differential equations including the KvN and Carleman embeddings of nonlinear systems, the semiclassical Koopman-van Hove (KvH) equation, as well as the advection and Liouville equations.
\end{abstract}
\maketitle


\section{Introduction}\label{sec:introduction}

\subsection{Motivation}
A variety of quantum algorithms (QAs) were developed\cite{Suzuki93, Berry15, Gilyen19, Martyn21, Martyn23} to model the Hamiltonian evolution of linear unitary systems on fault-tolerant quantum computers, potentially offering a significant speedup compared to classical simulations.
Yet, efficient quantum computation of nonlinear dynamics remains an open question and an ongoing challenge due to the intrinsically linear nature of quantum mechanics.
Nonlinear models are ubiquitous in computational fluid dynamics,\cite{Gaitan20} plasma dynamics,\cite{Dodin20, Joseph23} chemical kinetics,\cite{Akiba23} reaction-diffusion processes,\cite{An23-diffusion, Liu23} economics\cite{Hsieh91} and social sciences.\cite{Richards00}
Moreover, the majority of reduced models of dynamical systems include dissipation in some form.  
However, while quantum computers (QCs) achieve enormous parallel processing power through their ability to work efficiently with superposition states, 
they can only act linearly on states.
Still, the fact that QCs can handle exponentially large amounts of data and, in some cases, perform operations with a polynomial or even exponential speedup is a key reason for pursuing efficient QAs for solving non-conservative linear and nonlinear ordinary differential equations (ODEs).

The Koopman--von Neumann (KvN) approach \cite{Joseph20,Joseph23} and the closely related Carleman approach \cite{Liu21, Joseph23} are two of the most promising techniques for directly solving nonlinear differential equations on QCs.
KvN provides a linear unitary embedding that allows QCs to simulate both conservative and dissipative nonlinear systems by evolving the probability distribution function (PDF) of the solutions in phase space.
The KvN equation combines both the advection equation used for describing the dynamics of constants of motion and the Liouville equation for conservation of the phase-space PDF in order to achieve a unitary description of the dynamics.
For Hamiltonian dynamics, the KvN, advection, and Liouville equations are equivalent, but the measurement of physical observables is more straightforward using the KvN approach.\cite{Joseph20, Joseph23} 

While the original inspiration for the KvN equation was to look for a semiclassical analogue of quantum mechanics,\cite{Conde23} the more accurate semiclassical version is known as the Koopman-van Hove (KvH) equation.\cite{Bondar19, Tronci21, Joseph23JPA}
Both KvN and KvH correctly evolve the amplitude of the state vector along classical trajectories, but only the KvH equation correctly evolves the phase of the wavefunction in accordance with Feynman’s prescription for the path integral.\cite{Joseph23JPA}
An essential point is that, while the phase is not measurable classically, the KvH equation is necessary for self-consistent modeling of coupled quantum-semiclassical systems because the phase of the semiclassical system, which varies for different quantum states, also becomes important.\cite{Joseph23JPA}

Since both the KvN and KvH equations define an evolution of the PDF, it is natural to generalize these equations to also include diffusion in phase space.
Diffusion results from stochastic processes like collisions that generate Langevin dynamics and an effective Brownian motion. \cite{VanKampen1976}
This allows one to consider generalizing the problem of interest to a stochastic differential equation (SDE), \cite{Kleinert2006book, Oksendal2013book} where one is naturally interested in the evolution of the PDF over phase space in time.
SDEs are widely used in fluid and kinetic models as well as many other domains, including biology, chemistry, ecology, and finance.
As will be explained below, adding dissipation is, in fact, an effective approach for obtaining high accuracy results with standard discretizations of the advection equation.

The overarching goal of this work is to provide an efficient QA for simulating the KvN equation on QCs.
The algorithm is based on an explicit finite difference approximation that is either first or second order accurate in space, depending on whether upwinding is used.
Working in the Eulerian picture is feasible on QCs, even for high-dimensional spaces, because of their enormous processing power. 
It is also desirable from an algorithmic viewpoint because Eulerian methods are much more accurate than Lagrangian Monte-Carlo (MC) or particle-in-cell (PIC) methods. 
Another advantage with respect to classical MC techniques is that a superposition of all initial conditions that correspond to the initial statevector can be simulated simultaneously, effectively in parallel.
For quantum simulation of the PDF, this leads to overall savings in computational work supported by efficient block-encoding of the KvN model.

However, we should emphasize that a complete end-to-end complexity analysis for such quantum Eulerian solvers warrants further investigation.
In particular, the comparison of the quantum Eulerian algorithms with classical MC methods, which are often exponentially faster than classical Eulerian algorithms, for specific choices of end-to-end applications remains an interesting open question. 
For instance, in contrast to classical or quantum MC methods, the KvN algorithm returns a quantum state containing a detailed and highly-resolved picture of the simulated nonlinear system at a given time instant.
This could result in higher accuracy as well as greater flexibility in choosing quantum post-processing, such as the quantum Fourier transform, of computed fields.

\subsection{Quantum advantage of QAs}\label{sec:qu-adv}
A QA achieves exponential quantum speedup if it only requires $Q_{\rm A}$ operations while its classical counterpart needs $C_{\rm A} \geq 2^{\gamma Q_{\rm A}}$ operations where $\gamma > 0$.
Similarly, a QA achieves a polynomial speedup with a speedup index $\ypqc>0$ if the corresponding classical algorithm (CA) requires $C_{\rm A} \geq Q_{\rm A}^{\ypqc}$ operations.
For instance, classical Eulerian methods usually require $C_{\rm A} = N_x^{\dim} t$ number of operations where $t$ is the integration time, $N_x = 2^{n_x}$ is the number of spatial points in a single dimension, and $\dim$ is the problem dimensionality.
In contrast, QAs such as the Quantum Signal Processing (QSP),\cite{Low17, Low19} which can be used for emulating unitary Eulerian discretized models, require at least $Q_{\rm A} = \dim n_x t$ operations if an efficient block-encoding (BE) is provided.
In this case, one cannot express $C_{\rm A}$ as $2^{\gamma Q_{\rm A}}$ because of the linear dependence of the QAs on the integration time.
Therefore, strictly speaking, one cannot claim that these QAs provide an exponential speedup.

At the same time, one can also consider the ratio of the number of operations required by a classical algorithm, $C_{\rm A}$, and its quantum counterpart, $Q_{\rm A}$:
\begin{equation}\label{eq:speedup-coef}
    S_{\rm QC} =  \frac{C_{\rm A}}{Q_{\rm A}}.
\end{equation}
We refer to $S_{\rm QC}$ as the speedup factor.
For instance, in the case of the QSP algorithm for Hamiltonian simulation, one has 
\begin{equation}\label{eq:Csp-QSP}
    S_{\rm QC} =  \frac{2^{n_x \dim}}{ n_x \dim } .
\end{equation}
In other words, the number of operations in classical Eulerian algorithms grows exponentially faster with $n_x$ than the number of operations in the QSP algorithm.
To remain consistent with the speedup classification described above, we will still refer to speedup of QAs such as the QSP as polynomial, even though the QSP has an exponential speedup factor.
For instance, in Ref.~\onlinecite{Babbush23}, it was shown how to achieve an exponential speedup factor in modeling sparsely coupled oscillator networks by providing efficient block-encoding of the system onto quantum Hamiltonian simulations and by measuring only global quantities such as kinetic and potential energies at a desired time instant.
A similar result was obtained in Ref.~\onlinecite{Barthe24}, where it was demonstrated in detail how to recast the evolution of a bosonic system with exponentially many modes as a simulation of expectation values of the corresponding quadrature operators.

It should be emphasized that the quantum algorithm proposed in this work can solve linear partial differential equations (PDEs), e.g. the Liouville equation, polynomially faster than classical Eulerian methods but it does not provide any speedup in solving a nonlinear ODE with a particular initial condition.
However, this quantum algorithm can be beneficial in analyzing a nonlinear ODE with $N_x$ different initial conditions.
In this case, one needs to perform $N_x$ different classical simulations while the quantum algorithm simulates all $N_x$ initial conditions at once.
Indeed, if we consider a $\dimN$-dimensional nonlinear ODE with $N_x$ initial conditions, the cost of classical solvers is $\oO(N_x {\rm poly}[\dimN])$.
On the other hand, to model the same ODE, the KvN technique requires $N_x^{\dimN}$ degrees of freedom in its Eulerian discretization.  
If an efficient block-encoding of the KvN formulation is provided, then the KvN model is mapped onto a quantum circuit with $\oO({\rm poly}[\dimN, \log N_x])$ quantum operations.
In other words, the extreme dilation in dimension performed by the KvN algorithm is balanced by efficient block-encoding.
However, this also means that the main source of quantum speedup in the KvN technique comes from the fact that the KvN algorithm allows one to model multiple initial conditions in parallel.
In particular, as shown in the numerical KvN simulations in this work, one can use uniform initial conditions or any smooth initial profile that can be easily encoded without drastically influencing the overall complexity of the entire KvN algorithm.
That said, to avoid negating the overall quantum speedup, one should measure only robust variables integrated over simulated space that can be performed by using well-established amplitude estimation (AE) techniques.\cite{Brassard02}

\subsection{QAs for modeling nonlinear dynamics}
The key challenge in developing QAs for solving iterative nonlinear problems is the so-called no-cloning theorem which forbids the creation of an exact copy of an arbitrary unknown quantum state because of the linear unitary nature of quantum mechanics.\cite{Wootters82}
In order to perform nonlinear operations, one can generate multiple replicas of the initial state.
However, if a nonlinear operation must be iterated, then the number of replicas must grow exponentially quickly.
Due to this, the first QA proposed for integrating nonlinear differential equations\cite{Leyton08} suffered from an exponential growth of qubits with the integration time.
However, this fact stimulated a productive discussion on whether quantum advantage is possible at all in modeling nonlinear dynamics.
This resulted in a broad class of nonlinear QAs including the mapping of nonlinear dynamics into many-body quantum systems,\cite{Lloyd20, Engel23} as well as
hybrid quantum-classical techniques such as variational methods,\cite{Kyriienko21, Demirdjian22, Jaksch23} quantum amplitude-estimation-based algorithms,\cite{Kacewicz06, Gaitan20, Gaitan24, Oz23} and quantum annealers for modeling nonlinear dynamics.\cite{Criado23, Nguyen24}

Linearization algorithms such as the KvN\cite{Koopman31, Joseph20, Dodin20} and  Carleman\cite{Carleman32, Liu21} techniques can be used to directly solve nonlinear ODEs on quantum computers.
Both of them are based on linear representations of nonlinear ODEs (see Appendix in Ref.~\onlinecite{Joseph23}).
Carleman linearization was used in Ref.~\onlinecite{Liu21} to solve a dissipative problem with quadratic nonlinearity where the ratio of nonlinearity to dissipation is less than one.
In Refs.~\onlinecite{An23-diffusion, Liu23}, a similar approach was applied to model nonlinear reaction--diffusion equations.
The Carleman method is better for describing nonlinear dynamics in the vicinity of a single attractor, while the KvN technique can work accurately with multiple roots.

In Ref.~\onlinecite{Joseph20}, we proposed an Eulerian method to discretize a generalization of the KvN equation that encompasses the standard KvN, KvH, advection, and Liouville differential equations.
The same interpretation can be given to the Carleman method: it solves a complex-analytic version of the KvN equation in the complex-analytic basis of monomials (for the proof, see Appendix B and C of Ref.~\onlinecite{Joseph23}).
Now that the problem is linear, the quantum computer can evolve an initial wavefunction that represents a superposition of a large number of initial conditions at once. 
Moreover, the discretized KvN Hamiltonian is sparse. 
This allows one to compile an efficient mapping of the Hamiltonian onto quantum circuits.  
This is an important result because the high accuracy of Eulerian methods throughout all of phase space makes them the best choice for many applications. 

Although the resulting probabilistic description provided by the embedding techniques is expensive, it is essential for understanding the evolution of the system in phase space.
Moreover, since QCs can deal with exponentially large Hilbert spaces, it seems suitable to encode the extended linear space resulting from KvN and Carleman embeddings into quantum states.
Apart from this, these techniques do not suffer from expensive initialization and readout operations when information is transferred in-between classical and quantum computers.
For instance, by using Hamiltonian-simulation-based algorithms for modeling KvN systems, one can significantly reduce the number of calls to the initialization circuit.
In addition, once the KvN simulation is finished, there is no need to extract information about the entire PDF throughout phase-space.
For an efficient application, one can measure the expectation value of a desired phase-space observable by simply using the fact that the square of the KvN state vector is equal to the probability distribution function in the phase-space (the explicit algorithm is given in Ref.~~\onlinecite{Joseph20,Joseph23}).
Thus, the KvN technique, which turns nonlinear problems into a linear form suitable for QC, utilizes the main advantages of QCs while minimizing the drawbacks that usually accompany the implementation of most QAs.

Another example of an important application is to develop a nonlinear root finding algorithm.
The KvN method can serve as a root-finder for a nonlinear system with multiple fixed points, while the Carleman technique can accurately hone in on a single root.
In principle, these methods can be used together, starting with the KvN approach to identify the most interesting fixed point, and then using the Carleman approach to linearize the same system in the proximity of the chosen point of interest, allowing for the investigation of local dynamics with higher precision.
In fact, a unitary transfer operator (which can be considered to be the evolution operator of the KvN equation for a finite time interval) was used as a root-finder in Ref.~\onlinecite{Dodin21}, but found difficulties with oscillatory behavior that we will begin to address in this work.

\subsection{QAs for solving nonunitary problems}

Robust spatial discretization of the advection operator for the KvN, KvH, Liouville, and advection equations is an important issue. 
Godunov’s theorem\cite{Godunov59} states that numerical schemes for advection that preserve monotonicity must be first order accurate in space. 
It is well known that the upwind (UW) discretization method, which provides dissipation, is best for nonlinear computational fluid and plasma applications. 
Classical algorithms have also benefited from the development of higher order methods such as total variation diminishing (TVD)\cite{Harten83} and weak essentially non-oscillatory (WENO)\cite{Liu94WENO} schemes.
However, because these methods use a nonlinear computation of the numerical flux, which depends on the statevector itself, one cannot use such methods within a QA without incurring an exponential cost in memory and time. \cite{Lin22Koopman}
Therefore, we use the UW scheme for discretizing the KvN equation (called the chemical master equation in Ref.~\onlinecite{Lin22Koopman}).
This allows us to avoid spurious numerical oscillations that would otherwise be problematic for measuring classical information from quantum states.
However, since the resulting truncated discretized KvN equation becomes nonunitary, we need QAs that can simulate non-conservative linear models.

To model dissipative dynamics, one can use the dilation technique where the original ODEs are transformed into a system of linear equations characterized by some matrix $M$.
The system is then solved using QAs for matrix inversion.\cite{Ambainis12, Chakraborty19, Costa21}
For instance, this approach was used in Ref.~\onlinecite{Liu21} to deal with a nonlinear dissipative problem.
The complexity of these Quantum Linear System Algorithms (QLSAs) scales as $\oO(\kappa)$ at best where $\kappa$ is the condition number of $M$.
In particular, QLSAs require $\oO(\kappa)$ repetitions of the initialization operator, where for equations with second-order spatial derivatives such as the heat or advection-diffusion equations, the condition number grows quadratically\cite{Linden22} with $t$.
However, in some cases where fast-forwarding is possible with the condition number scaling sublinearly with time, one can obtain $\oO(t^{r})$ query complexity with $r < 1$.\cite{An24FF, Jennings24}
For instance, this is possible for linearized collisionless plasma problems.\cite{Jennings24}

A different technique was proposed in Ref.~\onlinecite{Fang23} where one splits the simulated temporal interval into a sequence of small intervals and simulates each of them by using the QSP algorithm\cite{Low17, Low19} where an output quantum state from a previous interval is used as an initial quantum state for the next temporal interval.
If one models the system with a non-Hermitian Hamiltonian, the norm of the output quantum state changes at each time step.
In particular, after several steps, the amplitude of the quantum state can become exponentially small.
To deal with this, Ref.~\onlinecite{Fang23} uses the so-called Uniform Singular Value Amplification (USVA) to rescale quantum state after each step.
Moreover, to reduce the number of ancillae in the quantum circuit, Ref.~\onlinecite{Fang23} applies the so-called compression gadget (CG) technique that makes the number of ancillae scale only logarithmically with the number of temporal steps.
This approach allows one to combine a sequence of quantum circuits that simulate short-time intervals into a longer circuit for modeling long-time evolution without exponentially increasing the number of ancilla qubits.
The advantage of this algorithm over the QLSAs is its high success probability, but this QA also scales quadratically with the integrated time, $\sim t^2$.

References ~\onlinecite{Jin22Sch, Jin23Sch, Hu24Sch, Lu24} proposed the so-called ``Schr\"odingerization'' algorithm for mapping any linear PDE into a system of Schr\"odinger equations by extending the PDE to a higher extra dimension. 
This allows one to solve the original PDE by quantum Hamiltonian simulation methods.
A similar approach was proposed in Ref.~\onlinecite{An23, An23impr}, where it was called Linear Combination of Hamiltonian Simulations (LCHS), and we use this somewhat more specific terminology in this work.
In fact, it can be shown that the LCHS and Schr\"odingerization algorithms are identical (for instance, one can see Remark 3.1 in Ref.~\onlinecite{Jin24}).

In particular, consider a linear differential equation with a non-Hermitian generator $A$
\begin{equation}\label{eq:initial-value-problem}
    \partial_t \psi(t,x) = - A \psi(t,x).
\end{equation}
Separate $A$ into Hermitian matrices
\begin{subequations}\label{eq:A-decomposition}
\begin{eqnarray}
    &&A = A_L + \yi A_H,\\
    &&A_L = (A + A^\dagger)/2,\label{eq:Ah}\\
    &&A_H = (A - A^\dagger)/ (2\yi) \label{eq:Aa}.
\end{eqnarray}
\end{subequations}
In this case, LCHS represents the solution as a weighted superposition of unitary Hamiltonian evolution operators
\begin{align}\label{eq:LCHS-theorem1}
    &e^{- A t} = \int_\mathbb{R} \frac{1}{\pi(1+k^2)} e^{-\yi (A_H + k A_L) t}\diff k,
\end{align}
which is valid as long as $A_L$, the Hermitian part of $A$, is positive semi-definite.\cite{An23}
Thus, the LCHS algorithm offers a natural interpretation of a dissipative linear evolution operator as a weighted sum of Hamiltonian evolution operators for different rescalings of time (which is equivalent to different rescalings of the Hamiltonian itself).

The LCHS algorithm has the advantage that, in the best case, the complexity of this method scales linearly with time, $\sim t$, just as is the case for classical time integration.
Another advantage of this method is its high success probability, which is independent of condition number.
This implies that amplitude amplification (AA), which must query the initial conditions many times, is not needed solely due to an increase in the simulated time interval.
Note that, if trotterization is used for time-integration, then the algorithm can easily be made second order in time, while if the optimal selector is used, the accuracy in time is only limited by the LCHS truncation error.

In this work, we propose an explicit quantum circuit for solving Eq.~\eqref{eq:LCHS-theorem1} and apply it to solving the non-conservative dynamics of the advection-diffusion equation and the UW discretization of the KvN equation.

\subsection{Block-encoding (BE)}
Block-encoding (BE) refers to a set of dilation techniques that are a central part of many state-of-the-art QAs, such as the QSP, the Quantum Singular Value Transformation (QSVT)\cite{Gilyen19}, and the LCHS, and is often the main source of their quantum speedup.
Usually, the BE oracles are considered black boxes, assuming that an efficient implementation can be provided by the user.
However, when one develops QAs for modeling specific problems, an efficient exact BE may not always be possible. 
Therefore, the explicit construction of the BE oracles must be included in the description and analysis of QAs. 

In most BE techniques,\cite{Clader22, Camps22, Camps23, Sunderhauf24} either the number of ancillary qubits grows logarithmically with the matrix size, or the number of gates grows linearly with the number of nonzero matrix elements.
However, by taking into account the structure of an encoded matrix, such as a banded or tridiagonal structure, one can reduce the number of ancillae and make it independent of matrix size.\cite{Novikau24-EVM} 
Additionally, one can significantly reduce the BE depth by encoding groups of matrix elements simultaneously instead of addressing each element separately.
As shown in Refs.~\onlinecite{Novikau22, Novikau23}, one can make the number of gates in a BE oracle scale logarithmically with the encoded matrix size, keeping the spectral norm of the block-encoded matrix close to one. 
A similar approach is also used in Ref.~\onlinecite{Liu24BE}.

We should note that the main difference between the various BE techniques lies mainly in the meaning attributed to ancillary qubits: either they encode column or row matrix indices, or they save the element positions relative to the matrix diagonal.
The choice between these two encoding techniques has a drastic effect on both the BE circuit depth and the number of required ancillary qubits.
 
The matrices that naturally appear in many problems usually contain elements whose values change gradually with the matrix row index.
For instance, in the case of the KvN operator, the elements store nonlinear functions.
To block-encode these nonlinear functions, one can use the signal-processing-based algorithms such as the QSVT or QETU\cite{Dong22} methods.
We dedicate a significant part of this work to the detailed description of the BE oracles in Appendices~\ref{app:be} and~\ref{app:be-kvn}.

\subsection{Outline}
This paper is organized as follows.
In Sec.~\ref{sec:main-results}, we summarize the main results of this paper and the complexity of the proposed algorithms.
In Sec.~\ref{sec:kvn}, we review the KvN technique for turning a nonlinear differential equation into a linear Schr\"odinger equation.
In Sec.~\ref{sec:LCHS-circuit}, we present the LCHS algorithm following Ref.~\onlinecite{An23}, discuss how different parts of the LCHS can be implemented, and propose an explicit LCHS quantum circuit.
In Sec.~\ref{sec:LCHS-complexity}, the scaling of the complexity of the algorithm is discussed.
In Sec.~\ref{sec:LCHS-simulations}, we perform LCHS simulations of linear and nonlinear dissipative systems.
In Sec.~\ref{sec:conclusions}, we present the main conclusions.
In the Appendices, we discuss the Linear Combination of Unitaries (LCU) with ancillary qubits (Appendix~\ref{app:LCU}), demonstrate how the success probability of LCHS is changed by AA on simple test cases (Appendix~\ref{app:num-LCHS-toy}), and provide detailed procedures for obtaining explicit BE oracles for the upwind KvN equation in Appendices~\ref{app:be} and~\ref{app:be-kvn}.

\section{Main results}\label{sec:main-results}

\begin{table*}
\caption{\label{table:res-scaling} 
The query complexity of the LCHS-based algorithm described in this paper for various cases. 
The LCHS approximation error $\varepsilon_{\rm LCHS}$ is determined according to Eq.~\eqref{eq:err-LCHS-scaling}.
The matrices $A_L$ and $A_H$ are defined in Eq.~\eqref{eq:A-decomposition}, $p$ is the trotterization order, $||A_C||$ is defined in Eq.~\eqref{eq:AC-p-trot}, and $||\bar{C}_{\rm max}||$ is defined in Eq.~\eqref{eq:Cmax-bar}.
The term $\xi_{\rm QSP}$ is related to the QSP error according to Eq.~\eqref{eq:QSP-scaling}, and $\varepsilon_w$ is the QSVT error of the computation of the LCHS weights described in Sec.~\ref{sec:Ow}.
Here, $Q_w = \varepsilon_{\rm LCHS}^{-3/2}\varepsilon_w^{-1}$ is the complexity due to the QSVT computation of the LCHS weights.
}
\begin{ruledtabular}
\begin{tabular}{ll}
 & \\
non-commuting $A_L$ and $A_H$ (with trotterization, Sec.~\ref{sec:selector-scaling}) 
        &$\oO\left( \frac{||\psi(0)||}{||\psi(t)||} \left[||A_H|| t + \frac{||A_L|||}{\varepsilon_{\rm LCHS} } t + \xi_{\rm QSP}   \left(\frac{||A_C||\, ||A_L|| t^2}{\varepsilon_{\rm LCHS}^3}\right)^{1+1/p} + Q_w \right]\right)$  \\
 & \\
\hline
 & \\
commuting $A_L$ and $A_H$ (Sec.~\ref{sec:LCHS-scaling-zero-commutator})                       
        &$\oO\left( \frac{||\psi(0)||}{||\psi(t)||} \left[ 
         ||A_H ||t + \left(1+\frac{\xi_{\rm QSP}}{\varepsilon_{\rm LCHS} }\right) \frac{||A_L || }{ \varepsilon_{\rm LCHS} }  t
        + Q_w \right]\right)$    \\
 & \\
\hline
 & \\
optimal selector (Sec.~\ref{sec:selector-optimal})                      
        &$\oO\left( \frac{||\psi(0)||}{||\psi(t)||} \left[\frac{||\bar{C}_{\rm max} ||}{\varepsilon_{\rm LCHS}} t + \xi_{\rm QSP} + Q_w \right]\right)$    \\
 & \\
\end{tabular}
\end{ruledtabular}
\end{table*}

\subsection{Key results}
In this work, we construct an explicit QA for simulating the KvN dynamics of a nonlinear ODE using the LCHS method for time integration.
The proposed combined KvN-LCHS method can be applied to a wide class of nonlinear differential equations that are posed as first order in time where only BE is the problem-specific part of the algorithm.
Moreover, the same approach based on the LCHS can be used for solving linear advection and Liouville equations, as well as the KvH equation.

First, the nonlinear ODE is recast into an effective linear KvN Schr\"odinger equation where the nonlinearity is hidden within the KvN Hamiltonian and the temporal evolution of the corresponding statevector remains linear.
Second, to get rid of numerical artifacts or spurious echos unavoidable for standard finite difference approximations of the advection operator,\cite{Dodin21, Lin22Koopman} we discretize the KvN Hamiltonian with the UW scheme thus turning it into a non-Hermitian operator.
Third, to simulate the resulting dissipative problem, we use the LCHS\cite{An23} algorithm which is a state-of-art method with optimal scaling with respect to the simulated time and the number of calls to the initialization oracle. 
We provide an explicit quantum circuit for the LCHS algorithm and analyse the query complexity of the overall algorithm.
Our LCHS implementation is verified by modeling a linear advection-diffusion equation (ADE) on a numerical emulator of fault-tolerant digital quantum computers.
Then, we explain how to perform the BE for the KvN equation using QSVT and quantum arithmetic circuits.
For our nonlinear test case, we provide BE oracles with $\oO(1)$ ancillae and $\oO(\log_2 N_x)$ gate complexity, using the technique introduced in Refs.~\onlinecite{Novikau22, Novikau23, Novikau24-EVM}.
Finally, the KvN approach is combined with the LCHS algorithm and the BE oracles to simulate the nonlinear test problem formulated as a linear dissipative Schr\"odinger equation. 
The numerical simulations are performed in the framework QuCF\cite{QuCF} based on the toolkit QuEST.\cite{Jones19}

In Table~\ref{table:res-scaling}, we summarize the query complexity of our LCHS-based algorithm for various cases; this corresponds to the complexity of time integration. 
The first method that we explore in detail relies on trotterization of the Hermitian and anti-Hermitian parts of the linear system within the LCHS algorithm.
This method uses fewer ancilla qubits, which makes it easier to simulate classically and the complexity is linear in time for a reasonable range of time scales.
However, the  complexity for large time scales is eventually dominated by the term $\oO(\xi_{\rm QSP} t^{2+2/p})$ in the LCHS scaling (Sec.~\ref{sec:selector-scaling}).
Here $p$ is the order of trotterization accuracy, so that the complexity eventually scales as $t^3$ for a second-order accurate trotterization.
This term disappears when the matrices $A_H$ and $A_L$ are commuting (Sec.~\ref{sec:LCHS-scaling-zero-commutator}).
Finally, we show that, in general, one can use LCU to combine the Hermitian and anti-Hermitian parts of the matrix in an optimal implementation of the selector. 
This reformulation of the LCHS equation~\eqref{eq:LCHS-theorem1}, using the simple change of variables given in Eq.~\ref{eq:LCHS-theorem1-ksinphi}, allows one to construct a near-optimal LCHS circuit with linear scaling with time (Sec.~\ref{sec:selector-optimal}).

\subsection{Overall Complexity}\label{sec:overall-complexity}
First, we consider the complexity of classical and quantum Eulerian algorithms.
For this, one must take into account the cost of both the spatial representation and the time integrator. 
Consider a grid of $\Nx=2^{\nx}$ points in each of $\dim$ spatial directions and $\Nt=2^{\nt}$ time steps for the integration in the time interval $t$.
According to the Courant-Friedrichs-Lewy (CFL) convergence condition, the number of time steps scales as $\oO(||A t||)$. Here, $||A t||=\oO(2^{\cfl n_x})$ for the fixed interval $t$ where $\cfl=1$ for advection and $\cfl=2$ for diffusion.
Thus, the overall classical cost is 
\begin{equation}
    C_{\rm Eu} \sim \oOrder{||A t|| 2^{\nx\cporder\dim} },
\end{equation}
where each time step could require a sparse update with $\cporder=1$, or it could require matrix inversion with $\cporder=3$ if Gaussian elimination is used.

Quantumly, the cost in space  $\oOrder{\poly{(\nx \dim)}}$ is only logarithmic, as long as an efficient BE of the Eulerian problem can be provided, and leads to an exponentially large savings in memory and computational time.
The overall quantum cost in integration time scales as $\oOrder{||A t||^\tporder}$.
Here, $\tporder=1$ for the QSP\cite{Low19}, LCHS\cite{An23, An23impr} (Sec.~\ref{sec:selector-optimal}), or the most advanced QLSA implementations,\cite{Ambainis12, Chakraborty19, Costa21} and $\tporder=2$ for the time-marching method\cite{Fang23} or QSVT\cite{Gilyen19} used as a QLSA.
Thus, the overall quantum cost in time and space is 
\begin{align}
    Q_{\rm Eu} =  \oOrder{||A t||^\tporder \poly{(\nx \dim)} }.
\end{align} 
In this case, the speedup factor $S_{\rm QC}$ becomes
\begin{equation}\label{eq:Csp-Eu}
    S_{\rm QC} = \frac{C_{\rm Eu}}{Q_{\rm Eu}} = \oO\left(
    \frac{||A t||^{1-\tporder} 2^{\cporder \nx \dim}}{\poly{}(\nx \dim)} \right).
\end{equation}
Thus, for the LCHS algorithm, which has $\tporder = 1$ (Sec.~\ref{sec:selector-optimal}), the speedup factor grows exponentially with $n_x\dim$ and does not depend on $\cfl$.

Considering the fact that the temporal complexity must grow with the condition number of $A$, the overall quantum speedup is ultimately polynomial at best.
If the total number of classical time steps is parameterized as $N_t=2^{\cfl\nx+n_{t0}}$, then the polynomial speedup index $\ypqc$ (Sec.~\ref{sec:qu-adv}) is
\begin{align}
    \ypqc = \frac{ \cfl\nx + n_{t0} + \nx\cporder\dim}{\tporder(\cfl\nx + n_{t0})}
    \leq 1 + \frac{\cporder\dim}{\cfl + n_{t0}\nx^{-1}},
\end{align}
where the upper bound follows from the optimal value of $\tporder=1$.
Clearly, the speedup index increases with spatial resolution $n_x$, spatial dimension $\dim$, and polynomial index $\cporder\geq1$.

\subsection{Comparison to Lagrangian PIC-MC Approaches}
Classical Lagrangian PIC-MC algorithms excel over Eulerian methods for high dimensional problems where the solution is not smooth and the location of the discontinuities is not known.\cite{HeinrichNovak01arxiv}  
The PIC-MC approach is to evolve the classical trajectories and to then bin the resulting PDF in real space and/or phase space as needed.  However, this requires iterating many nonlinear operations, which due to the no-cloning theorem, does not appear possible to perform efficiently using quantum computers.  
A classical PIC-MC algorithm costs $C_{MC}\sim \Nt \Ns$, where $\Ns$ is the number of samples.
In good scenarios, the error converges as the relative standard deviation, $\epsilon_{MC}=\sigma_{MC}/\Ns^{1/2}$, which leads to the overall cost $C_{MC}\sim\Nt (\sigma_{MC}/\epsilon_{MC})^2$, where $\sigma_{MC}^2$ is the variance in the measured values of the observable of interest.

Quantum Lagrangian PIC-MC algorithms for sampling the PDF also excel for high-dimensional problems and non-smooth solutions. \cite{HeinrichNovak01arxiv}
In fact, one could consider proposing a QA that emulates a CA for evolving the trajectories, in a manner similar to the algorithm proposed in Ref.~\onlinecite{DongAn21MC} for simulating SDEs that accelerates classical Monte Carlo algorithms using the techniques in Ref.~\onlinecite{Montanaro2015}. 
Directly sampling the probability in phase space has the same complexity as the classical PIC-MC algorithm. 
However, if the data can be encoded efficiently into the amplitude of a quantum state, then AE allows one to quadratically improve this scaling of the error to $\epsilon_{MC} = \sigma_{AE}/N_s$. 
In this case, the quantum cost has quadratically improved scaling with error $Q_{MC}\sim \Nt \sigma_{AE}/\epsilon_{MC}$.
 
Classical Lagrangian PIC-MC methods can also be exponentially faster than classical Eulerian methods for the specific task of estimating a robust observable.
Yet, for these problems, by using quantum AA techniques, both the quantum Eulerian KvN method  based on LCHS explored in detail in this work and a quantum emulator of a classical Lagrangian method can generically achieve up to a quadratic speedup factor \cite{HeinrichNovak01arxiv,Montanaro2015} over classical  PIC-MC algorithms. 
A key advantage of the quantum Eulerian KvN method over the classical emulator is that, by simulating the PDF directly, it simulates many trajectories in parallel,
and, as explained above, the quantum Eulerian KvN method has an exponential speedup factor [Eq.~\eqref{eq:Csp-Eu}] over a classical Eulerian algorithm.
In principle, the Eulerian method also has the potential for much higher accuracy in regions of low probability, but the challenge remains in effectively extracting this information with measurements.
\def\Ham{H}
\def\Lag{L}

\def\phat{{\hat P}}
\def\xhat{{\hat X}}

\section{Koopman--von Neumann (KvN) approach}\label{sec:kvn}

\subsection{Overview}
The KvN approach\cite{Koopman31, Joseph20} transforms the nonlinear dynamics of a variable $x(t)$ into the linear dynamics of a state vector $\psi(t,x)$ of the operator $\xhat$.
The spectrum of the latter determines the possible values of the original variable $x(t)$. 
The dynamics of $\psi(t,x)$ is described by a Schr\"odinger equation on an infinite Hilbert space.
Hence, the KvN formulation describes the propagation of the probability density $\rho = |\psi|^2$ in time similarly to the Liouville equation which evolves $\rho$ directly.

The KvN Schr\"odinger equation is characterized by a Hermitian Hamiltonian.
After the discretization and truncation of the Hilbert space, which must be done in order to perform actual numerical simulations, the KvN Hamiltonian can still be kept Hermitian.
This allows one to use standard QAs for Hamiltonian simulation such as the QSP\cite{Low17} for modeling Hamiltonian evolution.
However, the resulting unitary evolution is accompanied by numerical artifacts detrimental for the efficient measurement of classical information from quantum computations.\cite{Dodin21, Lin22Koopman}
Because of this growing numerical noise, the KvN unitary simulations can be useful only for relatively short time intervals.

To avoid this problem, the KvN Hamiltonian can be made non-Hermitian by introducing any form of
dissipation that acts to smooth oscillations.
Yet, in this case, it is necessary to apply more complex QAs designed for simulating nonunitary dynamics.
For example, QAs such as the LCHS\cite{An23} or Schr\"odingerisation\cite{Jin23Sch} can be used.
It is important to point out that the same algorithms can be applied to solve the advection, Liouville, and KvH equations, which are ultimately equivalent to solving the KvN Schr\"odinger equation.
These QAs have an exponential speedup factor on fault-tolerant quantum digital machines in comparison to classical Eulerian discretization as was discussed in Sec.~\ref{sec:overall-complexity}.

\subsection{KvN Hamiltonian}
Let us consider a nonlinear dynamical system 
\begin{equation}\label{eq:dyn-auto}
    \ydd{x}{t} = F(t,x(t))
\end{equation}
defined for a set of real-valued coordinates $x$ with the initial condition $x(t=0) = x_0$.
To solve Eq.~\eqref{eq:dyn-auto} on a quantum computer, we reformulate the problem using the KvN approach.\cite{Joseph20}
Rather than directly considering the evolution of the variable $x$, 
one considers the evolution of a wavefunction $\psi(t,x)$ that defines the probability distribution function (PDF) $\rho =\left|\psi(t,x)\right|^2$ for any time instant $t$. 
Hence, one introduces the operator $\xhat$, defined as multiplication by $x$, and the conjugate momentum operator, defined as $\phat := -\yi\partial_x$ which acts on all other operators to its right.
For simulating classical dynamics, one can choose the state vector $\psi(t,x)$ to satisfy the linear Schr\"odinger equation:
\begin{equation}\label{eq:kvn}
    \yi\partial_t\psi(t,x) = \yham\psi(t,x),
\end{equation}
where the KvN Hamiltonian is defined as \cite{Joseph20}
\begin{equation}\label{eq:kvn-operator}
    \yham(t,\xhat,\phat) = \frac{1}{2} \left[\phat \cdot F(x) + F(x)\cdot \phat\right].
\end{equation}
More explicitly, the KvN Hamiltonian acting on $\psi(t,x)$ is
\begin{equation}\label{eq:kvn-h}
    \yham\psi = -\frac{\yi}{2} [\partial_x \cdot ( F\psi) + F\cdot \partial_x \psi].
\end{equation}

\subsection{Koopman - von Hove (KvH) Hamiltonian}
If the equations of motion are themselves Hamiltonian, then one can locally define a set of canonical coordinates $x=\{q,p\}$, consisting of generalized coordinates, $q$, and canonical momenta, $p$.
Then, the velocity $F$ corresponding to Hamiltonian $\Ham(t,q,p)$ must take the canonical form 
\begin{align}
    F^q(t,q,p)=\partial_p\Ham && F^p(t,q,p)=-\partial_q\Ham .
\end{align}
Since the velocity is divergence free, $\partial_qF^q+\partial_pF^p=0$, the KvN, advection, and Liouville equations 
\begin{equation}\label{eq:kvn-general}
    \yham\psi = \left\{\Ham,\psi\right\} := (\partial_q\Ham) \cdot (\partial_p\psi) - (\partial_p \Ham)\cdot (\partial_q \psi)
\end{equation}
are equivalent.
Writing the equation  in terms of the Poisson bracket $\{\cdot,\cdot\}$ allows one to generalize to Hamiltonian dynamics in an arbitrary coordinate system and to Hamiltonian dynamics with a degenerate Poisson bracket. \cite{Joseph20}  

Because the wavefunction is complex, in general, one can add a term to the KvN equation that determines the phase evolution. 
This generalized KvN equation simply adds a function of $q$ and $p$ to the KvN Hamiltonian. \cite{Joseph20}
As proven in Ref.~\onlinecite{Joseph23JPA}, the correct semiclassical limit of the Schr\"odinger equation can be obtained if the phase changes with the classical action along the path.
This results in the KvH Hamiltonian
\begin{align} \label{eq:KvH}
\yham^{\rm KvH} := \yham  - \Lag(t,q,p),
\end{align}
where the classical Lagrangian
\begin{align}
\Lag(t,q,p) = p\cdot\dot q - \Ham(t,q,p)
\end{align}
is defined as a function of $q$ and $p$ via the substitution $\dot q(t,q,p)=\partial_p \Ham$.

While the KvH Hamiltonian, Eq.~\ref{eq:KvH}, is quite interesting from the physical point of view, 
in this work, we primarily focus on the KvN Hamiltonian, Eq.~\ref{eq:kvn-h}, for two overarching reasons.
Obtaining a high accuracy algorithm is most difficult when the dynamics contains strictly stable or unstable fixed points.\cite{Dodin21, Lin22Koopman}
The issues are related to the discretization of the advection operator in the KvN Hamiltonian and can be studied clearly and simply using a one-dimensional dissipative system with stable fixed points.
In this work, we solve this problem using the UW version of these operators, which makes the KvN (or KvH) Hamiltonian non-Hermitian.
For a full implementation, this requires the use of the LCHS algorithm as well as an explicit block encoding of the relevant advection operators, given in Appendix \ref{app:be-kvn}.

\subsection{Discretization of the KvN Hamiltonian}
The KvN formulation results in the Hamiltonian~\eqref{eq:kvn-operator} in an infinite Hilbert space.
To model Eq.~\eqref{eq:kvn}, the space should be truncated and discretized.
In the one-dimensional case, the discretized $\xhat$-operator has the following spectrum
\begin{equation}\label{eq:x-grid}
    \begin{split}
        &x_j = - x_{\rm max} + j h, \quad h = 2x_{\rm max}/(N_x-1),\\
        &N_x = 2^{n_x},\quad j=[0,N_x).
    \end{split}
\end{equation}
Here, the notation $[j_1,j_2)$ with integers $j_1$ and $j_2$ denotes the set of all integers from $j_1$ to $j_2$, including $j_1$ but excluding $j_2$.
After the discretization~\eqref{eq:x-grid}, where $x_j$ are a discrete set of possible values of the original single variable $x$, the functions $F$ and $\psi$ are now represented as vectors with $N_x$ elements.

Standard numerical approaches to 
discretization of the momentum operator $\phat$ or, equivalently, of the derivative $\partial_x$ can lead either to a Hermitian KvN Hamiltonian or a non-Hermitian one.
As noticed in Ref.~\onlinecite{Dodin21, Lin22Koopman}, a unitary formulation of an originally dissipative problem that is based on standard finite difference techniques is typically accompanied by numerical artifacts that corrupt statistics and make extraction of useful classical information from quantum computations difficult at sufficiently long times.
To eliminate these artifacts and significantly improve the statistics, we introduce dissipation into the KvN model by using a UW discretization scheme of the first-order advection operator.
In the next section, we consider both unitary and nonunitary discretizations of these advection terms.

From a numerical perspective, it is reasonable to add dissipation to the KvN or KvH Hamiltonians directly.
However, from a physical perspective, dissipation should generate a mixed state, and, hence, it might be more appropriate to work directly with the quantum Liouvillian, i.e. with Lindblad operators providing dissipation. \cite{Lindblad76cmp,Gorini76jmp,KraussBook,Manzano20aipa}
However, this is equivalent to doubling the dimension of the problem, requiring even greater computational resources, and, hence, is well beyond the scope of work here. 
Thus, we leave the exploration of the answers to these questions to future work.

For a multidimensional system, like a Hamiltonian system specified by Hamiltonian $\Ham(t,q_j,p_j)$, one would generalize this procedure by introducing multiple coordinates, e.g. with values $q_{j,r}$ and $p_{j,r}$ that are specified as a function of the index $r$ along the diagonal.
The discretization of the KvH Hamiltonian~\ref{eq:KvH} is also straightforward.
One simply uses a valid discretization for KvN and then subtracts the discretized Lagrangian, which is a local operator, from the KvN Hamiltonian.
For example, for finite difference schemes, one would simply subtract the local value of the Lagrangian along the diagonal, $\Lag(t,q_{j,r},p_{j,r})$.

\subsubsection{Central finite difference scheme}\label{sec:kvn-cfd}
In order to achieve a perfectly unitary evolution operator, one can use the central difference (CD) scheme for the discretization of the KvN Hamiltonian~\eqref{eq:kvn-h}:
\begin{equation}
    \yi\partial_t\psi_k = -\frac{\yi}{4h}\bigg[(F_{k+1}\psi_{k+1} - F_{k-1}\psi_{k-1}) + F_{k} (\psi_{k+1} - \psi_{k-1})\bigg],
\end{equation}
where $F_k = F(x_k)$.
By rearranging terms, one gets:
\begin{equation}
    \yi\partial_t\psi_k = -\frac{\yi}{4h}\bigg[(F_{k+1} + F_{k})\psi_{k+1} - (F_{k-1} + F_{k})\psi_{k-1}\bigg],
\end{equation}
where the nonvanishing elements of the resulting Hermitian Hamiltonian are
\begin{equation}\label{eq:kvn-h-cfd}
    \yhamc_{r,c} = -\frac{\yi}{4h}
    \left\{ \begin{aligned}
        -(F_{r} + F_{c})&,\quad c = r - 1,\\
         (F_{r} + F_{c})&,\quad c = r + 1,
    \end{aligned}\right.
\end{equation}
where $r = [0, N_x)$ and $c = [0, N_x)$.

If the equations are Hamiltonian, then one can use a conservative multi-dimensional generalization of Eq.~\ref{eq:kvn-h-cfd}.
For example, in canonical coordinates, one can use the Arakawa bracket.\cite{Arakawa1966jcp,Arakawa1977jcp}
To generalize even further to the KvH Hamiltonian in Eq.~\ref{eq:KvH}, one simply subtracts the Lagrangian along the diagonal
\begin{equation}
\Delta \yham^{\rm KvH}_{r,r}(t)=-\Lag(t,q_{j,r},p_{j,r}).
\end{equation}
where $q_{j,r}$ and $p_{j,r}$ are the values of the coordinates along the diagonal.
This same form would be used for the cases below.

\subsubsection{Upwind (UW) finite difference scheme}\label{sec:kvn-uw}

\begin{figure}[!t]
\centering
\subfloat{\includegraphics[width=0.25\textwidth]{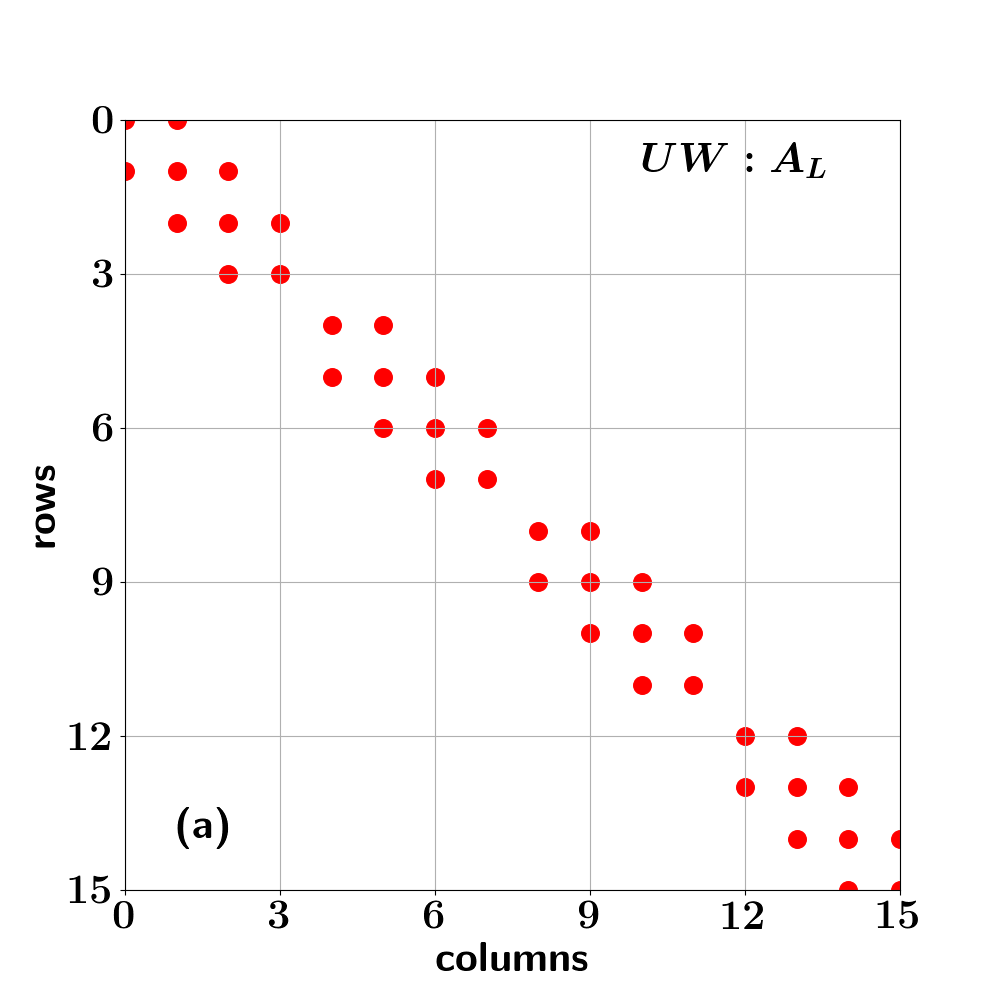}}
\subfloat{\includegraphics[width=0.25\textwidth]{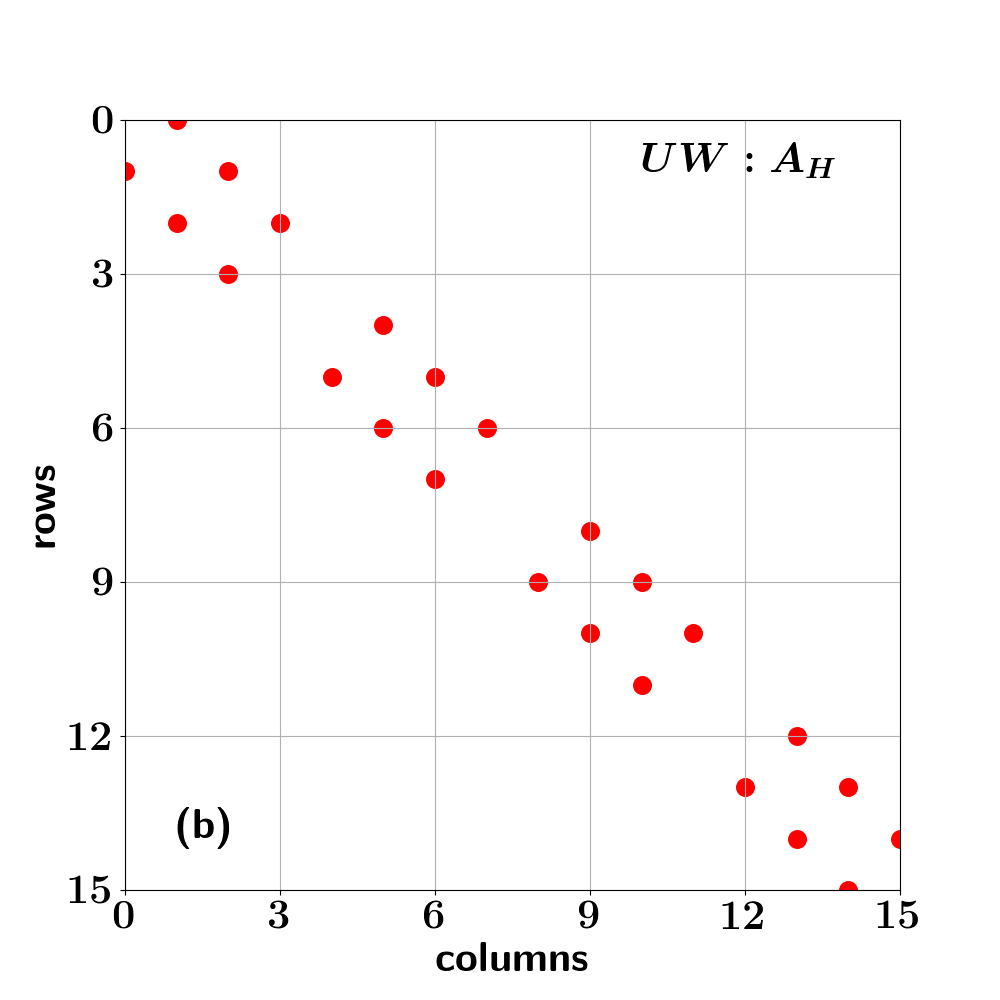}}
\caption{
    \label{fig:x2-matrices} 
    Schematic form of the matrices $A_L$ and $A_H$ computed using Eqs.~\eqref{eq:Ah-UW} and~\eqref{eq:Aa-UW}, respectively.
}
\end{figure}

In contrast, the majority of applications use the UW finite difference scheme where a forward or backward difference stencil is used depending on the sign of the local velocity $F_k$.
Due to Godunov's theorem,\cite{Godunov59} only first-order discretizations can preserve monotonicity, thus, this is the minimal amount of dissipation required to stabilize the oscillations.
If $F_k < 0$, then:
\begin{equation}
    \yi\partial_t \psi_k = -\frac{\yi}{2h}\bigg[- 2 F_k\psi_k + (F_{k+1} + F_k)\psi_{k+1}\bigg].
\end{equation}
If $F_k > 0$, then:
\begin{equation}
    \yi\partial_t \psi_k = -\frac{\yi}{2h}\bigg[- (F_{k-1} + F_k)\psi_{k-1} + 2 F_k\psi_k\bigg].
\end{equation}
Hence, the upwind KvN Hamiltonian is dissipative (non-Hermitian) and has the following nonzero elements:
\begin{equation}\label{eq:kvn-h-uw}
    \yhamu_{r,c} = -\frac{\yi}{2h}
    \left\{ \begin{aligned}
         2\xi_{r}F_{r}&,\quad c = r,\\
        -\xi_{r}(F_{r} + F_{c})&,\quad c = r - \xi_{r},
    \end{aligned}\right.
\end{equation}
where $c = [0, N_x)$, $r = [0, N_x)$, and 
\begin{equation}\label{eq:xi-sign}
    \xi_k = F_k/|F_k|.
\end{equation}

Since the UW Hamiltonian $\yhamu$ is not Hermitian, one cannot use standard QAs for Hamiltonian simulations like QSP\cite{Low17} for evolving the KvN Schr\"odinger equation~\eqref{eq:kvn} with $\yhamu$.
Instead, we use the LCHS algorithm which will be discussed in detail in the next section.
To apply the LCHS algorithm, one needs to specify the matrices $A_L$ and $A_H$ which can be found using Eqs.~\eqref{eq:Ah} and ~\eqref{eq:Aa}, respectively, with the matrix $A = \yi \yhamu$.
Using the Kronecker delta function
\begin{equation}\label{eq:delta-Kronecker}
   \delta_{k,l} = \left\{ \begin{aligned}
                &1,\quad k = l,\\
                &0,\quad k \ne l,
        \end{aligned} \right.
\end{equation}
one obtains the following explicit expression for the matrix nonzero elements of $A_H$
\begin{equation}\label{eq:Aa-UW}
    A_{H, r, c} = \frac{\yi}{4h}
        \left\{ \begin{aligned}
            - 2 \xi_{r} (F_{r} - F_{r}^\dagger),\quad &c = r,\\
            - \delta_{\xi_{r},-1} (F_{r} + F_{c}) 
                - &\delta_{\xi_{c},1} (F_{r}^\dagger + F_{c}^\dagger),\\
            &c = r + 1,\\
            \delta_{\xi_{r},1} (F_{r} + F_{c}) 
                + &\delta_{\xi_{c},-1} (F_{r}^\dagger + F_{c}^\dagger),\\
            &c = r - 1.\\
        \end{aligned}\right.
\end{equation}
The matrix nonzero elements of $A_L$ are
\begin{equation}\label{eq:Ah-UW}
    A_{L, r, c} = -\frac{1}{4h}
        \left\{ \begin{aligned}
            - 2 \xi_{r} (F_{r} + F_{r}^\dagger),\quad &c = r,\\
            - \delta_{\xi_{r},-1} (F_{r} + F_{c}) 
                + &\delta_{\xi_{c},1} (F_{r}^\dagger + F_{c}^\dagger),\\
            &c = r + 1,\\
            \delta_{\xi_{r},1} (F_{r} + F_{c}) 
                - &\delta_{\xi_{c},-1} (F_{r}^\dagger + F_{c}^\dagger),\\
            &c = r - 1.\\
        \end{aligned}\right.
\end{equation}
The structure of the matrices $A_L$ and $A_H$ is shown in Fig.~\ref{fig:x2-matrices}.
In the matrix $A_H$, the delta functions $\delta_{\xi_{r},-1}$ and $\delta_{\xi_{c},1}$ with $c = r + 1$, as well as $\delta_{\xi_{r},1}$ and $\delta_{\xi_{c},-1}$ with $c = r - 1$, do not overlap.
Because of this, only one of the sums $F_{r} + F_{c}$ or $F_{r}^\dagger + F_{c}^\dagger$ contributes to $A_{H,r,c}$ at each particular $r$.
The same is true for the matrix elements $A_{L,r,c}$.
In Eqs.~\eqref{eq:Aa-UW} and~\eqref{eq:Ah-UW}, the matrices are written in a general form for some complex function $F$. 
Yet, the KvN Hamiltonian is typically only used with real functions $F$.
Therefore, $F_k = F^\dagger_k$ for all $k$, and the main diagonal in $A_H$ is zeroed.

\subsubsection{Central finite difference scheme with a supplemental diffusivity}\label{sec:kvn-diff}
As an alternative to the UW scheme, one can use the CD scheme but with an explicit diffusivity in the KvN Hamiltonian.
For the sake of simplicity, we abbreviate this scheme as CDD (Central Difference with Diffusivity).
In this case, Eq.~\eqref{eq:kvn-h} becomes:
\begin{equation}\label{eq:kvn-h-diff}
    \yham\psi = -\frac{\yi}{2} \left[\partial_x(F\psi) + F \partial_x \psi \right] 
    +i D\partial^2_x\psi,
\end{equation}
where $D$ is a constant diffusivity.
The nonzero elements of the discretized KvN Hamiltonian with diffusivity and Dirichlet boundary conditions are
\begin{equation}\label{eq:kvn-h-diff-expl}
    \begin{aligned}
        &\yhamd_{r,c} = \\
        &-\frac{\yi}{2}\left\{ \begin{aligned}
             1&,\quad c = r = 0 \text{ and } N_x - 1,\\
             4D h^{-2}& , \quad c= r = [1,N_x-2], \\
             -\frac{2D}{h^2} \pm \frac{F_c+F_r}{2h} &, \quad c = r \pm 1,\, r = [1,N_x-2].
        \end{aligned}\right.
    \end{aligned}
\end{equation}

\subsubsection{UW discretization of the advection and Liouville equations}\label{sec:discr-Liouville-adv}
We should note that the technique described in this paper can also be easily applied to the advection equation:
\begin{equation}
    \partial_t \psi + F \partial_x \psi = 0,
\end{equation}
and the Liouville equation:
\begin{equation}
    \partial_t \psi + \partial_x (F \psi) = 0,
\end{equation}
which are widely used in fluid and kinetic simulations.
For completeness, we show here the discretized versions of these equations in the one-dimensional case.
The nonvanishing elements of the corresponding UW Hamiltonian for the advection equation are
\begin{equation}\label{eq:kvn-h-uw-A}
    \yhamuA_{r,c} = -\frac{\yi}{h}
    \left\{ \begin{aligned}
         \xi_{r}F_{r}&,\quad c = r,\\
        -\xi_{r}F_{r}&,\quad c = r - \xi_{r}.
    \end{aligned}\right.
\end{equation}
The nonvanishing elements of the corresponding UW Hamiltonian for the Liouville equation are
\begin{equation}\label{eq:kvn-h-uw-L}
    \yhamuL_{r,c} = -\frac{\yi}{h}
    \left\{ \begin{aligned}
         \xi_{r}F_{r}&,\quad c = r,\\
        -\xi_{r}F_{c}&,\quad c = r - \xi_{r}.
    \end{aligned}\right.
\end{equation}
In fact, from Eqs.~\eqref{eq:kvn-h-uw},~\eqref{eq:kvn-h-uw-A}, and~\eqref{eq:kvn-h-uw-L}, one can see that the UW Hamiltonian of the KvN Schr\"odinger equation is simply the average of the advection and Liouville Hamiltonians:
\begin{equation}
    \yhamu = \left(\yhamuA + \yhamuL\right)/2.
\end{equation}
For instance, the LCHS algorithm can be applied to modeling the Liouville equation in the same manner as it is used for simulating the KvN Schr\"odinger equation. 
The only difference with the KvN case will be a different BE of the matrices $A_L$ and $A_H$.
For the Liouville equation, the nonzero elements of these matrices are 
\begin{equation}\label{eq:Aa-UW-Liouville}
    A_{H, r, c}^{\rm Liouville} = \frac{\yi}{2h}
        \left\{ \begin{aligned}
            - \xi_{r} (F_{r} - F_{r}^\dagger),\quad &c = r,\\
            - \delta_{\xi_{r},-1} F_{c} - \delta_{\xi_{c},1} F_{c}^\dagger,\quad
            &c = r + 1,\\
            \delta_{\xi_{r},1} F_{c} + \delta_{\xi_{c},-1} F_{c}^\dagger,\quad
            &c = r - 1.
        \end{aligned}\right.
\end{equation}
and
\begin{equation}\label{eq:Ah-UW-Liouville}
    A_{L, r, c}^{\rm Liouville} = -\frac{1}{2h}
        \left\{ \begin{aligned}
            - \xi_{r} (F_{r} + F_{r}^\dagger),\quad &c = r,\\
            - \delta_{\xi_{r},-1} F_{c} + \delta_{\xi_{c},1} F_{c}^\dagger,\quad
            &c = r + 1,\\
            \delta_{\xi_{r},1} F_{c} - \delta_{\xi_{c},-1} F_{c}^\dagger,\quad
            &c = r - 1.
        \end{aligned}\right.
\end{equation}

\begin{figure*}[!t]
\centering
\subfloat{\includegraphics[width=0.36\textwidth]{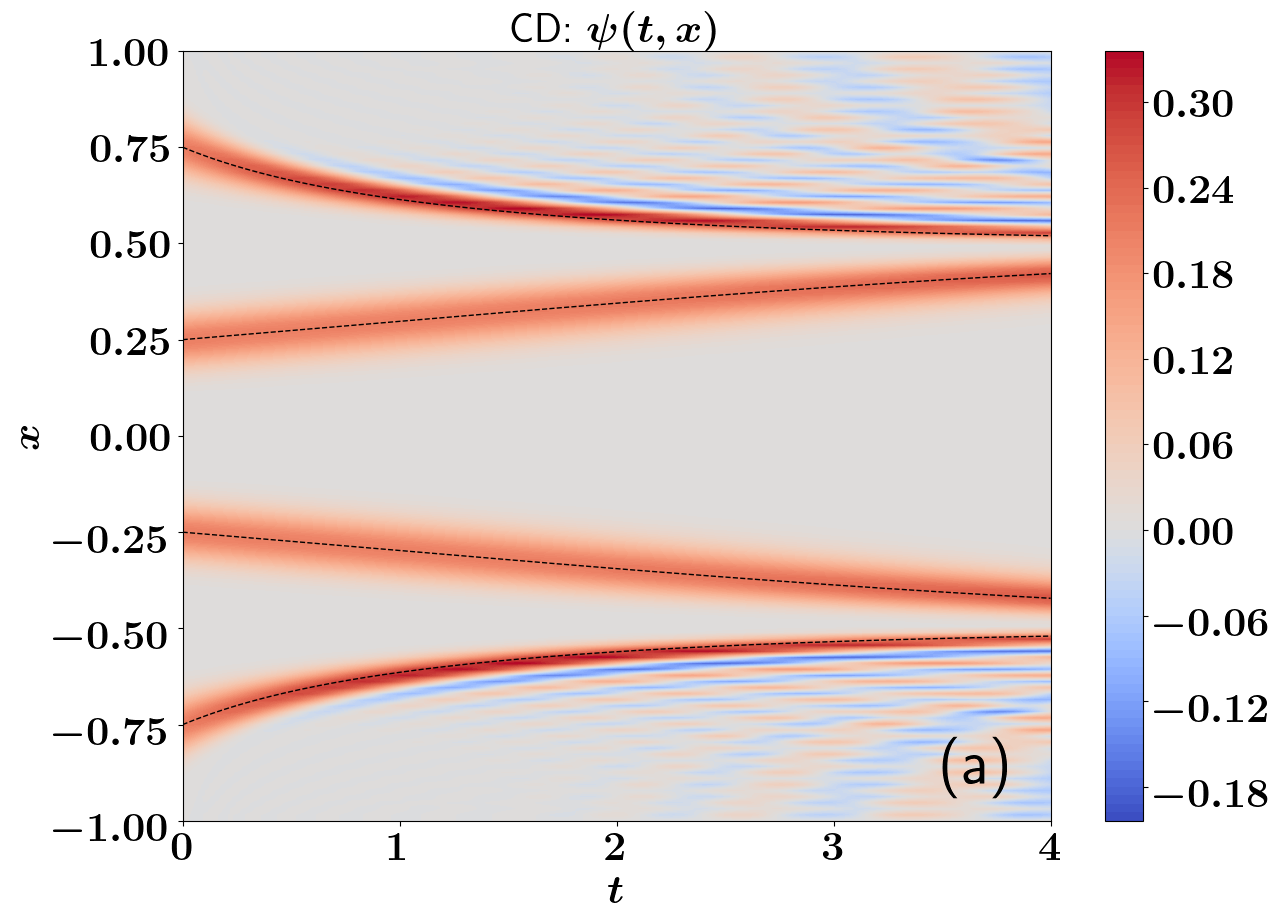}}
\subfloat{\includegraphics[width=0.36\textwidth]{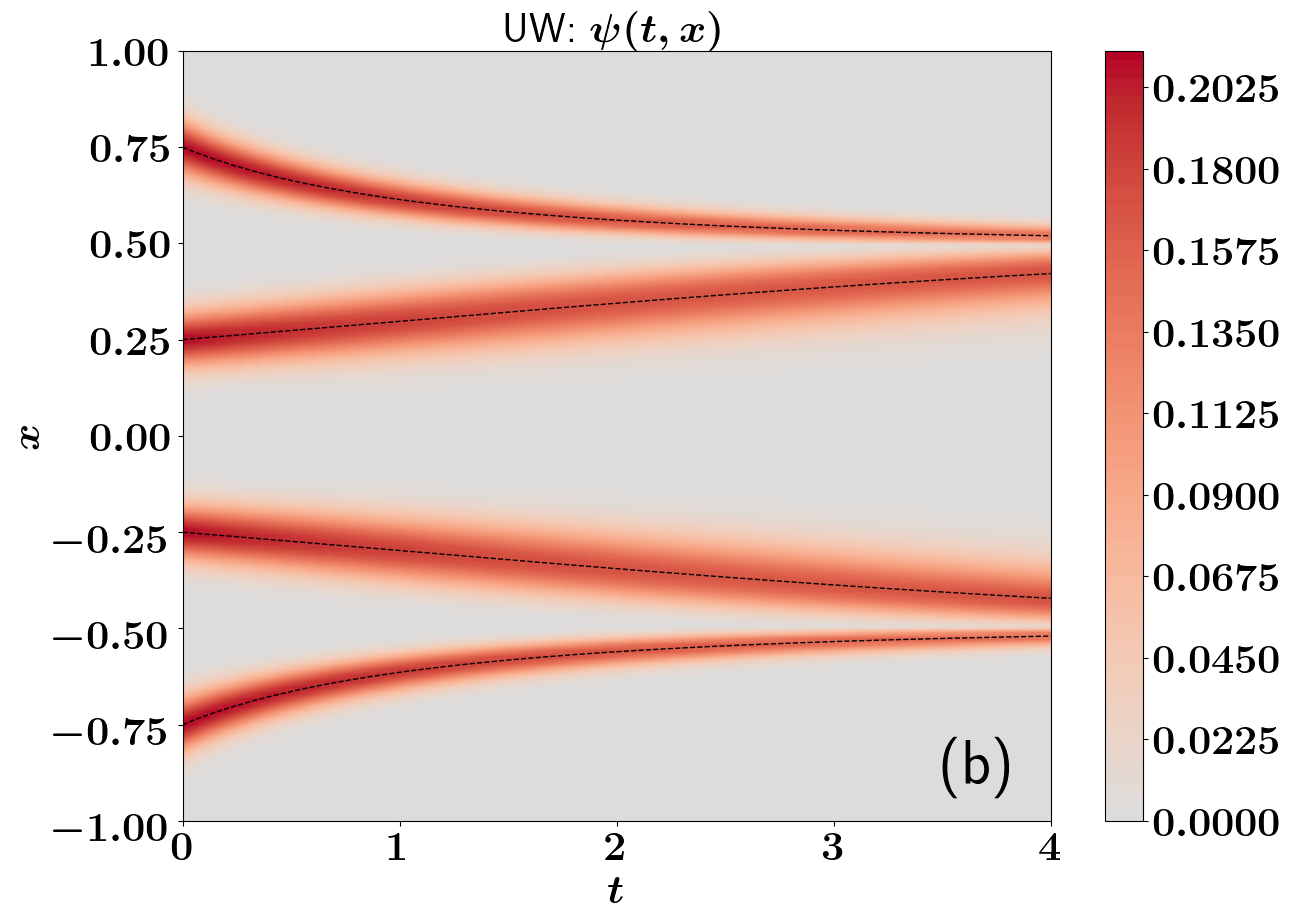}}
\subfloat{\includegraphics[width=0.295\textwidth]{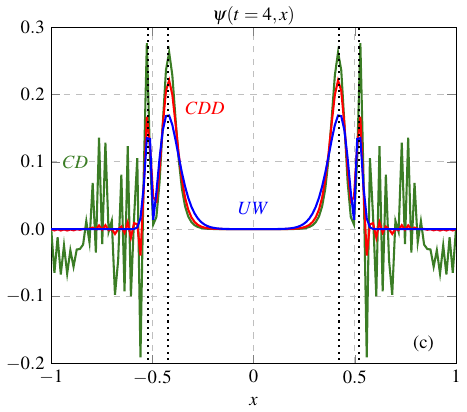}}
\caption{
    \label{fig:KvN-xt} 
    Contour plots showing the state vector $\psi(t,x)$ from the KvN simulations of Eq.~\eqref{eq:x2a2} performed on a classical computer with $n_x = 7$ and $n_t = 11$ with the CD scheme (a) and the UW scheme (b).
    The black dotted lines are the trajectories of the original system~\eqref{eq:x2a2} starting from different initial conditions~\eqref{eq:x0}. 
    (c): The plot showing the comparison between the CD (green solid line), UW (blue solid line) and CDD with $D = 10^{-4}$ (red solid line) schemes at $t = 4.0$. 
    Here, the black dotted vertical lines correspond to the actual $x$ at $t = 4.0$ computed using original equation~\eqref{eq:x2a2} with various initial conditions~\eqref{eq:x0}. 
}
\end{figure*}
\begin{figure*}[!t]
\subfloat{\includegraphics[width=0.33\textwidth]{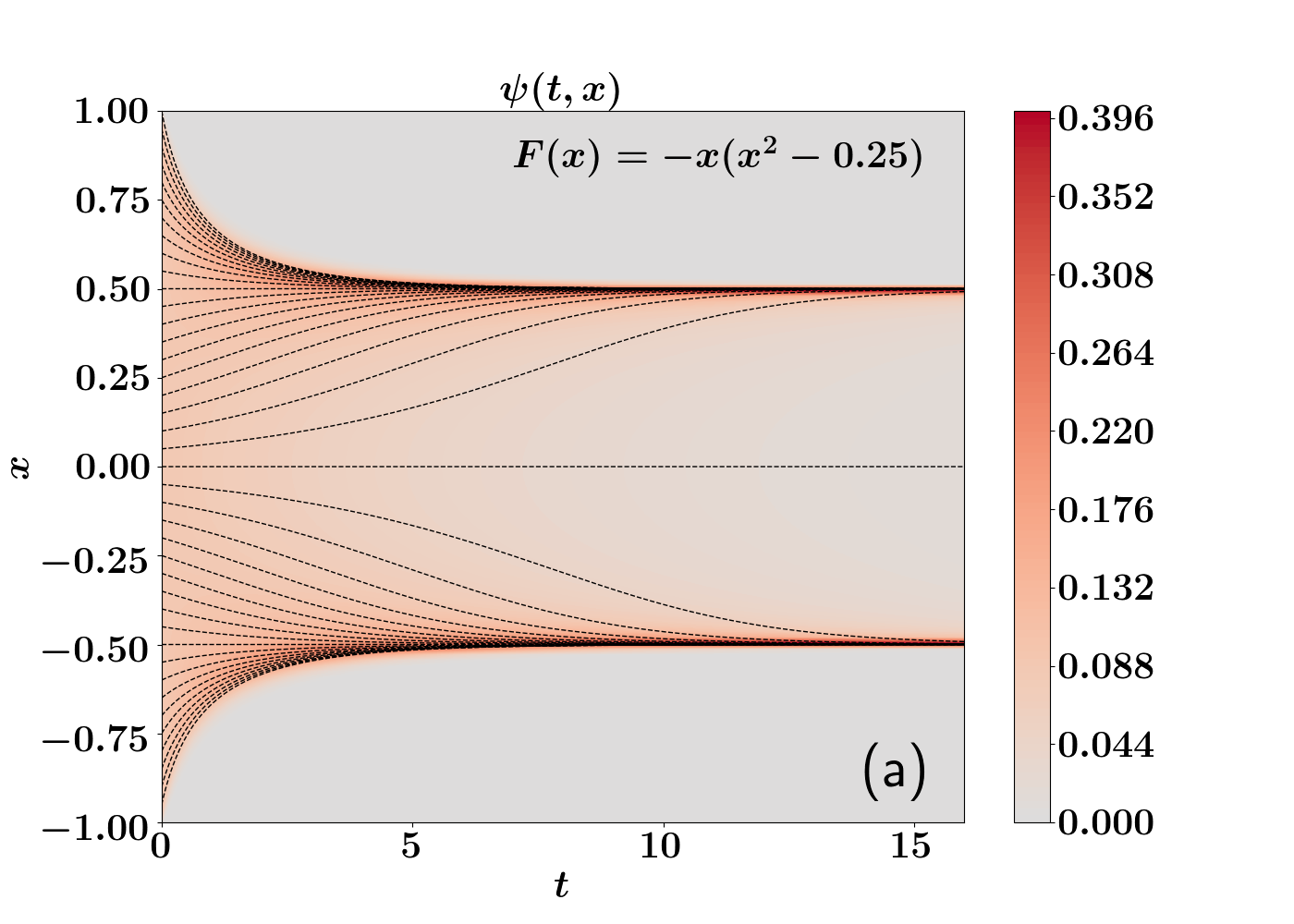}}
\subfloat{\includegraphics[width=0.33\textwidth]{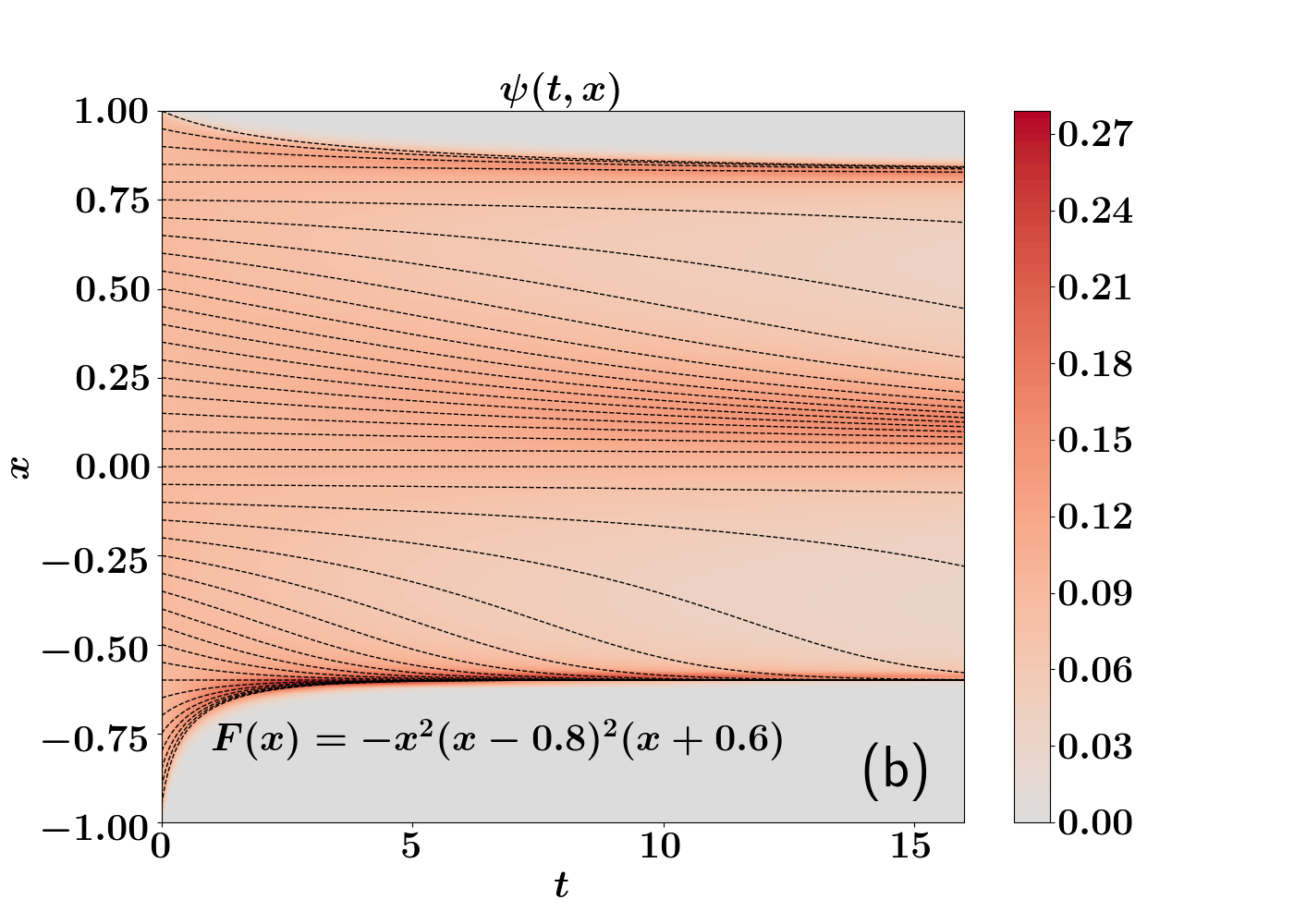}}
\subfloat{\includegraphics[width=0.33\textwidth]{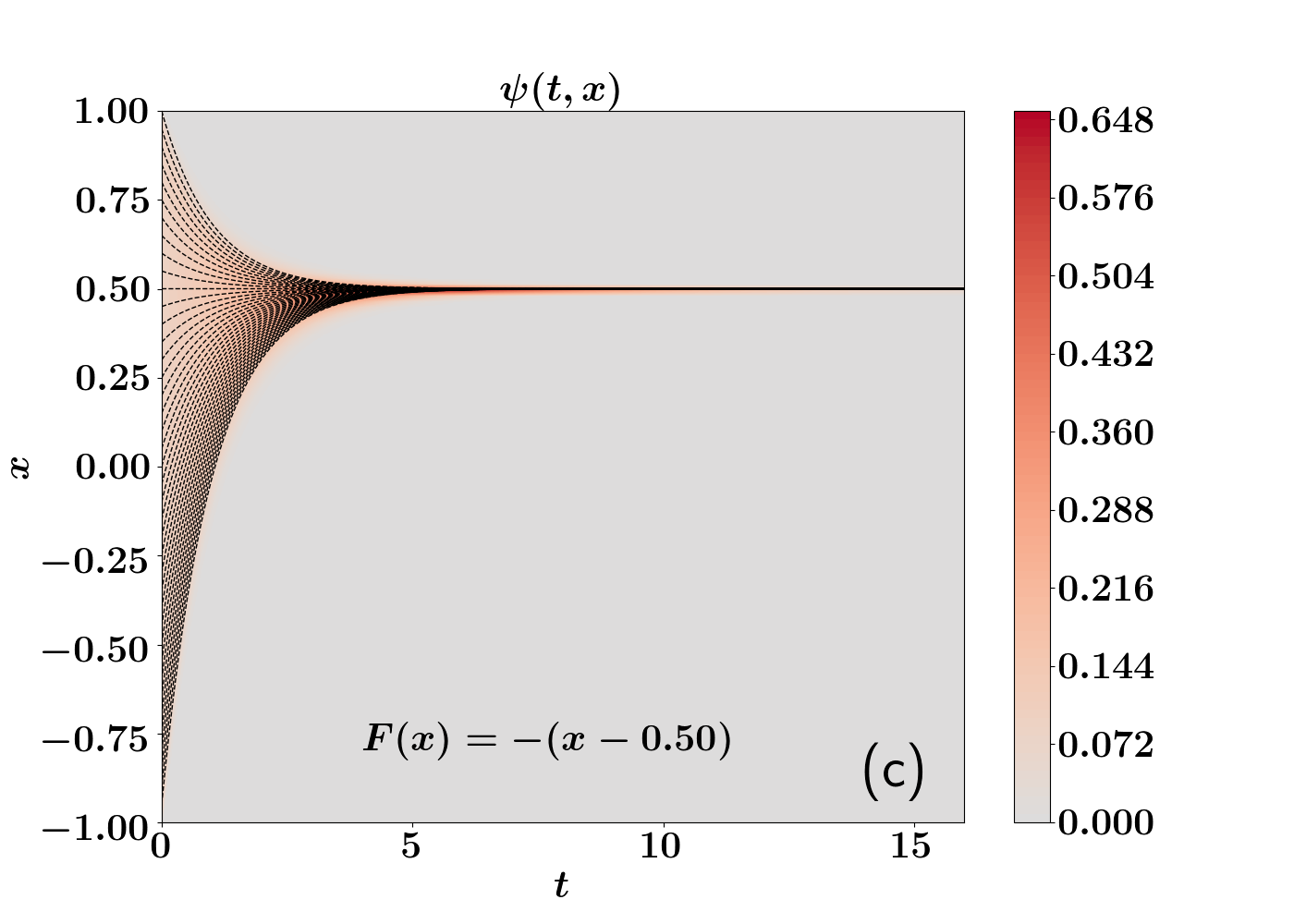}}
\caption{
    \label{fig:KvN-LONG} 
    Contour plots showing the state vector $\psi(t,x)$ from classical KvN simulations of UW-discretized nonlinear problems with (a) $F(x) = -x(x^2 - 0.25)$, (b) $F(x) = -x^2(x-0.8)^2(x+0.6)$, and (c) $F(x) = -(x-0.50)$.
    In all these cases, the initial condition is the uniform superposition $\psi(0,x) = 2^{-n_x/2}$.
    The dashed black lines are obtained from the direct integration of the original nonlinear systems using various initial conditions.
}
\end{figure*}

\subsection{Examples of initial conditions}\label{sec:kvn-init}
If equation~\eqref{eq:dyn-auto} has the initial condition $x(t=0) = x_0$, then the state vector can be initialized as $\psi_j(0) = G_j$ where $G_j$ is the Gaussian with the width $\sigma_0$ and centered at $x_0$:
\begin{equation}\label{eq:Gaussian}
    G_j = \frac{1}{\sqrt{2\pi} \sigma_0}\exp\ylb-\frac{(x_j-x_0)^2}{2 \sigma_0^2}\yrb.
\end{equation}
One can also initialize the state as a superposition of several Gaussians centered at various $x_0$.
In this case, a single computation of the KvN system provides information about multiple possible paths of the temporal evolution of $x(t)$ started from different initial conditions. 

As an alternative option, to guarantee fully-coherent initialization, one can start from a superposition of equally weighted states.
This corresponds to a flat initial profile
\begin{equation}\label{eq-psi-init-flat}
    \psi_k(0) = 2^{-n_x/2},
\end{equation}
which can be done by using just $n_x$ Hadamard gates.

\begin{figure}[!t]
\includegraphics[width=0.48\textwidth]{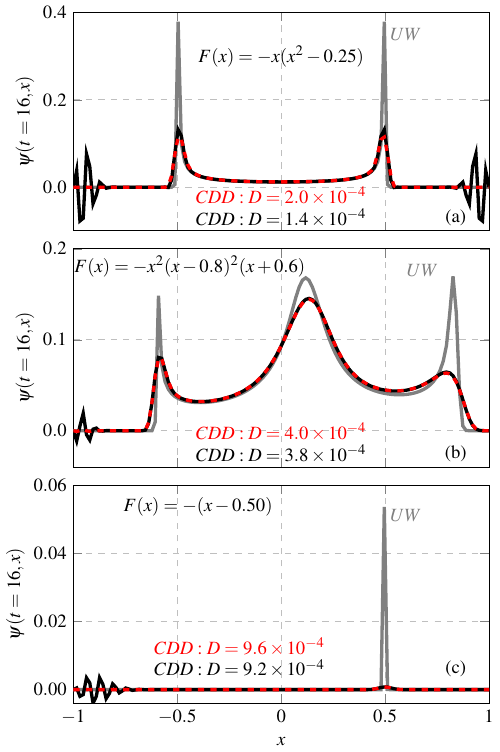}
\caption{
    \label{fig:KvN-LONG-t1} 
    Plots showing $\psi(t_1,x)$ at $t_1 = 16$ from classical KvN simulations of various nonlinear problems using the UW and CDD schemes.
}
\end{figure}

\subsection{Examples of classical KvN simulations}\label{sec:case-CL}
Once the nonlinear equation~\eqref{eq:dyn-auto} is formulated in the form of the discretized linear Schr\"odinger equation, the variable $x(t)$ is discretized onto the fixed grid~\eqref{eq:x-grid}.
Then, one constructs the KvN Hamiltonian using the preferred discretization scheme.
Using the discretized Hamiltonian, one performs the KvN simulation by integrating Eq.~\eqref{eq:kvn} with a chosen initial state vector $\psi(0,x)$.
As a result, the solution of Eq.~\eqref{eq:kvn} provides the temporal evolution of the state vector $\psi$ whose components $\psi_j(t) \equiv \psi(t,x_j)$ for $j = [0, N_x)$ are related to the probability to observe the variable $x(t)$ equal $x_j$ at the time instant $t$.  

Now, as an example, we consider the following nonlinear equation:
\begin{equation}\label{eq:x2a2}
    \ydd{x}{t} =  - x (x^2 - 0.25).
\end{equation}
This ODE can be easily integrated to find $x(t)$ by considering various initial conditions: 
\begin{equation}\label{eq:x0}
    x(t = 0) = -0.25,\ \ 0.25,\ \ -0.75,\ \ 0.75.
\end{equation}
Note that for each $x(t=0)$, one should perform a separate calculation of Eq.~\eqref{eq:dyn-auto}.
The results of these simulations are shown by black lines in Fig.~\ref{fig:KvN-xt}.a and~\ref{fig:KvN-xt}.b. 
One can see that the variable $x(t)$ evolves towards two fixed points at $\pm 0.5$.

Another option is to apply the KvN technique.
The KvN simulation is performed by integrating Eq.~\eqref{eq:kvn} with the initial state vector $\psi(t=0)$ encoding four Gaussian peaks as in Eq.~\eqref{eq:Gaussian} with $\sigma_x = 0.05$ centered at the points specified in Eq.~\eqref{eq:x0}.
The contour plot in Fig.~\ref{fig:KvN-xt}.a shows the evolution of the state vector using the CD Hamiltonian $\yhamc$ [Eq.~\eqref{eq:kvn-h-cfd}].
One can see that this unitary evolution is characterized by parasitic oscillations that grow in time and can eventually hide the evolution of the Gaussian wave packets.
For instance, at $t = 4$, the amplitudes of these numerical oscillations become comparable to the amplitudes of the Gaussians (Fig.~\ref{fig:KvN-xt}.c).
On the other hand, the nonunitary evolution described by $\yhamu$ [Eq.~\eqref{eq:kvn-h-uw}] and shown by the contour plot in Fig.~\ref{fig:KvN-xt}.b smooths this noise.
In both cases, the initialized Gaussians evolve consistently with the direct integration of Eq.~\eqref{eq:x2a2}.
In particular, one can see from Fig.~\ref{fig:KvN-xt}.c that, at $t = 4$, the state vector has the highest amplitudes at those $x_j$ which are close to the values $x(t)$ (indicated by vertical black dotted lines) obtained by the direct integration of Eq.~\eqref{eq:x2a2}.

As discussed in Sec.~\ref{sec:kvn-init}, one can start from a flat initial state instead of a superposition of Gaussians.
In this case, the KvN evolution will still find all fixed points in the simulated system as one sees from Fig.~\ref{fig:KvN-LONG} where different nonlinear ODEs are considered.
Clearly, by making an appropriate choice of $F(x)$, evolving the KvN Hamiltonian can be used as a nonlinear equation solver, aka root-finder, or as an optimizer.

The comparison between the UW and the CDD schemes is shown in Fig.~\ref{fig:KvN-xt}.c and Fig.~\ref{fig:KvN-LONG-t1}.
The CDD significantly reduces the numerical oscillations and correctly localizes fixed points.
However, over longer time intervals, the CDD not only damps the numerical artifacts but may also significantly reduce the useful signal. 
According to Fig.~\ref{fig:KvN-LONG-t1}, the CDD produces a statevector with lower amplitudes than the UW scheme.
Moreover, the UW statevector is more strongly localized at fixed points than the CDD statevector.
Thus, the UW scheme provides a more precise prediction of the fixed points.
Nevertheless, the UW scheme is more difficult for block-encoding into quantum circuits than the CDD scheme.
The CDD scheme is appropriate for near integrable systems or for short time-intervals.
Therefore, in some cases, the CDD scheme can be a good alternative to the UW scheme if one wants to simplify the mapping of the KvN formulation onto quantum circuits.

\section{Linear Combination of Hamiltonian Simulations (LCHS) Algorithm}\label{sec:LCHS-circuit}

\subsection{Overview}\label{sec:LCHS-basics}
One of the most efficient methods proposed for dealing with nonunitary dynamics is the Linear Combination of Hamiltonian Simulations (LCHS) algorithm~\cite{An23, An23impr} which is equivalent to Schr\"odingerization.~\cite{Jin22Sch, Jin23Sch, Hu24Sch, Lu24}
This technique models nonunitary systems such as Eq.~\eqref{eq:initial-value-problem} by representing them as a superposition of unitary Hamiltonian simulations [Eq.~\eqref{eq:LCHS-theorem1}].
In the general case, the matrix $A$ can be time-dependent, $A(t)$, but in this work, we deal only with time-independent matrices, $A = \yi \yham$. 

\begin{figure}[!t]
\centering
\includegraphics[width=0.49\textwidth]{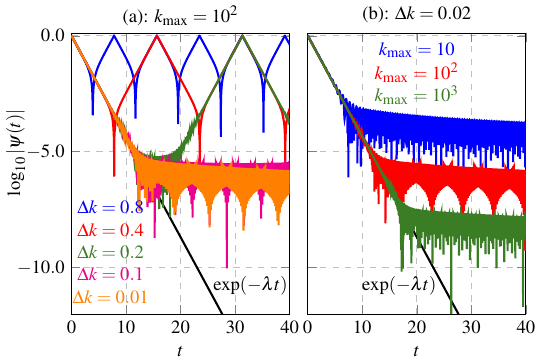}
\caption{
    \label{fig:LCHS-scalar-errors} 
    Plots showing $\psi(t) = e^{-\lambda t}$ for $\lambda = 1$ (black lines) and the corresponding LCHS simulation results (colored lines).
    (a) Results for fixed $k_{\rm max} = 100$ and various values of $\Delta k$.
    (b) Results for fixed $\Delta k = 0.02$ and various values of $k_{\rm max}$.
}
\end{figure}

The integral in Eq.~\eqref{eq:LCHS-theorem1} can be considered a Fourier transformation with the Fourier coordinate $k = (-\infty, \infty)$.
While Eq.~\eqref{eq:LCHS-theorem1} is exact, in order to compute this transformation numerically, one needs to truncate the range of $k$, i.e. $k = [-k_{\rm max}, k_{\rm max}]$ with some real $k_{\rm max}$, and to discretize the $k$-grid with a small enough step $\Delta k$:
\begin{equation}\label{eq:k-grid}
    k_j = -k_{\rm max} + j\Delta k,\quad \Delta k = \frac{2k_{\rm max}}{N_k-1}, \quad j = [0,N_k). 
\end{equation}
In the discretized Fourier space, the integral~\eqref{eq:LCHS-theorem1} becomes a sum of $N_k$ weighted unitaries.
After the discretization, the state vector $\psi(t)$ at a particular time instant $t$ can be approximately computed as 
\begin{equation}\label{eq:LCHS}
    \psi(t) = \left(\sum_{j=0}^{N_k-1} w_j V_j(t)\right)\psi(0) + \varepsilon_{\rm LCHS},
\end{equation}
where $\varepsilon_{\rm LCHS}$ is the discretization error of the truncated LCHS algorithm, 
and $V_j(t)$ are unitaries corresponding to Hamiltonian simulations with Hermitian matrices $(A_H + k_j A_L)$ dependent on the discretized Fourier coordinate $k_j$:
\begin{subequations}\label{eq:Vj}
\begin{eqnarray}
    &&V_j(t) = e^{-\yi C_j t},\label{eq:LCHS-Vj}\\
    &&C_j = A_H + k_j A_L,\label{eq:C}\\
    &&w_j = \frac{\Delta k}{\pi(1+k_j^2)} \times
        \left\{ \begin{aligned}
                &1/2,\quad j = 0 \text{ and } N_k-1,\\
                &1,\quad j = [1, N_k-2].
        \end{aligned} \right.\label{eq:wj}
\end{eqnarray}
\end{subequations}
According to the analysis of Ref.~\onlinecite{An23}, which is verified by Fig.~\ref{fig:LCHS-scalar-errors}.b, the error $\varepsilon_{\rm LCHS}$ is inversely proportional to the parameter $k_{\rm max}$:
\begin{equation}\label{eq:err-LCHS-scaling}
    \varepsilon_{\rm LCHS} = \oO(k_{\rm max}^{-1}).
\end{equation}
We should note that the dependence of $\varepsilon_{\rm LCHS}$ on $k_{\rm max}$ can be improved by using a different functional dependence of the weights $w_j$ on $k_j$ as investigated in Ref.~\onlinecite{An23impr}.

The Fourier step $\Delta k$ determines the maximum temporal interval $t$ that can be simulated by the discretized LCHS.
As seen from Fig.~\ref{fig:LCHS-scalar-errors}.a, the time interval is linear with $\Delta k^{-1}$.
For instance, if one decreases the step $\Delta k$ from $0.4$ to $0.2$, one doubles the maximum simulated time interval.

When the LCHS is mapped onto a quantum circuit, Eq.~\eqref{eq:LCHS} becomes
\begin{equation}\label{eq:LCHS-circuit}
    \psi(t) = \left(\sum_{j=0}^{N_k-1} (w_j + \varepsilon_w) [V_j(t) + \varepsilon_t]\right)\psi(0) + \varepsilon_{\rm LCHS},
\end{equation}
where $\varepsilon_w$ is the error in the computation of the LCHS weights, and $\varepsilon_t$ is the error of the Hamiltonian simulations.
In this section and Sec.~\ref{sec:LCHS-complexity}, we discuss the different sources of these errors and how this impacts the overall performance of our algorithm.
In our simulations discussed in Sec.~\ref{sec:LCHS-simulations}, the discretization error $\varepsilon_{\rm LCHS}$ is the dominant contributor to the inaccuracy of the LCHS algorithm.

\subsection{Implementation using LCU-like circuit}\label{sec:LCHS-LCU}

\begin{figure*}[!t]
\centering
\includegraphics[width=0.98\textwidth]{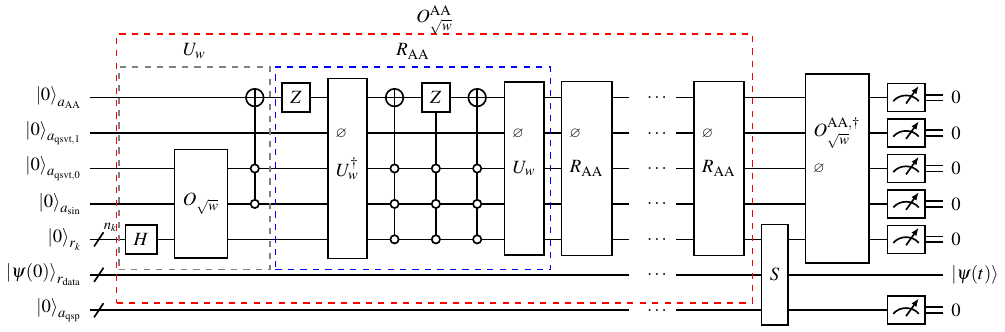}
\caption{
    \label{circ:LCHS-AA}
    LCU-like circuit for the LCHS algorithm with amplitude amplification (AA)\cite{Brassard02} of the LCHS weights. 
    The circuit for $O_{\sqrt{w}}$ is shown in Fig.~\ref{circ:LCHS-Ow}.
    The AA operator $R_{\rm AA}$ is repeated $N_{\rm AA}$ [Eq.~\eqref{eq:N_AA}] times within the operator $O_{\sqrt{w}}^{\rm AA}$. 
    The circuit for the selector $S$ is shown in Fig.~\ref{circ:LCHS-selector}.
    Here, we do not take into account the ancillary registers that are needed for the BE of the matrices $A_H$, $B_k$, and $B_{\rm max}$ (Appendix~\ref{app:be-kvn}).
    Here, $\varnothing$ means that the corresponding gates do not act on the indicated qubits.
}
\end{figure*}

The implementation of LCHS is based on the well-known LCU technique\cite{Childs12} which allows one to compute a weighted sum of unitaries as in Eq.~\eqref{eq:LCHS}.
A schematic LCU circuit that can be used for LCHS simulations is shown in Fig.~\ref{circ:LCHS-AA}.
There, the initial state $\ket{\psi(0)}$ saved in the register $r_{\rm data}$ is computed by an initialization circuit not shown here. 
For instance, to encode the uniform initial condition~\eqref{eq-psi-init-flat}, one needs to apply a single Hadamard gate to each qubit in the register $r_{\rm data}$ initialized in the zero state.
The output state $\psi(t)$ at a required $t$ is returned in the register $r_{\rm data}$ if all ancillae and the register $r_k$ are measured in the zero state.

In Fig.~\ref{circ:LCHS-AA}, the oracle $O_{\sqrt{w}}$ and its Hermitian-adjoint version $O_{\sqrt{w}}^\dagger$, which is hidden inside the subcircuit $O_{\sqrt{w}}^{{\rm AA,\dagger}}$,
together compute the weights $w_j$ [Eq.~\eqref{eq:wj}].
On the other hand, the selector $S$ computes the unitaries $V_j$. 
The ancilla register $a_{\rm qsp}$ is used for the implementation of the selector $S$.
The ancillae $a_{\sin}$ and $a_{\rm qsvt}$ are used to construct the oracles $O_{\sqrt{w}}$ and $O_{\sqrt{w}}^\dagger$. 
The register $r_k$ with 
\begin{equation}
    n_k = \log_2(N_k) 
\end{equation}
qubits is necessary to create the sum over $k_j$ using Hadamard gates.
The bitstrings saved in $r_k$ encode the index $j$ used in the sum~\eqref{eq:LCHS}.
The oracles $O_{\sqrt{w}}$ and $O_{\sqrt{w}}^\dagger$ use these bitstrings to address the LCHS weights $w_j$.
The selector uses these bitstrings to address the unitaries $V_j$.
Taking into account the newly introduced registers, the action of $O_{\sqrt{w}}$ is described as 
\begin{align}\label{eq:O_w_sqrt-schematic}
    O_{\sqrt{w}}&\ket{0}_{a_{\rm qsvt,0}}\ket{0}_{a_{\rm sin}}\ket{j}_{r_k} = \nonumber\\
        &(\sqrt{w_j}\ket{0}_{a_{\rm qsvt,0}}\ket{0}_{a_{\rm sin}} + \dots\ket{\neq 0}_{a_{\rm qsvt,0},\, a_{\rm sin}})\ket{j}_{r_k},
\end{align}
and the selector $S$ performs the following operation:
\begin{align}\label{eq:S-schematic}
    S\ket{0}_{a_{\rm qsp}}&\ket{j}_{r_k}\ket{\psi(0)}_{r_{\rm data}} = \nonumber\\
        &(V_j\ket{0}_{a_{\rm qsp}} + \dots\ket{\neq 0}_{a_{\rm qsp}})\ket{j}_{r_k}\ket{\psi(0)}_{r_{\rm data}}.
\end{align}
We should also remark that the Hermitian-adjoint oracle $O_{\sqrt{w}}^\dagger$ (and the corresponding AA operator $O_{\sqrt{w}}^{{\rm AA}, \dagger}$) acts on the ancilla $a_{\rm qsvt,1}$ instead of $a_{\rm qsvt,0}$.
This is necessary for the correct entanglement of the register $r_{\rm data}$ with the zero state of the ancilla $a_{\rm qsvt}$ (Appendix~\ref{app:LCU}).

In our work, to construct the LCHS circuit and perform the BE of the matrices necessary for the LCHS computations, we use so-called Single-Target Multi-controlled (STMC) gates.
We define STMC gates as single-target quantum operators such as Hadamard, Pauli, phase, and rotation gates that are controlled by multiple qubits.  
The STMC gates can be used as an intermediate step before mapping a quantum circuit to actual quantum hardware.
Since each STMC gate controlled by $n_c$ qubits can be decomposed into a circuit with $\oO(n_c)$ elementary gates without using extra ancillary qubits,\cite{Barenco95, Claudon24} the decomposition of the STMC gates does not drastically change the overall scaling of the algorithm.
Clearly, the number of STMC gates in a circuit is a lower limit on the number of gates that one should expect in the circuit.
The total number of gates in the circuit on a specific quantum computer will depend on the available elementary gates and on the topology of qubit connectivity provided by the quantum hardware platform.
Finally, taking into account the complexity and size of the circuits, we consider only fault-tolerant error-corrected quantum computation with logical qubits.

\subsection{Computation of the LCHS weights}\label{sec:Ow}

\begin{figure*}[!t]
\centering
\subfloat[Circuit for computing LCHS weights.]{\includegraphics[width=0.64\textwidth]{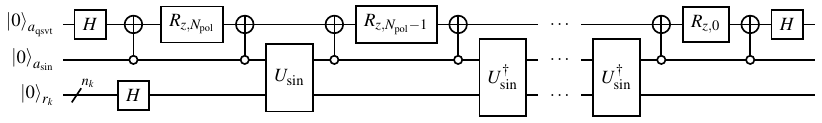}\label{circ:qsvt-even}}
\hspace{0.2cm}
\subfloat[Circuit for computing $\sin{(k/k_{\rm max})}$.]{\includegraphics[width=0.34\textwidth]{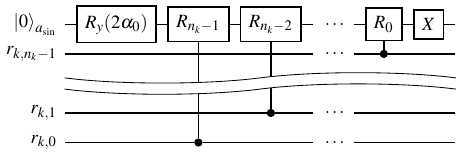}\label{circ:sin}}
\caption{
    \label{circ:LCHS-Ow}
    (a) QSVT circuit for encoding an even function used for constructing $O_{\sqrt{w}}$. Here, $R_{z,j} = R_z(2\phi_{{\rm qsvt},j})$.
    (b) Quantum circuit for the BE oracle $U_{\rm sin}$ computing $y_k$ [Eq.~\eqref{eq:y}]. Here, $R_j = R_y(2\alpha_1/2^j)$. The parameters $\alpha_0$ and $\alpha_1$ are defined in Eq.~\eqref{eq:pars-sin}.
}
\end{figure*}

The oracle $O_{\sqrt{w}}$ can be implemented by using the QSVT\cite{Gilyen19} or the QETU\cite{Dong22} algorithms.
For that, without taking into account the constant multiplicative factor $\Delta k / \pi$ [Eq.~\eqref{eq:wj}], we use the following expression for the square root of the LCHS weights:
\begin{subequations}
\begin{eqnarray}
    &&p_w(k_j) = \eta_{\rm qsvt} [1 + k_{\rm max}^2 \arcsin^2(y_j)]^{-1/2},\label{eq:pol-weights-k}\\
    &&y_j = \sin(k_j/k_{\rm max}),\label{eq:y}
\end{eqnarray}
\end{subequations}
where $\eta_{\rm qsvt} \lesssim 1$ is a scalar which rescales the maximum amplitude of the polynomial $p_w$ to a value less than one and is necessary for a successful computation of the QSVT parameters.
The even polynomial $p_w$ is computed by the QSVT circuit shown in Fig.~\ref{circ:qsvt-even} where the parameters $\phi_{{\rm qsvt},j}$ (the so-called QSVT angles or phase-factors) can be calculated using a variety of methods.\cite{Dong21, Ying22, Dong23}
The QSVT circuit addresses the variable $y_j$ using the BE oracle $U_{\rm sin}$ shown in Fig.~\ref{circ:sin} where the following parameters are used
\begin{equation}\label{eq:pars-sin}
    \alpha_0 = -1.0,\quad \alpha_1 = N_k / (N_k - 1).
\end{equation}
(In the general case, when one needs to encode $\sin\varphi_j$ with $\varphi_j = -\varphi_{\rm max} + 2\varphi_{\rm max} j /(N_k-1)$ with $j = [0, N_k)$, one takes $\alpha_0 = -\varphi_{\rm max}$ and $\alpha_1 = \varphi_{\rm max} N_k / (N_k - 1)$.)
The BE oracle $U_{\rm sin}$ scales as $\oO(n_k)$, and the query complexity (i.e. the number of calls $N_{w,D}$ to the BE oracle) of the QSVT circuit scales as 
\begin{eqnarray}\label{eq:Ow-query-complexity}
    N_{w,D} = \oO(k_{\rm max}/\varepsilon_{w}),
\end{eqnarray}
where $\varepsilon_{w}$ is the error of the QSVT approximation of the polynomial $p_w$ which corresponds to the computational error of the LCHS weights discussed in Eq.~\eqref{eq:LCHS-circuit}.
For a different shape of the weights, the scaling may differ.
Moreover, if one needs to encode complex weights, e.g. to improve the scaling of $\varepsilon_{\rm LCHS}$ with $k_{\rm max}$ as proposed in Ref.~\onlinecite{An23impr}, one needs two QSVT circuits to compute the real and imaginary components of the weights separately.
Then, the QSVT circuits are combined with LCU (e.g. see Fig. 3 in Ref.~\onlinecite{Dong21}). 
Thus, one needs one extra qubit for the computation of the complex weights proposed in Ref.~\onlinecite{An23impr}.

We should note that because of the poor scaling with $\varepsilon_{w}$ in Eq.~\eqref{eq:Ow-query-complexity}, it is better to use a direct brute-force computation of the LCHS weights to achieve better gate complexity of the oracle $O_{\sqrt{w}}$ in the case of low $n_k$.
In the direct computations, the $N_k$ weights are calculated by $N_k$ quantum gates, e.g. controlled $R_y(\theta_k)$ gates with $\theta_k = 2 \arccos(w_k)$. 
The direct computation of the weights is exact, requires one ancilla fewer compared to the QSVT-version but needs $\oO(N_k)$ STMC gates.
At the same time, the QSVT computation of the weights scales as Eq.~\eqref{eq:Ow-query-complexity} but has the approximation error $\varepsilon_{w}$.
In Appendix~\ref{app:num-LCHS-toy} and Sec.~\ref{sec:num-LCHS-ADE}, we use the QSVT computation with $\varepsilon_{w} \approx 10^{-4}$ which requires around $N_{w,D} n_k$ STMC gates with $N_{w,D} \approx 350$ to build $O_{\sqrt{w}}$.
In Sec.~\ref{sec:LCHS-KvN}, we use the direct computation of the weights for $n_k = 7$ to reduce the gate complexity of $O_{\sqrt{w}}$.

To guarantee that the success probability of the LCHS circuit is $\oO(1)$, one needs to apply AA\cite{Brassard02} to the computation of the LCHS weights. 
According to Fig.~\ref{circ:LCHS-AA}, the operator $U_w$ (gray box) computes the non-amplified LCHS weights and encodes their values as the amplitude of  $\ket{1}_{a_{\rm AA}}$.
The success probability of this operator is
\begin{equation}
    \gamma_w = |\bra{1}_{a_{\rm AA}} U_w \ket{0}_{a_{\rm AA}}|^2.
\end{equation}
To boost this probability and bring it to $\oO(1)$, we apply the operator $R_{\rm AA}$, indicated by the blue box in Fig.~\ref{circ:LCHS-AA}, $N_{\rm AA}$ times:
\begin{equation}\label{eq:N_AA}
    N_{\rm AA} = \floor{\pi/(4\theta_{\gamma_w})}, \quad \theta_{\gamma_w} = \arcsin{\sqrt{\gamma_w}}.
\end{equation}
The operator $R_{\rm AA}$ is a standard AA operator.\cite{Brassard02}
Apart from the operators $U_w$ and $U_w^\dagger$, it includes a reflector on the ``useful'' state.
In our case, the ``useful'' state is $\ket{1}_{a_{\rm AA}}$, therefore, the reflector is a single $Z$ gate applied to the ancilla $a_{\rm AA}$.
The operator $R_{\rm AA}$ also includes a reflector on the initial state.
Since the latter is $\ket{0}_{a_{\rm AA}}\ket{0}_{a_{\rm qsvt,0}}\ket{0}_{a_{\rm sin}}\ket{0}_{r_k}$ for the operator $O_{\sqrt{w}}^{\rm AA}$ and $\ket{0}_{a_{\rm AA}}\ket{0}_{a_{\rm qsvt,1}}\ket{0}_{a_{\rm sin}}\ket{0}_{r_k}$ for the operator $O_{\sqrt{w}}^{{\rm AA}, \dagger}$, the reflector is a combination of controlled Pauli $X$ and $Z$ gates.
After applying the rotation $R_{\rm AA}$ $N_{\rm AA}$ times, one boosts the success probability of the combined subcircuit $O_{\sqrt{w}}^{\rm AA}$ to $\oO(1)$.
This brings the success probability of the entire LCHS circuit to $\oO(1)$.
The influence of AA on the circuit scaling is discussed later in Sec.~\ref{sec:LCHS-scaling}.
Without AA, the subcircuit $O_{\sqrt{w}}^{\rm AA}$ is reduced to the Hadamard gates on the register $r_k$ and the oracle $O_{\sqrt{w}}$.

\subsection{Selector}\label{sec:selector-main}

\begin{figure*}[!t]
\centering
\subfloat[Selector $S$]{\includegraphics[width=0.28\textwidth]{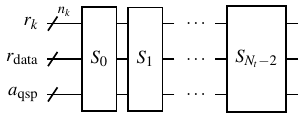}\label{circ:LCHS-prod-t}}
\hspace{0.2cm}
\subfloat[Subcircuit $S_l$]{\includegraphics[width=0.36\textwidth]{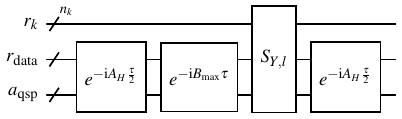}\label{circ:LCHS-LTS}}
\hspace{0.2cm}
\subfloat[Subselector $S_{Y,l}$]{\includegraphics[width=0.28\textwidth]{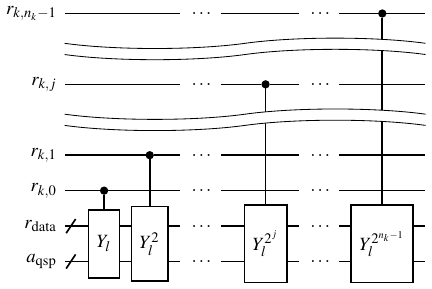}\label{circ:LCHS-subselector}}
\caption{
    \label{circ:LCHS-selector}
    Quantum circuits showing the structure of the selector $S$ used in circuit~\ref{circ:LCHS-AA}.
    (a) The selector $S$ as a sequence of subcircuits $S_l$ for simulating the product~\eqref{eq:LCHS-Vj-prod} for all $j = [0, N_k)$.
    (b) The subcircuit $S_l$ for performing trotterization~\eqref{eq:Gj-decomposition} in a short time interval $[t_l, t_l + \tau]$.
    The unitaries $e^{-\yi A_H \tau/2}$ and $e^{-\yi B_{\rm max}\tau}$ are computed by the QSP.\cite{Low19,Haah20}
    (c) The circuit for $S_{Y,l}$ [Eq.~\eqref{eq:SYl}] for a given $l$. The unitary $Y_l$ [Eq.~\eqref{eq:Yl}] is computed by the QSP.
}
\end{figure*}

To construct the selector $S$, we split each $V_j$ into several Hamiltonian simulations and, in this way, remove the dependence on $k_j$ from the BE.
More precisely, Eq.~\eqref{eq:LCHS-Vj} describes the temporal evolution of a sum of two generally non-commuting although Hermitian matrices $A_L$ and $A_H$.
To perform the corresponding Hamiltonian simulations in a long time interval $[0, t]$, we discretize the time:
\begin{equation}\label{eq:t-grid}
    t_l = l \tau,\quad \tau = \frac{t}{N_t - 1},\quad l = [0, N_t), 
\end{equation}
and decompose Eq.~\eqref{eq:LCHS-Vj} into an ordered product of unitaries:
\begin{subequations}
\begin{eqnarray}
    &&V_j(t) = G_j(t_{N_t-2})G_j(t_{N_t-3})\dots G_j(t_0),\label{eq:LCHS-Vj-prod}\\
    &&G_j(t_l) = e^{-\yi C_j \tau},\quad j = [0, N_k).\label{eq:LCHS-Gj}
\end{eqnarray}
\end{subequations}
After that, we use the second-order Lie-Trotter-Suzuki (LTS) decomposition\cite{Suzuki93} of ordered exponentials: 
\begin{subequations}\label{eq:G}
\begin{eqnarray}
    &&\begin{aligned}
        G_j(t_l) &\approx e^{-\yi A_H \tau/2} Y_l^j e^{ - \yi B_{\rm max} \tau} e^{-\yi A_H \tau/2} + \varepsilon_{\rm trot},
    \end{aligned}\label{eq:Gj-decomposition}\\
    &&Y_l =  e^{-\yi B_k \tau},\label{eq:Yl}
\end{eqnarray}
\end{subequations}
where the matrices $B_{\rm max}$ and $B_k$ are defined as:
\begin{subequations}\label{eq:B}
\begin{eqnarray}
    &&B_{\rm max} = - k_{\rm max} A_L,\label{eq:Bmax}\\
    &&B_k = \Delta k A_L.\label{eq:Bk}
\end{eqnarray}
\end{subequations}
(Although, we consider only time-independent matrices, here and below, we keep the index $l$ to show explicitly that the operator $Y_l$ deals with a separate short-time interval.)

For the second order LTS decomposition, the local trotterization error scales as\cite{Childs21} 
\begin{equation}\label{eq:trot-error-init}
    \varepsilon_{\rm trot} = \oO\left[ \frac{(\tau ||A_C||)^3}{\varepsilon^2_{\rm LCHS}}\right],
\end{equation}
where 
\begin{subequations}
\begin{eqnarray}
    &&||A_C|| =  \left( ||T_1|| + k_{\rm max}^{-1}||T_2|| \right)^{1/3},\\
    &&T_1 = [A_L,[A_L,A_H]],\\
    &&T_2 = [A_H,[A_H,A_L]],
\end{eqnarray}
\end{subequations}
where $[M_1, M_2]$ is the commutator of two matrices.

In the more general case of the $p$-order LTS decomposition, the trotterization error becomes:
\begin{equation}\label{eq:trot-error-p}
    \varepsilon_{\rm trot} = \oO\left[\frac{(\tau||A_C||)^{p+1}}{\varepsilon_{\rm LCHS}^p}\right],
\end{equation}
where now $||A_C||$ is the normalized to $k_{\rm max}^p$ sum of the norms of all possible $(p+1)$-order commutators of the matrices $A_H$ and $k_{\rm max} A_L$:
\begin{align}\label{eq:AC-p-trot}
    ||A_C|| &= \big(||[A_L,[A_L,[\dots ,[A_L,A_H]]\dots]|| + \nonumber\\
        &k_{\rm max}^{-1}||[A_H,[A_L,[\dots ,[A_L,A_H]]\dots]|| + \dots +\nonumber\\
        &k_{\rm max}^{-(p-1)}||[A_H,[A_H,[\dots ,[A_H,A_L]]\dots]||
    \big)^{\frac{1}{p+1}}.
\end{align}

Consistently with Eq.~\eqref{eq:LCHS-Vj-prod}, the selector $S$, which is the circuit representation of $V_j$ for all $j = [0,N_k)$, includes $N_t-1$ operators $S_l$ (Fig.~\ref{circ:LCHS-prod-t}) where each $S_l$ simulates the $l$-th time interval, $l = [0, N_t - 2]$.
More precisely, each $S_l$ shown in Fig.~\ref{circ:LCHS-LTS} computes $G_j(t_l)$ for all $j$ at once:
\begin{equation}\label{eq:Sl}
    S_l =  \left(I\otimes e^{-\yi A_H \tau/2}\right) S_{Y,l} \left(I\otimes e^{-\yi B_{\rm max}\tau}\right) \left(I\otimes e^{-\yi A_H \tau/2}\right).
\end{equation}
The dependence on $k_j$ is now localized within the subselector $S_{Y,l}$ shown in Fig.~\ref{circ:LCHS-subselector}.
In particular, $S_{Y,l}$ simulates the operators $Y_l^j$ [Eq.~\eqref{eq:Yl}] for all $j$ at once:
\begin{equation}\label{eq:SYl}
    S_{Y,l} = \sum_{j = 0}^{N_k-1} \ket{j}_{r_k} \bra{j}_{r_k} \otimes Y_l^j.
\end{equation}

\section{Complexity analysis of the LCHS Algorithm}\label{sec:LCHS-complexity}

\subsection{Query complexity of the selector}\label{sec:selector-scaling}
The operators $e^{-\yi A_H \tau/2}$, $e^{-\yi B_{\rm max}\tau}$, and $Y_l$ are implemented using the QSP algorithm\cite{Low17, Low19} whose circuit representation can be found in Ref.~\onlinecite{Novikau22}.
The QSP for computing $e^{-\yi M\tau}$ for some Hermitian matrix $M$ scales linearly with time as 
\begin{equation}\label{eq:QSP-scaling}
    \oO(||M||\tau + \xi_{\rm QSP}),\quad \xi_{\rm QSP} = \log(\varepsilon_{\rm QSP}^{-1}),
\end{equation}
where $\varepsilon_{\rm QSP}$ is the QSP approximation error which is the second part of the total error $\varepsilon_t$:
\begin{equation}
    \varepsilon_t = \varepsilon_{\rm trot} + \varepsilon_{\rm QSP}.
\end{equation}
One can make $\varepsilon_{\rm QSP}$ small enough (e.g. comparable to $\varepsilon_{\rm trot}$) by increasing the number of QSP angles in the short-time-interval Hamiltonian simulations.\cite{Haah20, Haah20code}
In our numerical simulations, we keep $\varepsilon_{\rm QSP} = 10^{-12}$.
In this case, one can assume that $\varepsilon_t \simeq \varepsilon_{\rm trot}$.
Since $\varepsilon_t$ enters the sum over $k$ in Eq.~\eqref{eq:LCHS-circuit}, we assume that $\varepsilon_t$ accrues with $N_k$.
Moreover, $\varepsilon_t$ is accumulated with $N_t$. 
Taking into account Eqs.~\eqref{eq:t-grid} and~\eqref{eq:trot-error-p} and requiring the global error $N_t N_k \varepsilon_t$ to be of order of $\varepsilon_{\rm LCHS}$, one obtains the following condition on the trotterization time step:
\begin{equation}\label{eq:trot-step}
    \tau = \oO\left[\frac{\varepsilon_{\rm LCHS}}{||A_C||}\left(\frac{\varepsilon_{\rm LCHS}}{||A_C||t N_k}\right)^{1/p}\right].
\end{equation}

Although the subselector $S_{Y,l}$ [Fig.~\ref{circ:LCHS-subselector}] formally calls only $n_k$ operators, the depth of each operator $Y_l^{2^j}$ changes linearly with $2^j$ because of Eq.~\eqref{eq:QSP-scaling}.
Therefore, without fast-forwarding, the query complexity of $S_{Y,l}$ is linear with $N_k$:
\begin{equation}\label{eq:SY-scaling-init}
    \oO[N_k(\Delta k ||A_L||\tau + \xi_{\rm QSP})] = \oO[\varepsilon_{\rm LCHS}^{-1} ||A_L||\tau + N_k \xi_{\rm QSP}],
\end{equation}
where we also took into account Eq.~\eqref{eq:err-LCHS-scaling}.
The query complexities of the QSP computation of the unitaries $e^{-\yi B_{\rm max}\tau}$ and $e^{-\yi A_H\tau}$ are
\begin{equation}
    \oO[\varepsilon_{\rm LCHS}^{-1} ||A_L||\tau + \xi_{\rm QSP}],
\end{equation}
and
\begin{equation}
    \oO[||A_H||\tau + \xi_{\rm QSP}],
\end{equation}
correspondingly.
Note that these unitaries are evaluated only $\oO(N_t)$ times, whereas the unitary $Y_l$ is called $\oO(N_k N_t)$ times (Eqs.~\eqref{eq:Yl} and~\eqref{eq:SYl}).

As a result, the complexity of the entire selector $S$ is
\begin{equation}\label{eq:S-complexity-init}
    \oO(||A_H|| t + \varepsilon_{\rm LCHS}^{-1} ||A_L|| t + N_t N_k \xi_{\rm QSP}),
\end{equation}
where we took into account the fact that the QSP simulation of each unitary is repeated $N_t-1$ times.
To recast the above complexity only in terms of $t$ and $\varepsilon_{\rm LCHS}$, one needs to rewrite the last additive term.
First of all, due to the trapezoidal rule used for the discretization of the LCHS integral, one has (see Sec. II in Supplemental Material in Ref.~\onlinecite{An23})
\begin{equation}\label{eq:Nk-cond}
    N_k = \oO\left(\frac{||A_L||t}{\varepsilon^2_{\rm LCHS}}\right),
\end{equation}
where the requirement~\eqref{eq:err-LCHS-scaling} was taken into account.
Because of this, equation~\eqref{eq:trot-step} can be rewritten as
\begin{equation}\label{eq:trot-step-2}
    \tau = \oO\left[\frac{\varepsilon_{\rm LCHS}}{||A_C||}\left(\frac{\varepsilon_{\rm LCHS}^3}{||A_C||\,||A_L||t^2}\right)^{1/p}\right].
\end{equation}
From the above equation, one easily obtains that
\begin{equation}
    N_t = \oO\left[\frac{||A_C|| t}{\varepsilon_{\rm LCHS}} \left(\frac{||A_C||\,||A_L||t^2}{\varepsilon_{\rm LCHS}^3}\right)^{1/p}\right].
\end{equation}
Hence, the product $N_k N_t$, now denoted $Q_{\xi}$, can be recast as 
\begin{equation}\label{eq:Q-xi}
    Q_{\xi} = \oO\left[
    \left(\frac{||A_C||\, ||A_L|| t^2}{\varepsilon_{\rm LCHS}^3}\right)^{1+1/p}\right].
\end{equation}
With that, the query complexity of $S$ becomes:
\begin{equation}\label{eq:selector-query-complexity}
    Q_S = \oO(||A_H|| t + \varepsilon_{\rm LCHS}^{-1} ||A_L|| t + Q_{\xi} \xi_{\rm QSP}).
\end{equation}
For the second order trotterization used here, $p=2$, the term $Q_{\xi} \xi_{\rm QSP}$ scales as the third order of time, $t^3$.
This term appears due to the additive term $\xi_{\rm QSP}$ in QSP scaling~\eqref{eq:QSP-scaling}.
Yet, numerical tests show that for order unity timescales the term linear with time is the dominant one in the QSP scaling, and the term linear with $\xi_{\rm QSP}$ has significantly smaller contribution.
For example, see Fig.~16 in Ref.~\onlinecite{Novikau22}.
Therefore, for a broad range of time intervals, the term $Q_{\xi}$ in Eq.~\eqref{eq:selector-query-complexity} has a smaller contribution to the depth of the LCHS circuit than the terms linear with time.

\subsection{The case of commuting matrices $A_L$ and $A_H$}\label{sec:LCHS-scaling-zero-commutator}
If the matrices $A_H$ and $A_L$ commute with each other, then there is no need to split the time interval into small time intervals as done in Eq.~\eqref{eq:t-grid}.
In this case, the selector $S$ is reduced to 
\begin{subequations}\label{eq:S-zero-commutator}
\begin{eqnarray}
    &&S = \left(I\otimes e^{-\yi A_H t}\right) S_Y \left(I\otimes e^{-\yi B_{\rm max}t}\right),\\
    &&S_Y = \sum_{j = 0}^{N_k-1} \ket{j}_{r_k} \bra{j}_{r_k} \otimes e^{-\yi B_k j t}.
\end{eqnarray}
\end{subequations}
Because of this, the query complexity becomes:
\begin{equation}\label{eq:S-complexity-init-zero-commutator}
    \oO(||A_H|| t + \varepsilon_{\rm LCHS}^{-1} ||A_L|| t + N_k \xi_{\rm QSP}).
\end{equation}
One can compare Eq.~\eqref{eq:S-complexity-init-zero-commutator} with Eq.~\eqref{eq:S-complexity-init}.
More precisely, the complexity~\eqref{eq:S-complexity-init-zero-commutator} does not have an extra multiplicative term $N_t$ in front of $\xi_{\rm QSP}$.
Taking into account Eq.~\eqref{eq:Nk-cond}, one obtains a linear dependence on overall simulation time for the complexity of the selector for the case with commuting matrices $A_H$ and $A_L$:
\begin{equation}\label{eq:selector-query-complexity-zero-commutator}
    Q_S = \oO\left[ t(||A_H|| + \varepsilon_{\rm LCHS}^{-1} ||A_L|| + \varepsilon_{\rm LCHS}^{-2} ||A_L|| \xi_{\rm QSP})\right].
\end{equation}

\subsection{An alternative version of the selector}\label{sec:selector-alternative}
In circuit~\ref{circ:LCHS-LTS}, trotterization is performed in order to isolate the dependence on $k$ within the subselector $S_{Y,l}$ (Fig.~\ref{circ:LCHS-subselector}).
Because of this, to perform Hamiltonian simulations, one needs to block-encode the matrices $A_H$ and $A_L$ separately.
The advantage of this approach is that in the case when fast-forwarding is possible, i.e. when the query complexity of Hamiltonian simulations does not depend on time, the subselector $S_{Y,l}$ scales as $n_k$ instead of $N_k$.

Another option for the LCHS circuit is to avoid trotterizing the system and perform Hamiltonian simulations using the matrices $C_j$ [Eq.~\eqref{eq:C}], whose spectral norm is $||A_H + k_j A_L||$.
Since $||A_H + k_j A_L|| \leq ||A_H|| + k_j ||A_L||$, the query complexity of the selector $S$ becomes 
\begin{equation}\label{eq:S-scaling-init-alternative}
    \oO\left[N_k (||A_H|| t + \xi_{\rm QSP}) + k_{\rm max} ||A_L||t\right].
\end{equation}

As a result, even in the case with fast-forwarding, the query complexity~\eqref{eq:S-scaling-init-alternative} contains an additional dependence on $N_k$.
In terms of $\varepsilon_{\rm LCHS}$, the complexity~\eqref{eq:S-scaling-init-alternative} is
\begin{equation}\label{eq:S-scaling-init-alternative-err}
    Q_S = \oO\left[\frac{||A_L|| t}{\varepsilon^2_{\rm LCHS}}(||A_H|| t + \xi_{\rm QSP}) + \varepsilon_{\rm LCHS}^{-1} ||A_L||t\right],
\end{equation}
where one obtains a quadratic scaling with respect to the simulated time.
This happens because one Hamiltonian evolution in time $t$ is performed for each of $N_k$ matrices $C_j$, where each matrix $C_j$ contains $A_H$.
In other words, one addresses the elements of the $A_H$ matrix  $N_k t$ times where $N_k$ is also linear with time.

\subsection{Optimal implementation of the selector $S$}\label{sec:selector-optimal}

\begin{figure}[!t]
\centering
\includegraphics[width=0.40\textwidth]{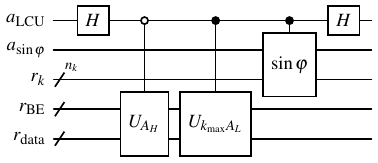}
\caption{
    \label{circ:BE-eff-selector}
    Circuit showing the BE oracle used by the QSP circuit in the optimal implementation of the LCHS selector.
    The oracles $U_{A_H}$ and $U_{k_{\rm max} A_L}$ encode the matrices $A_H$ and $k_{\rm max} A_L$, respectively, normalized to $||C_{\rm max}||$ [Eq.~\eqref{eq:Cmax}].
    The subcircuit $\sin\varphi$ is shown in Fig.~\ref{circ:sin}.
}
\end{figure}

To achieve an optimal scaling of the selector, it is necessary to find an efficient BE of $C_j$ where the dependence on $k_j$ is isolated from the matrix elements of $A_H$ without performing the trotterization.
In this case, one avoids the term linear with $Q_\xi$ in Eq.~\eqref{eq:selector-query-complexity} and the term quadratic with time in Eq.~\eqref{eq:S-scaling-init-alternative-err}.
This can be achieved by applying the transformation $k = k_{\rm max} \sin\varphi$ where $\varphi\in[-\pi/2, \pi/2]$ so that Eq.~\eqref{eq:LCHS-theorem1} becomes
\begin{align}\label{eq:LCHS-theorem1-ksinphi}
    &e^{- A t} = \int_{-\pi/2}^{\pi/2} \frac{k_{\rm max} \cos{\varphi} }{\pi(1+k_{\rm max}^2 \sin^2{\varphi})} e^{-\yi (A_H +  k_{\rm max} \sin\varphi A_L) t}  \diff \varphi .
\end{align}
In this case, Eqs.~\eqref{eq:Vj} are recast into
\begin{subequations}\label{eq:Vj-eff}
\begin{eqnarray}
    &&C_j = A_H + k_{\rm max} \sin\varphi_j A_L,\label{eq:C-eff}\\
    &&w_j = \frac{\Delta \varphi k_{\rm max}\cos\varphi_j}{\pi(1+k_{\rm max}^2\sin^2\varphi_j)}, \label{eq:wj-eff}
\end{eqnarray}
\end{subequations}
where we neglect the scalar factor $1/2$ in $w_j$ at $j = 0$ and $j = N_k-1$.
In this case, the selector becomes a single QSP circuit for the entire time interval $t$ where the BE oracle shown in Fig.~\ref{circ:BE-eff-selector} encodes the matrices $C_j$ for all $\varphi_j$ at once.
There, the register $r_k$ is not considered by the QSP circuit as an ancillary register, meaning that $r_k$ is not taken into account in the qubitization.
With this encoding, the query complexity of the selector becomes
\begin{equation}\label{eq:selectoropt-scaling}
    Q_S = \oO(\varepsilon_{\rm LCHS}^{-1}||\bar{C}_{\rm max}||t + \xi_{\rm QSP}),
\end{equation}
where
\begin{subequations}
\begin{eqnarray}
    &&||\bar{C}_{\rm max}|| = k_{\rm max}^{-1}||C_{\rm max}||,\label{eq:Cmax-bar}\\
    &&||C_{\rm max}|| = ||A_H + k_{\rm max} A_L||.\label{eq:Cmax}
\end{eqnarray}
\end{subequations}
This version of the selector is optimal with respect to the scaling with time, because the circuit~\ref{circ:BE-eff-selector} encodes the dependence on the Fourier coordinate $k$ using only $n_k$ simple rotations (Fig.~\ref{circ:sin}).
However, this implementation of the selector requires two extra ancillary qubits, $a_{\sin\varphi}$ and $a_{\rm LCU}$.
In our numerical tests of the LCHS algorithm discussed in Sec.~\ref{sec:LCHS-simulations}, we implement the first version of the LCHS circuit, specified in Sec.~\ref{sec:selector-main}, where the subselector $S_{Y,l}$ is used, which is cheaper with respect to the number of qubits in the LCHS circuit.

\subsection{Query complexity of the LCHS circuit}\label{sec:LCHS-scaling}
If one computes the LCHS weights without AA, then the success probability of the LCHS circuit decreases quadratically with $k_{\rm max}$:
\begin{equation}
    \gamma = \oO(\mu^{-2} k_{\rm max}^{-2}).
\end{equation}
where $\mu$ is the ratio of statevectors at the initial and final time instants:
\begin{equation}\label{eq:mu}
    \mu = ||\psi(0)||/||\psi(t)||.
\end{equation}
Because of this, one needs to repeat the entire LCHS and initialization circuits $\oO(\mu k_{\rm max})$ times to bring the probability close to one.
In this case, the LCHS query complexity is
\begin{equation}\label{eq-LLCHS-scaling-entire-AA}
    \oO\left[\mu \varepsilon_{\rm LCHS}^{-1}\left(Q_S + \varepsilon_{\rm LCHS}^{-1}\varepsilon_w^{-1}\right)\right],
\end{equation}
where we take into account Eqs.~\eqref{eq:err-LCHS-scaling} and ~\eqref{eq:Ow-query-complexity}, and the form of $Q_S$ depends on the chosen encoding of the selector $S$.

Instead, one can amplify the weights separately as shown in Fig.~\ref{circ:LCHS-AA}.
Since the success probability of the oracle $O_{\sqrt{w}}$ scales as $\oO(k_{\rm max}^{-1})$, this oracle should be repeated $N_{\rm AA} = \oO(\sqrt{k_{\rm max}})$ times to bring its probability to $\oO(1)$.
After that, the success probability of the entire LCHS circuit increases to $\oO(\mu^{-2})$.
In this case, the LCHS query complexity becomes
\begin{equation}\label{eq-LLCHS-scaling-AAw}
    \oO\left[ \mu \left(Q_S + \varepsilon_{\rm LCHS}^{-3/2}\varepsilon_w^{-1} \right)\right].
\end{equation}
Thus, with AA of the LCHS weights, the number of repetitions of the LCHS and initialization circuits is reduced to $\oO(\mu)$.
We should also note that the recent work\cite{An23impr} has shown that it is possible to achieve an exponential improvement in the LCHS scaling with respect to the error $\varepsilon_{\rm LCHS}$ by using a different form of the LCHS weights.

\subsection{Measurement}\label{sec:meas}
We do not provide an explicit cost for measurement, as this is a problem-specific aspect that depends heavily on the particular output the user of the LCHS algorithm is interested in.
A variety of possible measurement techniques such as AE\cite{Brassard02} can be employed for extracting classical data from quantum states returned by the LCHS circuit. 
We refer the reader to Ref.~\onlinecite{Joseph20, Novikau22, Novikau23, Joseph23} where several efficient approaches based on the AE and quantum Fourier transformation were discussed for measuring classical information and observables in fluid plasma and electrodynamic wave problems simulated by quantum circuits. 
Similar methods can be applied for extracting classical data from the LCHS circuit.\cite{Joseph20, Joseph23}
In particular, one can use AE to evaluate various integrals over phase space, such as the mean value of an observable in the proximity of a fixed point.
Typically, such measurements will add a global multiplicative factor $\epsilon^{-1}$ to Eq.~\eqref{eq-LLCHS-scaling-AAw}, where $\epsilon$ is the desired error in the measured observable.
Another possibility is to employ the LCHS to solve the KvN equation for emulating a nonlinear root-finder in a nonlinear system. 
This is an efficient way to use the KvN equation if the simulated nonlinear problem has a limited number of roots.
In this case, due to the high success probability of the LCHS algorithm, a small number of measurements will identify the most prominent fixed points in the considered nonlinear system.

\begin{figure}[!t]
\centering
\subfloat[]{\includegraphics[width=0.36\textwidth]{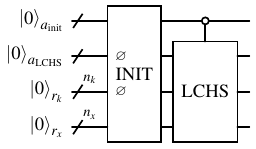}\label{circ:ADE-LCHS}}\\
\vspace{0.1cm}
\subfloat[]{\includegraphics[width=0.46\textwidth]{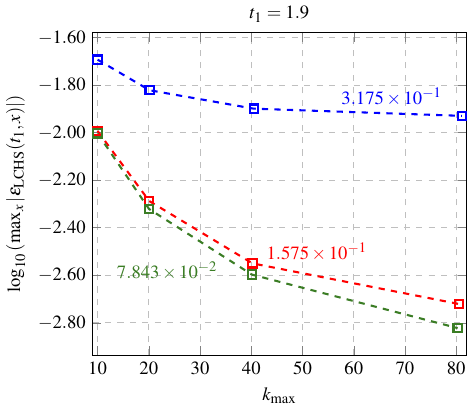}\label{fig:ADE-err-kmax}}
\caption{
    \label{circ:ADE-circuit-PY}
    (a) Schematic circuit for modeling the ADE [Eq.~\eqref{eq:ADE}].
    The INIT block is the initialization circuit which computes $\ket{\psi(0)}_{r_x}$ and entangles it with $\ket{0}_{a_{\rm init}}$.
    Here, $\varnothing$ means that the initialization circuit does not act on the qubits $a_{\rm LCHS}$ and $r_k$.
    (b) Plot showing the dependence of the maximum absolute error in the LCHS simulation of the ADE (without invoking quantum circuit~\ref{circ:ADE-LCHS}) on $k_{\rm max}$ for various $\Delta k$ whose values are indicated in different colors.
}
\end{figure}

\section{LCHS simulations of Advection-Diffusion and KvN examples}\label{sec:LCHS-simulations}

\subsection{Numerical emulation of the LCHS circuit}\label{sec:num-LCHS-circuit}
In this section, we use the LCHS algorithm for modeling the advection-diffusion equation (ADE) (Sec.~\ref{sec:num-LCHS-ADE}) and the KvN formulation of a nonlinear differential equation (Sec.~\ref{sec:LCHS-KvN}).
Also, we test the LCHS on toy problems with weak and strong dissipation in Appendix~\ref{app:num-LCHS-toy} to demonstrate how the success probability of the LCHS circuit changes after introducing AA of the oracles computing the LCHS weights.

As mentioned previously, we use trotterization to reduce the number of qubits in the LCHS circuit and we use the QSP algorithm\cite{Low17, Low19} to perform Hamiltonian simulations with Hermitian matrices.
The success probability of a QSP circuit simulating a single time interval is $[1 - \oO(\varepsilon_{\rm qsp})]^2$.
To avoid using CGs\cite{Fang23}, which are necessary for dealing with incoherent subcircuits\cite{Fang23}, we impose $\varepsilon_{\rm qsp} = 10^{-12}$ while computing the QSP angles.\cite{Haah20}
This allows us to assume that the QSP circuits are fully coherent.
However, the QSP circuits are called $N_t N_k$ times and this leads to an accumulation of the error related to the $\oO(\varepsilon_{\rm qsp})$ QSP failure probability.
Because of this, in general, for large $N_t$ and $N_k$, one might need to increase the QSP precision even further or include CGs into the LCHS circuit~\ref{circ:LCHS-AA}.
The latter will require an additional $\oO(\log_2 N_t)$ ancillae.
The QSP simulations also lose full coherence when the simulated system has strong dissipation.
In this case, one might also need to use CGs to increase the overall precision of the LCHS circuit. 
However, in this work, we avoid using CGs even in the case of strong dissipation to keep the numerical resources necessary for the emulation of the LCHS circuit at a reasonable level.
Indeed, the LCHS circuit without CGs for modeling even the toy problems discussed in Appendix~\ref{app:num-LCHS-toy} requires $n_k + 6$ ancillae (without taking into account the block encoding of the matrices $A_H$, $B_k$, and $B_{k_{\rm max}}$).
Moreover, even toy linear problems require more than $10^5$ STMC gates.

Simulations of the LCHS circuit are performed on NERSC using the C++ emulator QuCF\cite{QuCF} based on the toolkit QuEST\cite{Jones19} with GPU-parallelization on a single GPU device.
For instance, the KvN case described in Sec.~\ref{sec:LCHS-KvN} and simulated for the temporal interval $t_1 = 2$ took around $6$ hours on a $80$ GB GPU device.
Detailed circuits of all discussed cases can be found in Ref.~\onlinecite{code-KvN-LCHS}.

\subsection{Linear advection-diffusion equation}\label{sec:num-LCHS-ADE}
To test the LCHS algorithm numerically, we solve the one-dimensional linear ADE:
\begin{equation}\label{eq:ADE}
    \partial_t\psi = - v\partial_x\psi + D \partial_x^2 \psi,
\end{equation}
where $v$ is a constant velocity, $D$ is a constant diffusivity.
The field $\psi = \psi(t,x)$ is discretized on the following spatial grid:
\begin{equation}
    x_j = j\, \Delta x, \quad \Delta x = x_{\rm max}/(N_x - 1),\quad j = [0,N_x).
\end{equation}
and initialized as 
\begin{equation}\label{eq:ADE-init}
    \psi(0, x) = \exp\left[- \frac{(x - x_{\rm max}/2)^2}{2\sigma_0^2} \right],
\end{equation}
where $\sigma_0 = 0.05$ and $x_{\rm max} = 1.0$.
After applying the central finite difference scheme to the first and second spatial derivatives and using periodic boundary conditions, one obtains
\begin{subequations}\label{eq:ADE-discr}
\begin{eqnarray}
    &&\partial_t \psi_j = c_{-1}\psi_{j-1} + c_{0} \psi_j + c_{+1} \psi_{j+1},\\
    &&c_{0} = - \frac{2 D}{\Delta x^2},\quad c_{\pm 1} = \left(\frac{D}{\Delta x^2} \mp \frac{v}{2\Delta x}\right),
\end{eqnarray}
\end{subequations}
where $\psi_{-1} = \psi_{N_x-1}$, $\psi_{N_x} = \psi_0$, and $j = [0,N_x)$.
In the ADE, we use $D = 0.01$, $v = 1.0$, and $n_x = 6$ (for classical simulations, $N_t = 2^8$ is used).
To model the ADE with the LCHS algorithm, we recast the system~\eqref{eq:ADE-discr} into form~\eqref{eq:initial-value-problem} with the following matrix:
\begin{equation}\label{eq:ADE-A}
    A_{i_r i_c} = -
    \left\{ \begin{aligned}
        c_{-1}&,\quad i_c = i_r - 1,\\
        c_{0}&,\quad i_c = i_r,\\
        c_{+1}&,\quad i_c = i_r + 1,
    \end{aligned}\right.
\end{equation}
where $i_r, i_c = [0,N_x)$, $A_{(N_x-1), N_x} \equiv A_{(N_x-1),0}$, and $A_{0, -1} \equiv A_{0,(N_x-1)}$.

The circuit for performing the quantum computation of the ADE is shown schematically in Fig.~\ref{circ:ADE-LCHS} where extra ancillae $a_{\rm init}$ are used for the circuit initialization.
The initial condition~\eqref{eq:ADE-init} is encoded using QSVT circuit~\ref{circ:qsvt-even}.
To increase the success probability of the initialization, the AA technique, analogous to the algorithm in Fig.~\ref{circ:LCHS-AA} is applied.
Therefore, one needs three qubits in the register $a_{\rm init}$ in Fig.~\ref{circ:ADE-LCHS} where two qubits are needed for the QSVT computation and one extra qubit is used by AA.
To construct the LCHS block in Fig.~\ref{circ:ADE-LCHS}, we follow the algorithm described in Sec.~\ref{sec:LCHS-LCU} where the matrices $A_H$, $B_{\rm max}$, and $B_k$ are block-encoded into QSP circuits following the algorithm briefly described in Appendix~\ref{app:be}.
Since matrix~\eqref{eq:ADE-A} has only constant elements, one can perform the BE using techniques described in detail in Refs.~\onlinecite{Novikau22, Novikau23, Novikau24-EVM}.

The matrices $A_H$ and $A_L$ computed according to Eqs.~\eqref{eq:A-decomposition} for the ADE commute with each other. 
Because of this, there is no need to perform the trotterization and one can take $\tau = t$. 
However, since the computation of the QSP angles for large $t$ can be numerically challenging, one might still need to split $t$ into shorter temporal intervals, but now the length of these intervals does not have to satisfy the rule~\eqref{eq:trot-step-2}.

The maximum absolute approximation error of the LCHS simulations of the ADE is shown in Fig.~\ref{fig:ADE-err-kmax} as a function of $k_{\rm max}$ for various choices of $\Delta k$.
Here, the results are computed without invoking circuit~\ref{circ:ADE-LCHS} but just by using LCHS equation~\eqref{eq:LCHS}. 
One can see from this figure that, for the chosen time instant, it is enough to take $\Delta k \approx 0.16$ since any further decrease of the Fourier step does not significantly improve the LCHS precision.
Instead, one can still reduce the LCHS error by using larger values of $k_{\rm max}$.

\begin{figure}[!t]
\centering
\includegraphics[width=0.49\textwidth]{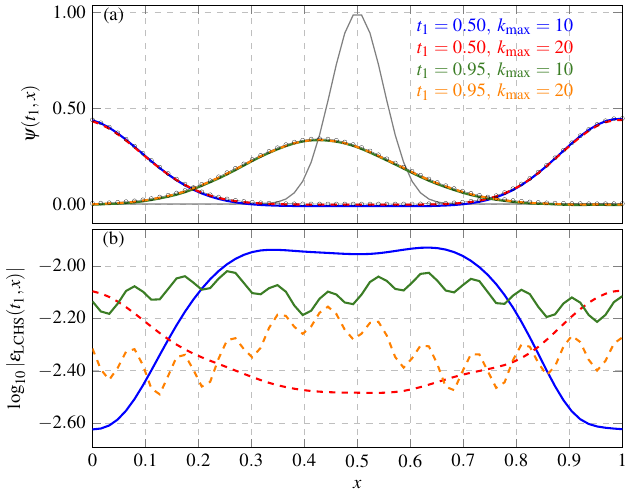}
\caption{
    \label{fig:ADE-QuCF}
    Results from the LCHS simulations of the ADE [Eq.~\eqref{eq:ADE-discr}] using circuit~\ref{circ:ADE-LCHS} with $\Delta k = 1.575\times 10^{-1}$.
    Simulations with $k_{\rm max} = 10$ and $k_{\rm max} = 20$ are performed with $n_k = 7$ and $n_k = 8$, respectively.
    (a): The spatial distribution $\psi(t_1,x)$ obtained by the LCHS for various $t_1$ and $k_{\rm max}$ are shown by colored lines. 
    The initial condition~\eqref{eq:ADE-init} is shown in gray, and the classical simulations are shown by black markers.
    (b): The log of the difference between classical and LCHS simulations for the same values of $t_1$ and $k_{\rm max}$.
}
\end{figure}

\begin{figure}[!t]
\centering
\includegraphics[width=0.49\textwidth]{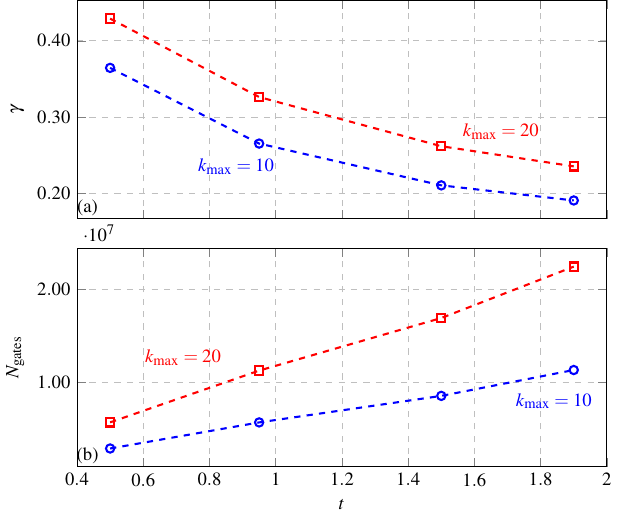}
\caption{
    \label{fig:ADE-QuCF-Ng-ssp}
    (a): Success probability of circuit~\ref{circ:ADE-LCHS} used for modeling the ADE.
    (d): The number of STMC gates in this circuit.
}
\end{figure}

We also model the ADE by using circuit~\ref{circ:ADE-LCHS}.
The results are shown in Fig.~\ref{fig:ADE-QuCF}.
The approximation errors in these simulations are comparable with those discussed above, and the quantum state returned by the LCHS circuit correctly reproduces the wave function $\psi(t_1,x)$.
One can also see from Fig.~\ref{fig:ADE-QuCF-Ng-ssp} that the success probability of the circuit is close to one but decreases in time due to the damping of the simulated wave function by the diffusivity $D$.
For the chosen $k_{\rm max}$, the number of STMC gates grows linearly with the length of the simulated time interval and is of the order of $10^7$.


\begin{figure*}[!t]
\centering
\includegraphics[width=0.90\textwidth]{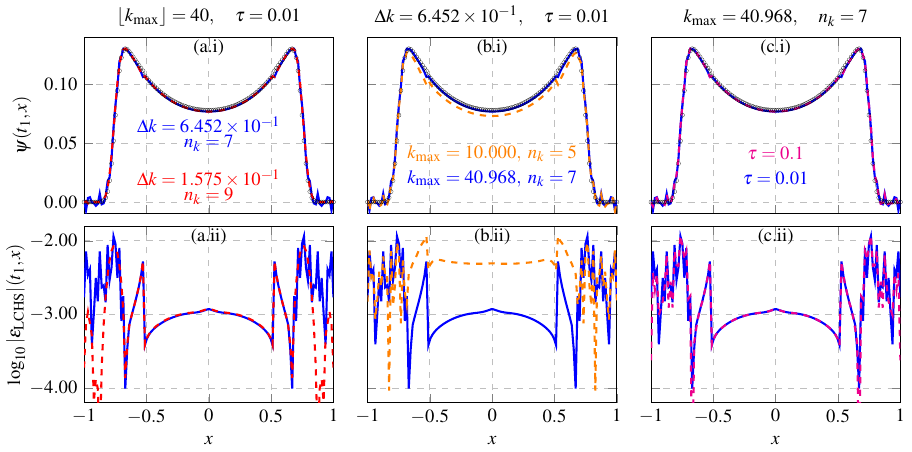}
\caption{
    \label{fig:KvN-LCHS-py}
    Plots showing the comparison of the classical KvN simulations of Eq.~\eqref{eq:KvN-LCHS-problem} with the LCHS simulations of the same problem (without invoking the LCHS circuit).
    Here, $t_1 = 1.0$.
    The black markers in all plots of the upper row indicate the classical simulation.
    The colored lines show the LCHS simulations.
    (a.i) and (a.ii): $\psi(t_1, x)$ and $\varepsilon_{\rm LCHS}(t_1, x)$, correspondingly, for various $\Delta k$ keeping $\floor{k_{\rm max}} = 40$. 
    (b.i) and (b.ii): $\psi(t_1, x)$ and $\varepsilon_{\rm LCHS}(t_1, x)$ for various $k_{\rm max}$ keeping $\Delta k = 0.65$.
    (c.i) and (c.ii): $\psi(t_1, x)$ and $\varepsilon_{\rm LCHS}(t_1, x)$ for various $\tau$ keeping $\floor{k_{\rm max}} = 40$ and $\Delta k = 0.65$.
}
\end{figure*}

\subsection{LCHS simulation of a nonlinear problem}\label{sec:LCHS-KvN}

\begin{figure}[!t]
\centering
\includegraphics[width=0.46\textwidth]{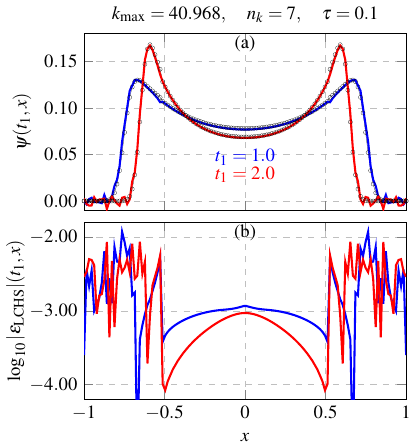}
\caption{
    \label{fig:KvN-LCHS-QuCF}
    Comparison of the classical KvN simulations of Eq.~\eqref{eq:KvN-LCHS-problem} with the LCHS simulations of the KvN system using quantum circuit~\ref{circ:LCHS-AA}.
    The black markers indicate the classical simulations at $t_1 = 1.0$ and $t_1 = 2.0$.
    The colored lines correspond to the LCHS simulations.
    The state vectors $\psi(t_1, x)$ and the absolute error of the LCHS simulations are shown in plots (a) and (b), correspondingly.
}
\end{figure}

In this section, we apply the LCHS algorithm for modeling the KvN formulation~\eqref{eq:kvn} with UW Hamiltonian~\eqref{eq:kvn-h-uw} for the following nonlinear problem:
\begin{equation}\label{eq:KvN-LCHS-problem}
    F(x) = -\sin(x) [\sin^2(x) - 1/4],
\end{equation}
where we use $\sin(x)$ to simplify the BE of the system which is discussed in detail in Appendix~\ref{app:be-kvn}.
For the discretization of the nonlinear problem, we take $n_x = 7$. 
After constructing the matrices $A_H$ and $A_L$ using Eqs.~\eqref{eq:Aa-UW} and~\eqref{eq:Ah-UW}, one solves the LCHS equation~\eqref{eq:LCHS}.
First, as a test, we perform this simulation without invoking the LCHS circuit~\ref{circ:LCHS-AA}.
In Fig.~\ref{fig:KvN-LCHS-py}, one can see the comparison between classical and LCHS computations of the KvN formulation of Eq.~\ref{eq:KvN-LCHS-problem}.
These simulations demonstrate that one can keep relatively large trotterization step $\tau = 0.1$ and have an absolute error around $10^{-2}$.
Also, to reduce spurious numerical oscillations near the boundary, one should take $\Delta k \leq 0.16$ for the shown time interval, $t_1 = 1.0$. 
To decrease the LCHS error even further, one should take a larger $k_{\rm max}$.

The next step is simulating our nonlinear problem by emulating LCHS quantum circuit~\ref{circ:LCHS-AA} and using $n_x$ Hadamard gates to encode initial conditions according to Eq.~\eqref{eq-psi-init-flat}.
For this task, we take $k_{\rm max} = 40.968$ and $n_k = 7$ which corresponds to $\Delta k = 6.452\times 10^{-1}$. 
The chosen trotterization step is $\tau = 0.1$.
For these simulations, we use explicit circuits of the BE oracles encoding the matrices $A_H$, $B_k$, and $B_{\rm max}$.
The BE is described in Appendix~\ref{app:be-kvn}.
Here, we also use the direct computation of the LCHS weights, as described in Sec.~\ref{sec:Ow}, instead of the calculation by QSVT which allows us to remove one ancillary qubit.

In these simulations, the LCHS circuit requires $25$ logical qubits where $n_x = 7$ qubits are used for encoding the spatial distribution of the state vector $\psi$, $n_k = 7$ qubits are used for encoding the Fourier space, $6$ ancillary qubits are used for the BE, $3$ ancillae are necessary for computing the amplified LCHS weights, and two qubits are used by the QSP.
For $t_1 = 1.0$, the LCHS circuit consists of around $10^8$ STMC gates and returns $\psi(t_1,x)$ with success probability $0.88$.
For $t_1 = 2.0$, the LCHS circuit consists of around $2\times 10^8$ STMC gates and returns $\psi(t_1,x)$ with success probability $0.84$.

The results from the emulation of the LCHS circuit are shown in Fig.~\ref{fig:KvN-LCHS-QuCF}.
One can see some spurious numerical oscillations at $t_1 = 2$ whose amplitudes are smaller at $t_1 = 1$.
According to Fig.~\ref{fig:KvN-LCHS-py}.a, one needs to decrease $\Delta k$ to remove these oscillations, in a manner consistent with Fig.~\ref{fig:LCHS-scalar-errors}, because simulating longer temporal intervals requires smaller Fourier steps. 
For the chosen LCHS parameters, the emulation of LCHS circuit~\ref{circ:LCHS-AA} demonstrates a precision similar to that observed in the direct computation (Fig.~\ref{fig:KvN-LCHS-py}) of LCHS equation~\eqref{eq:LCHS}.
This proves that our BE described in Appendix~\ref{app:be-kvn} is correct and that the LCHS circuit combined with the KvN technique can be successfully used for modeling nonlinear problems.

\section{Conclusions}\label{sec:conclusions}
We proposed an explicit quantum algorithm (QA) for simulating the advection-diffusion equation with optimal scaling in time.
This same QA allows one to simulate the advection, Liouville, Koopman-von Neumann (KvN), and Koopman-van Hove (KvH) equations efficiently.
Thus, we provide an explicit QA for the
KvN formulation of nonlinear differential equations and an explicit implementation of the corresponding quantum circuit.
By using a nonunitary version of the truncated KvN model, we managed to reduce the level of numerical noise in the KvN simulations significantly.
To handle the nonunitary dynamics, we combined the KvN approach with the LCHS algorithm, resulting in a method that can be used for modeling a broad class of nonlinear problems.
The resulting KvN-LCHS circuit features an LCU-like structure. 
In this approach, QSVT circuits enhanced by AA are used to compute the LCHS weights, while QSP circuits are applied for performing unitary Hamiltonian simulations.
The BE oracles used by the QSP are constructed as a combination of quantum arithmetic operations and QSVT circuits.
The quantum circuits were constructed with an explicit BE for both the ADE and KvN cases and have been verified through simulation on the QuCF digital emulator of fault-tolerant quantum computers.

We have shown that by using AA to boost the success probability of the subcircuits computing the LCHS weights, one can eliminate the global multiplicative factor $\oO(k_{\rm max})$ in the algorithm's scaling while still guaranteeing a high overall success probability of the QA.
We have also demonstrated that the algorithm's scaling strongly depends on the chosen encoding of the dependence on the Fourier coordinate $k$.
By performing the proposed coordinate transformation in the LCHS equation [Eqs.~\eqref{eq:LCHS-theorem1-ksinphi} and \ref{eq:Vj-eff}], one can efficiently encode the dependence on $k$, resulting in a linear LCHS complexity with time.

Due to the high success probability, this algorithm has a significant advantage over the QLSAs in the number of calls to the initialization circuit.
Yet, as is the case for most of the QAs developed for solving differential equations, the KvN-LCHS circuit requires significant resources and can be implemented only on fault-tolerant quantum computers.
Nevertheless, the proposed LCHS-based algorithm, in which the construction of the BE oracles is the only problem-specific component of the method, is universal enough to be applied to simulating both linear and nonlinear dissipative problems that are of practical interest. 
In particular, if one is interested in measuring robust observables, e.g. smooth integrals over phase-space, the algorithm can have an exponential speedup factor in modeling KvN, KvH, Liouville and advection-diffusion equations widely used in fluid and kinetic simulations.

\section*{Acknowledgments}
The authors thank Dong An and Lin Lin for proposing the use of AA in the computation of the LCHS weights to improve the LCHS success probability.
The authors also thank Frank Graziani, Vasily Geyko, and Roger Minich for valuable discussions.
This work, LLNL-JRNL-869590, was supported by the U.S. Department of Energy (DOE) Office of Fusion Energy Sciences “Quantum Leap for Fusion Energy Sciences” Project No. FWP-SCW1680 at Lawrence Livermore National Laboratory (LLNL). 
Work was performed under the auspices of the U.S. DOE under LLNL Contract DE-AC52–07NA27344.
This research used resources of the National Energy Research Scientific Computing Center, a DOE Office of Science User Facility supported by the Office of Science of the U.S. Department of Energy under Contract No. DE-AC02-05CH11231 using NERSC award FES-ERCAP0028618.

\appendix
\section{Linear Combination of Unitaries (LCU)}\label{app:LCU}

Let us consider a general LCU circuit to construct the following sum 
\begin{equation}\label{eq:app-sum}
    \sum_{k=0}^{2^{n_k}-1} w_k S_k
\end{equation}
where $S_k$ are unitary operators whose behaviour depends on the index $k$, and $w_k$ are scalar coefficients.
We assume that we have an oracle $O_{\sqrt{w}}$ computing the weights $w_k$ as
\begin{equation}
    O_{\sqrt{w}}\ket{0}_{r_{a,0}}\ket{0}_{r_k} = \frac{1}{2^{n_k/2}}\sum_k (\sqrt{w_k} \ket{0}_{r_{a,0}} + \eta_k\ket{1}_{r_{a,0}}) \ket{k}_{r_k},
\end{equation}
and that the selector $S$ is defined by
\begin{equation}
    S \ket{k}_{r_k}\ket{\psi}_{r_s} = \ket{k}_{r_k} S_k\ket{\psi}_{r_s}.
\end{equation}
Here, $\eta_k$ is a scalar introduced for each $k$ to keep the state of the ancilla $r_{a,0}$ normalized to one, and $\ket{\psi}_{r_s}$ is the input state to be modified by the LCU circuit.
Now, we introduce an extra ancilla $r_{a,1}$ that will be the target qubit for the Hermitian-adjoint operator $O_{\sqrt{w}}^\dagger$, and the resulting LCU circuit can be described as
\begin{align}
    O_{\sqrt{w}}^\dagger\, &S\, O_{\sqrt{w}}\ket{00}_{r_a}\ket{0}_{r_k}\ket{\psi}_{r_s} = \nonumber\\
        \frac{1}{2^{n_k}}\sum_k &\left[
            w_k \ket{00}_{r_a} + \sqrt{w_k}^\dagger\eta_k\ket{01}_{r_a} -  \eta_k^\dagger\sqrt{w_k}\ket{10}_{r_a} - \eta_k^2\ket{11}_{r_a}
        \right]\nonumber\\
        &\ket{0}_{r_k} S_k\ket{\psi}_{r_s}.
\end{align}
In other words, the LCU returns the required sum~\eqref{eq:app-sum} as a superposition of state amplitudes of the register $r_s$ entangled with the zero state of the ancillary registers $r_a$ and $r_k$.

\section{Testing LCHS on linear systems with weak and strong dissipation}\label{app:num-LCHS-toy}

\begin{figure}[!t]
\centering
\includegraphics[width=0.48\textwidth]{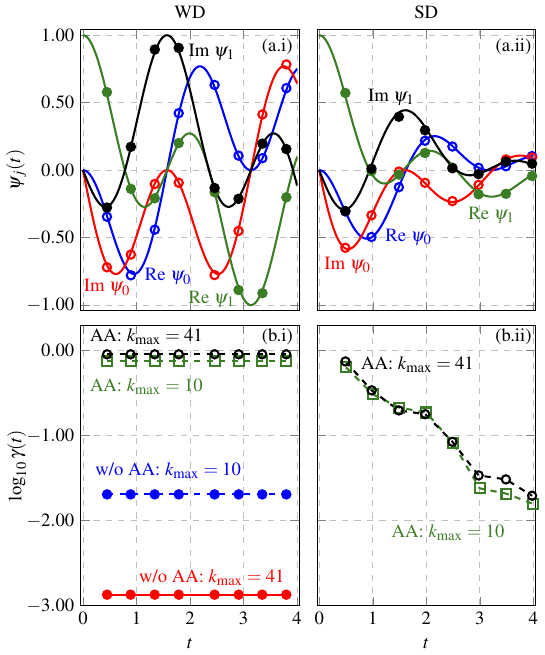}
\caption{
    \label{fig:LCHS-WD-SD-signal-ssp} 
    (a.i) and (a.ii): Time evolution of the systems~\eqref{eq:test-WD} and~\eqref{eq:test-SD}.
    The solid lines are classical simulations. 
    The markers are the LCHS simulations using circuit~\ref{circ:LCHS-AA} with $k_{\rm max} = 41$ and $n_k = 7$.
    (b.i) and (b.ii): Dependence of the LCHS success probability, $\gamma$, on the simulated time $t$ for various $k_{\rm max}$ without AA (filled markers) and with AA (hollow markers) of the LCHS weights.
}
\end{figure}
\begin{figure}[!t]
\centering
\includegraphics[width=0.48\textwidth]{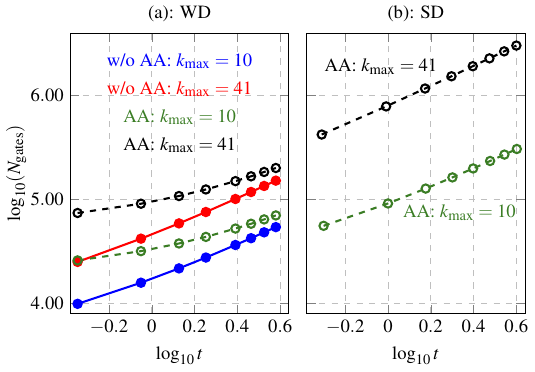}
\caption{
    \label{fig:LCHS-WD-SD-Ng} 
    (a): Dependence of the number of STMC gates, $N_{\rm gates}$, in the LCHS circuit on $t$ for various $k_{\rm max}$ with and without AA of the LCHS weights for simulating system~\eqref{eq:test-WD} with weak dissipation (WD).
    (b): $N_{\rm gates}$ in the LCHS circuit with AA for simulating system~\eqref{eq:test-SD} with strong disipation (SD).
}
\end{figure}

To analyse the influence of AA on the success probability of the LCHS circuit, we consider two linear zero-dimensional nonunitary problems.
The first case is a near-unitary system with weak dissipation (WD):
\begin{equation}\label{eq:test-WD}
    \partial_t 
        \begin{pmatrix}
            \psi_0 \\
            \psi_1
        \end{pmatrix} = -\yi
        \begin{pmatrix}
            1 - \yi\times 10^{-4} & 2 \\
            2 & 1
        \end{pmatrix} 
        \begin{pmatrix}
            \psi_0 \\
            \psi_1
        \end{pmatrix}.
\end{equation}
The second system has strong dissipation (SD):
\begin{equation}\label{eq:test-SD}
    \partial_t 
        \begin{pmatrix}
            \psi_0 \\
            \psi_1
        \end{pmatrix} = -\yi
        \begin{pmatrix}
            1 - \yi & 2 \\
            2 & 1
        \end{pmatrix} 
        \begin{pmatrix}
            \psi_0 \\
            \psi_1
        \end{pmatrix}.
\end{equation}
In both cases, the initial state is $\psi_0 = 0$ and $\psi_1 = 1$.

In these tests, we fix $\Delta k$ keeping $N_k$ relatively low to avoid modeling large quantum circuits because $N_k$ increases the circuit width logarithmically.
After that, we calculate the trotterization step according to
\begin{equation}\label{eq:tau-SD}
    \tau = \eta_{\tau} / (k_{\rm max} ||A_L||),
\end{equation}
or
\begin{equation}\label{eq:tau-WD}
    \tau = \eta_{\tau} / ||A_H||,
\end{equation}
whichever gives the smallest $\tau$.
The positive scalar $\eta_{\tau}$ is a multiplicative factor whose variation influences the trotterization error in Eq.~\eqref{eq:Gj-decomposition}.

Then, the time step $\tau$ is used to find the number of trotterization steps, $N_t - 1$, in the LCHS circuit via
\begin{equation}
    N_t = 1 + t/\tau.
\end{equation} 
The choice between Eq.~\eqref{eq:tau-SD} and Eq.~\eqref{eq:tau-WD} is determined by the strength of the dissipation in the system.
If the dissipation is strong, the time step is computed using Eq.~\eqref{eq:tau-SD}.

Ideally, one should have $\eta_{\tau} \leq 1$.
Yet, numerical tests with relatively low $k_{\rm max}$ show that one can take much larger $\eta_{\tau}$ since the trotterization error [Eq.~\eqref{eq:Gj-decomposition}] remains smaller than the error imposed by the truncation of the Fourier space. 
In these simulations whose results are shown in Fig.~\ref{fig:LCHS-WD-SD-signal-ssp}, we achieved an absolute approximation error around $10^{-2}$.
To reduce the error even further, one needs to increase $k_{\rm max}$ keeping $\Delta k$ unchanged.
Clearly, when $k_{\rm max}$ becomes large enough, one needs to decrease $\eta_{\tau}$ as well to keep the trotterization error small enough.

We model systems~\eqref{eq:test-WD} and~\eqref{eq:test-SD} using LCHS circuit~\ref{circ:LCHS-AA} with $\eta_{\tau} = 1$ and $\Delta k = 6.452\times 10^{-1}$ for various time points.
The temporal evolution of the WD system simulated classically (solid lines) and by the LCHS circuit (markers) are shown in Fig.~\ref{fig:LCHS-WD-SD-signal-ssp}.a.i for the case with $k_{\rm max} = 41$ and $n_k = 7$.
Here, the maximum absolute error is $\varepsilon_{\rm LCHS} = 2.405\times 10^{-2}$.
In a similar case with $k_{\rm max} = 10$ and $n_k = 5$, the error becomes $\varepsilon_{\rm LCHS} = 7.578\times 10^{-2}$.
In Fig.~\ref{fig:LCHS-WD-SD-signal-ssp}.b.i, one can see how the success probability of the LCHS circuit changes when one includes AA of the LCHS weights.
For instance, with $k_{\rm max} = 41$, the probability is boosted from around $10^{-3}$ (red markers) to $\oO(1)$ (black markers).
On the other hand, the inclusion of AA significantly increases the number of STMC gates in the LCHS circuit in simulations with relatively small $t$ as seen from Fig.~\ref{fig:LCHS-WD-SD-Ng}.a.
This happens due to the contribution of the subcircuit $O_{\sqrt{w}}^{\rm AA}$ and its Hermitian adjoint to LCHS circuit~\ref{circ:LCHS-AA}. 
However, the relative contribution of $O_{\sqrt{w}}^{\rm AA}$ drops with $t$ because the number of gates in the selector $S$, which is not touched by AA, grows with $t$. 
This behavior is consistent with Eq.~\eqref{eq-LLCHS-scaling-AAw} where AA influences only the additive factor in the LCHS scaling.

The temporal evolution of the SD system simulated classically (solid lines) and by the LCHS circuit (markers) are shown in Fig.~\ref{fig:LCHS-WD-SD-signal-ssp}.a.ii for the case with $k_{\rm max} = 41$ and $n_k = 7$.
Due to the strong dissipation, the ratio $\mu$ [Eq.~\eqref{eq:mu}] grows in time which results in the decrease of the LCHS success probability even in the case with AA (Fig.~\ref{fig:LCHS-WD-SD-signal-ssp}.b.ii).
In addition, the number of the gates in the SD case is higher (Fig.~\ref{fig:LCHS-WD-SD-Ng}.b) because the time step $\tau$ computed according to Eq.~\eqref{eq:tau-SD} is smaller than in the WD case. 
Because of this, $N_t$ is higher in the SD case. 
This increases the last additive term in the query complexity~\eqref{eq:S-complexity-init} of the selector $S$.
\section{Block Encoding (BE) Procedure}\label{app:be}

\begin{figure*}[!t]
\centering
\includegraphics[width=0.9\textwidth]{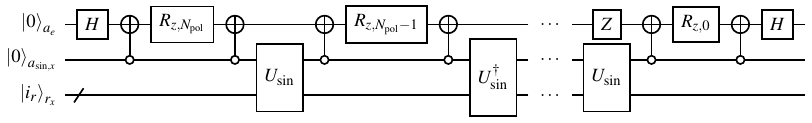}
\caption{
    \label{circ:qsvt} 
    The QSVT circuit encoding a real polynomial of order $N_{\rm pol}$, where $N_{\rm pol}$ is odd, by using $N_{\rm pol} + 1$ angles $\phi_k$ pre-computed classically.\cite{Gilyen19, Dong21, Ying22}
    The gates $R_{z,k}$ represent the rotations $R_z(2\phi_k)$.
    The ancilla $a_{\rm sin}$ is used to compute $y$ using the oracle $U_{\rm sin}$ shown in Fig.~\ref{circ:sin}.
}
\end{figure*}

To block-encode a Hermitian sparse matrix $M$ of size $N_M\times N_M$, we follow the methodology described in Refs.~\onlinecite{Berry12, Novikau22, Novikau23}.
According to Ref.~\onlinecite{Berry12}, the BE can be performed by using the following oracle:
\begin{equation}\label{eq:UA-Berry}
    U_A = O_F^\dagger O_{\sqrt{H}}^\dagger O_M O_{\sqrt{H}} O_F,
\end{equation}
where the suboracles $O_F$, $O_F^\dagger$, and $O_M$ encode the matrix structure, i.e. the positions of nonzero elements within the matrix, and the suboracles $O_{\sqrt{H}}$ and $O_{\sqrt{H}}^\dagger$ encode the values of the matrix elements into state amplitudes.
Since the state amplitudes are normalized to one, the matrix $M$ must be properly normalized as well:
\begin{equation}\label{eq:norm}
    M \to M/\ylb||M||_{\rm max}\varsigma\yrb,
\end{equation}
where the scalar $\varsigma$ is of the order of the matrix nonsparsity as explained in Eq.~\eqref{eq:nonsparsity}, and $||M||$ is the matrix spectral norm.

To implement $U_A$, one needs a state register, $r_x$, also called an input register, with $n = \yceil{\log_2 N_M}$ qubits to encode the row indices $i_r = [0, N_M)$ of the matrix $M$.
For example, in the LCHS circuit~\ref{circ:LCHS-AA}, $r_{\rm data}$ serves as the input register.
One also needs a register $a_x$ with $n_c = n$ ancilla qubits to encode the column indices $i_c = [0, N_M)$, and at least two supplemental ancillae serving as target qubits for $O_{\sqrt{H}}$ and $O_{\sqrt{H}}^\dagger$.
Because of $n_c$, the number of ancillae in $U_A$ grows logarithmically with the matrix size $N_M$.

To reduce $n_c$, one must take into account the structure of the target matrix $M$.
In particular, instead of addressing each column of $M$, bitstrings of $a_x$ can flag the nonzero diagonals of $M$. 
In other words, for a chosen matrix row, $a_x$ can store the positions of nonzero matrix elements as distances from the main diagonal (let us denote these distances as $l_d$).
This is particularly helpful when dealing with matrices that naturally arise after discretization in orthogonal coordinates.\cite{Novikau22, Novikau23, Novikau24-EVM}
In the case of sparse well-structured matrices, $l_d$ and the number of nonzero diagonals $N_{d}$ in the matrix $M$ do not change with $N_M$. 
Moreover, $N_{d}$ usually is much smaller than the matrix size, i.e. $N_d \ll N_M$.
Because of this, the register $a_x$ contains $n_c = \yceil{\log_2{N_d}}$ qubits instead of $n$ qubits, and $n_c$ does not depend on $N_M$ anymore.

However, when $a_x$ is used for flagging nonzero diagonals in $M$, one generally cannot use the BE in the form of Eq.~\eqref{eq:UA-Berry} any longer.
Instead, one needs to replace $O_F^\dagger$, which is just the Hermitian-adjoint version of $O_F$, with another oracle denoted $O_B^\dagger$ (where `B' stands for `backward').
The structure of $O_B^\dagger$ is usually similar to that of $O_F^\dagger$ but slightly modified so that, together with $O_F$ and $O_M$, the oracle $O_B^\dagger$ correctly encodes the structure of $M$.

The goal of the oracle $O_M$ is to map $l_d$ stored in the register $a_x$ to the column indices $i_c$ of the corresponding nonzero matrix elements and write $i_c$ back to the input register $r_x$.
Thus, one can schematically understand the action of the oracles $O_F$ and $O_M$ in the following way:
\begin{subequations}\label{eq:OFF-OM}
\begin{eqnarray}
    &&O_F: \ket{0}_{a_x}\ket{i_r}_{r_x} \to \sqrt{d_{i_r, i_c}}\ket{l_d}_{a_x}\ket{i_r}_{r_x},\\
    &&O_M: \sqrt{d_{i_r, i_c}}\ket{l_d}_{a_x}\ket{i_r}_{r_x} \to \sqrt{d_{i_r, i_c}}\ket{l_d}_{a_x}\ket{i_c}_{r_x},
\end{eqnarray}
\end{subequations}
where the nonzero scalars $d_{i_r, i_c} < 1$, which are usually real and positive, are necessary to keep the quantum states normalized to one. 
The purpose of the oracle $O_B^\dagger$ is to entangle the zero state of the ancillae $a_x$ with $\ket{i_c}_{r_x}$:
\begin{equation}\label{eq:OFB}
    O_B^\dagger: \sqrt{d_{i_r, i_c}}\ket{l_d}_{a_x}\ket{i_c}_{r_x} \to d_{i_r, i_c}\ket{0}_{a_x}\ket{i_c}_{r_x}.
\end{equation}

The oracles $O_{\sqrt{H}}$ and $O_{\sqrt{H}}^\dagger$ encode the square roots of the matrix elements and their complex-conjugates, respectively. 
The main reason behind the introduction of the Hermitian-adjoint $O_{\sqrt{H}}^\dagger$ is to return all ancilla qubits used by $O_{\sqrt{H}}$ back to the zero state.
Since we operate with different matrix diagonals, we consider functions $f(x_{i_r})$ that lie on these diagonals. 
These functions can be constant and do not depend on $i_r$, except possibly at a few boundary points, as in the matrix~\eqref{eq:ADE-A} used for modeling the ADE.
In this case, one can encode all values of nonzero matrix elements by a few rotation gates at once.\cite{Novikau23, Novikau24-EVM}
If $f(x_{i_r})$ is a continuous function whose values gradually change with $i_r$, then it can be encoded by the QSVT.
For example, this is the case for the KvN problem discussed in Appendix~\ref{app:be-kvn}.
Even if $f(x_{i_r})$ is real, its square root can be a complex function, which is more complicated for encoding than real functions.
To reduce the number of ancillae by at least one and to avoid computing complex functions, we replace $O_{\sqrt{H}}$ and $O_{\sqrt{H}}^\dagger$ with a single oracle $O_H$, which directly computes the element values instead of computing their square roots.
Taking into account the above modifications, the BE oracle~\eqref{eq:UA-Berry} becomes
\begin{equation}\label{eq:UA}
    U_A = O_B^\dagger O_M O_H O_F.
\end{equation}

Using QSVT to encode $f(x_{i_r})$ requires the following transformation:
\begin{equation}\label{eq:f-to-arcsin}
    f(x) \rightarrow f(\arcsin(y))
\end{equation}
where $y = \sin(x)$, and we assume that $x_{\rm max} = 1$, otherwise, it is also necessary to rescale $x$ by $x_{\rm max}$.
The oracle $O_H$ computes $f(\arcsin(y))$ instead of $f(x)$.
The reason for the transition from $x$ to $y$ is the same as in Sec.~\ref{sec:Ow}, i.e. it is much easier to efficiently encode $\sin(x_{i_r})$ than $x_{i_r}$. 
In the general case, the latter requires $N_x$ multi-controlled rotation gates, while the former needs only $n_x$ single-controlled gates, according to Fig.~\ref{circ:sin}.

If the function $f$ does not have definite parity, it should be decomposed into its odd and even components.
If $f$ is originally complex, it should be also decomposed into its real and imaginary components where each component is then decomposed into its odd and even parts. 
In this case, each component is computed by a separate QSVT circuit (Figs.~\ref{circ:qsvt} and~\ref{circ:qsvt-even}), and then the terms are added to each other by using the LCU circuit which requires one or two extra ancillae depending on the number of terms.

To compute the real function $f(\arcsin(y))$ with definite parity using the QSVT, one needs to represent the function as a sequence of Chebyschev polynomials, which is then used for finding the QSVT angles. 
We should emphasize that even if the original function $f(x)$ can be represented with high precision by just a few Chebyschev polynomials, its modified version $f(\arcsin(y))$ usually requires a significantly higher number of them.

Let us denote all ancilla qubits necessary for the implementation of $O_H$ as $a_h$.
Since the oracles $O_F$, $O_B^\dagger$, and $O_M$ act on the ancillae $a_x$, while the oracle $O_H$ acts on the register $a_h$, we can separately consider the following operator:
\begin{equation}\label{eq:UD}
    U_D = O_B^\dagger O_M O_F.
\end{equation}
This oracle does not include $O_H$ and encodes only the structure of $M$, excluding the actual values of the matrix elements.
The projection of the matrix representation of $U_D$ onto the space entagled with $\ket{0}_{a_x}$ is a matrix $D$ with elements $d_{i_r, i_c}$.
The positions of the nonzero elements in $D$ correspond to the positions of the nonzero elements in $M$.
According to Eqs.~\eqref{eq:OFF-OM} and~\eqref{eq:OFB}, the action of $U_D$ can be schematically written as 
\begin{equation}\label{eq:UD-action}
    U_D: \ket{0}_{a_x}\ket{i_r}_{r_x} \to d_{i_r i_c}\ket{0}_{a_x}\ket{i_c}_{r_x},
\end{equation}
where the register $a_h$ is kept untouched.
Therefore, the goal of the oracle $O_H$ is to perform the following mapping:
\begin{equation}\label{eq:OH-action}
    O_H: d_{i_r i_c}\ket{0}_{a_h}\ket{0}_{a_x}\ket{i_c}_{r_x} \to M_{i_r i_c}\ket{0}_{a_h}\ket{0}_{a_x}\ket{i_c}_{r_x}.
\end{equation}
To allow one to perform the mapping~\eqref{eq:OH-action}, one must rescale the matrix elements of $M$,
\begin{equation}\label{eq:rescaling}
    M_{i_r i_c}^\prime = \frac{M_{i_r i_c}}{|d_{i_r i_c}|}
\end{equation}
Since $d_{i_r i_c} \leq 1$, one has $M_{i_r i_c}^\prime \geq M_{i_r i_c}$, and $M_{i_r i_c}^\prime$ can be larger than one.
Because of this, the following scalar factor
\begin{equation}\label{eq:nonsparsity}
    \varsigma = \max_{i_r,i_c, d_{i_ri_c}\neq 0}|d_{i_ri_c}|^{-1}
\end{equation}
must be included in the normalization~\eqref{eq:norm} to guarantee that $M_{i_r i_c}^\prime \leq 1$ for any $i_r$ and $i_c$.
Typically, the oracle $U_D$ is constructed using only multi-controlled Hadamard $H$ and Pauli $X$ gates.
Therefore, the values $d_{i_ri_c}$ are usually equal to integer powers of $2^{-1/2}$ (e.g. one can see Eq.~\eqref{eq:D-Aa}), and the upper bound of the scalar $\varsigma$ can be estimated without computing the matrix $D$ as
\begin{equation}
    \varsigma = 2^{\yceil{\log_2\varsigma^\prime}},
\end{equation}
where $\varsigma^\prime$ is defined as the maximum number of nonzero elements in any row or column of the target matrix $M$.

Hence, to perform BE, and before constructing the BE quantum circuit, one should normalize the original matrix $M$ according to Eq.~\eqref{eq:norm} using the scalar~\eqref{eq:nonsparsity}, and then each element of the normalized matrix should be rescaled according to Eq.~\eqref{eq:rescaling}.
The structure of the target matrix is encoded by the oracle $U_D$ and the matrix elements of $M^\prime$ are used to compute parameters of the oracle $O_H$.
Finally, the matrix representation of the BE oracle $U_A$, after projection onto the space entangled with the zero state of all BE ancillae, is the matrix $M$ normalized according to Eq.~\eqref{eq:norm}.
One can find more details about BE in Refs.~\onlinecite{Novikau23, Novikau24-EVM}.

\begin{figure}[!t]
\centering
\subfloat[][Circuit $U_A$ for $A_H$.]{\includegraphics[width=0.42\textwidth]{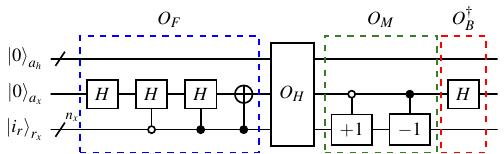}\label{circ:UA-Aa}}\\
\vspace{0.3cm}
\subfloat[][Circuit $U_A$ for $B_{\rm max}$ and $B_k$.]{\includegraphics[width=0.46\textwidth]{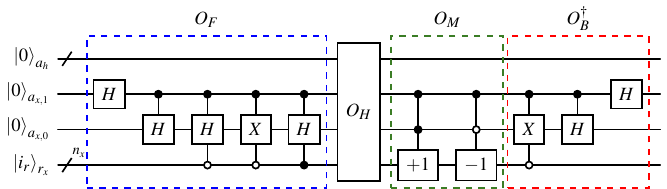}\label{circ:UA-Ah}}
\caption{
    \label{circ:BE}
    Schematic circuits of the BE oracles $U_A$ [Eq.~\eqref{eq:UA}] for encoding (a) the matrix $A_H$ [Eq.~\eqref{eq:Aa-UW}] and (b) the matrices $B_{\rm max}$ and $B_k$ [Eq.~\eqref{eq:B}]. 
    The detailed circuits of $U_A$ can be found in Ref.~\onlinecite{code-KvN-LCHS}.
    The colored dashed boxes indicate different parts of the oracle $U_D$ [Eq.~\eqref{eq:UD}].
    The matrices $B_{\rm max}$ and $B_k$ have the same structure, which is defined by the matrix $A_L$ [Fig.~\ref{fig:x2-matrices}a], and, therefore, have the same oracle $U_D$.
    The circuits of the incrementor and decrementor indicated here as `$+1$' and `$-1$', correspondingly, can be found in Ref.~\onlinecite{Novikau22}.
}
\end{figure}
\begin{figure}[!t]
\centering
\includegraphics[width=0.46\textwidth]{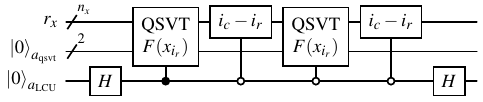}
\caption{
    \label{circ:LCU-QSVT}
    Linear combination of two QSVT circuits to compute the sum $F_{i_r} + F_{i_c}$.
    Here, the subcircuits indicated as `$i_c - i_r$' correspond to either an incrementor, or decrementor for $i_c - i_r = 1$ and $i_c - i_r = -1$, correspondingly.
}
\end{figure}

\section{Block-encoding matrices for the KvN example}\label{app:be-kvn}

To perform LCHS simulations with circuit~\ref{circ:LCHS-AA}, one needs to block-encode the Hermitian matrices $A_H$, $B_k$, and $B_{\rm max}$ [Eqs.~\eqref{eq:A-decomposition} and~\eqref{eq:B}] normalized according to Eq.~\eqref{eq:norm}.
The general procedure for the BE is briefly described in Appendix~\ref{app:be}. 
In this section, we focus on the particular steps that must be performed for block-encoding the nonlinear system~\eqref{eq:KvN-LCHS-problem}.

After normalization~\eqref{eq:norm}, the matrices $B_k$ and $B_{\rm max}$ are equal up to a sign, and their structure coincides with that of $A_L$, whose explicit expression is given in Eq.~\eqref{eq:Ah-UW}.
According to Fig.~\ref{fig:x2-matrices}a, this matrix is tridiagonal.
To address each of its nonzero diagonals, we introduce the ancilla register $a_x$ with two qubits.
In particular, the state $\ket{00}_{a_x}$ flags matrix elements on the main diagonal.
The state $\ket{10}_{a_x}$ marks the elements shifted by one cell to the left from the main diagonal (left sideband).
The state $\ket{11}_{a_x}$ marks the elements shifted by one cell to the right from the main diagonal (right sideband).
To summarize, the coding of the register $a_x$ is 
\begin{subequations}\label{eq:ax-Bk}
\begin{eqnarray}
    &&\ket{00}_{a_x} \to i_{c} = i_{r},\\
    &&\ket{10}_{a_x} \to i_{c} = i_{r} - 1,\\
    &&\ket{11}_{a_x} \to i_{c} = i_{r} + 1,
\end{eqnarray}
\end{subequations}
where $i_r$ and $i_c$ are the row and column indices of the matrix $A_L$.
Similarly, for the two-diagonal matrix $A_H$ [Fig.~\ref{fig:x2-matrices}b], we introduce the ancilla register $a_x$ with a single qubit:
\begin{subequations}\label{eq:ax-Aa}
\begin{eqnarray}
    &&\ket{0}_{a_x} \to i_{c} = i_{r} + 1,\\
    &&\ket{1}_{a_x} \to i_{c} = i_{r} - 1.
\end{eqnarray}
\end{subequations}
Quantum circuits performing the above transformations are shown in Fig.~\ref{circ:BE}.
The nonzero elements $d_{i_r i_c}$ of the matrix representation of the oracle $U_D$ [Eq.~\eqref{eq:UD}] encoding the structure of $A_H$ are
\begin{eqnarray}\label{eq:D-Aa}
    d_{i_r i_c} = \left\{ \begin{aligned}
                1/2,        &\qquad i_r = [0,N_x-2], &\quad i_c = i_r - 1,\\
                1/\sqrt{2}, &\qquad i_r = N_x-1,     &\quad i_c = i_r - 1,\\
                1/2,        &\qquad i_r = [1,N_x),   &\quad i_c = i_r + 1,\\
                1/\sqrt{2}, &\qquad i_r = 0,         &\quad i_c = i_r + 1.
        \end{aligned} \right.
\end{eqnarray}
The corresponding nonzero values $d_{i_r i_c}$ of $U_D$ encoding the matrix $A_L$ are 
\begin{eqnarray}\label{eq:D-Ah}
    d_{i_r i_c} = \left\{ \begin{aligned}
                1/4,        &\qquad i_r = [0,N_x-2], &\quad i_c = i_r - 1,\\
                1/\sqrt{8}, &\qquad i_r = N_x-1,     &\quad i_c = i_r - 1,\\
                1/2,        &\qquad i_r = [0,N_x),   &\quad i_c = i_r,\\
                1/4,        &\qquad i_r = [1,N_x),   &\quad i_c = i_r + 1,\\
                1/\sqrt{8}, &\qquad i_r = 0,         &\quad i_c = i_r + 1.
        \end{aligned} \right.
\end{eqnarray}
These values should be used in Eq.~\eqref{eq:rescaling} for rescaling the matrix elements of the normalized matrices $A_H$, $B_k$ and $B_{\rm max}$ before constructing the oracle $O_H$.

\begin{figure*}[!t]
\centering
\includegraphics[width=0.90\textwidth]{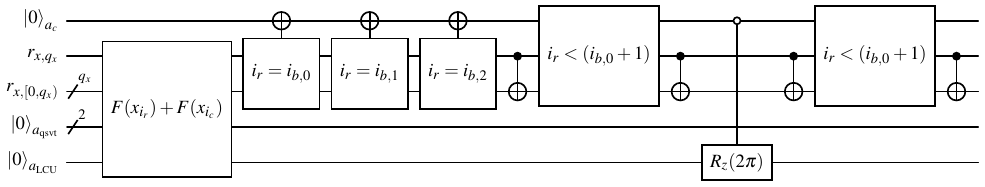}
\caption{
    \label{circ:KvN-BE-OH}
    Schematic circuit for encoding the left or right sidebands of the matrix $B_k$. Here, $q_x = n_x-1$, and the index $i_r = [0, N_x-1]$ is encoded in the register $r_x$.
    The subcircuit denoted as `$F(x_{i_r}) + F(x_{i_c})$' is shown in Fig.~\ref{circ:LCU-QSVT}.
    Controlled Pauli $X$ gates, whose controlling nodes are determined by indices $i_{b,k}$, are denoted here as `$i_r = i_{b,k}$'. These gates invert the state of the ancilla $a_c$ if $i_r = i_{b,k}$.
    An example of such a gate is shown in Fig.~\ref{circ:comp-equl}.
    The comparator denoted as $i_r < (i_{b,0}+1)$ is shown in Fig.~\ref{circ:comp-non-equl}.
    Here, the comparators use the most significant qubit of the register $r_x$ as the carry bit.
}
\end{figure*}
\begin{figure}[!t]
\centering
\subfloat[]{\includegraphics[width=0.07\textwidth]{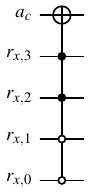}\label{circ:comp-equl}}
\hspace{0.6cm}
\subfloat[]{\includegraphics[width=0.26\textwidth]{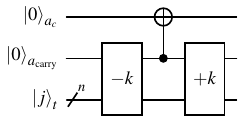}\label{circ:comp-non-equl}}\\
\vspace{0.2cm}
\subfloat[]{\includegraphics[width=0.44\textwidth]{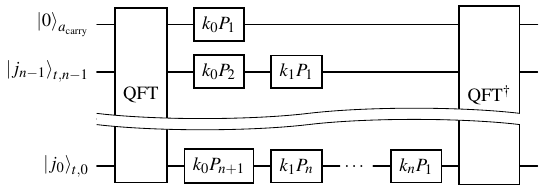}\label{circ:adder}}
\caption{
    \label{circ:arithm-vomp}
    (a): Circuit of an operator inverting the qubit $a_c$ if the index encoded in $r_x$ is equal to $12$.
    (b): Circuit of an operator comparing the integer $j \geq 0$ with the predefined unsigned index $k$. 
    If $a_{\rm carry}$ is initialized in the zero state, the circuit inverts the state of $a_c$ if $j < k$.
    If $\ket{1}_{a_{\rm carry}}$, the qubit $a_c$ is inverted if $j \geq k$. 
    (c): Circuit of the adder `$+k$'. 
    The operator adds the integer $k\geq 0$, represented as a string of $n+1$ bits $\left[k_n\dots k_1 k_0\right]$ where $k_0$ is the least significant bit, to the index $j$ represented as a string of $n$ bits $\left[j_{n-1}\dots j_1 j_0\right]$.
    The ancilla $a_{\rm carry}$ contains the carry bit of the resulting sum $k+j$.
    Here, $k_j P_q$ applies the phase gate $P(2\pi/2^q)$ [Eq.~\eqref{eq:phase-gate}] if $k_j = 1$.
    The subtractor `$-k$' is the Hermitian-adjoint version of the adder.
}
\end{figure}

\begin{figure*}[!t]
\centering
\includegraphics[width=0.90\textwidth]{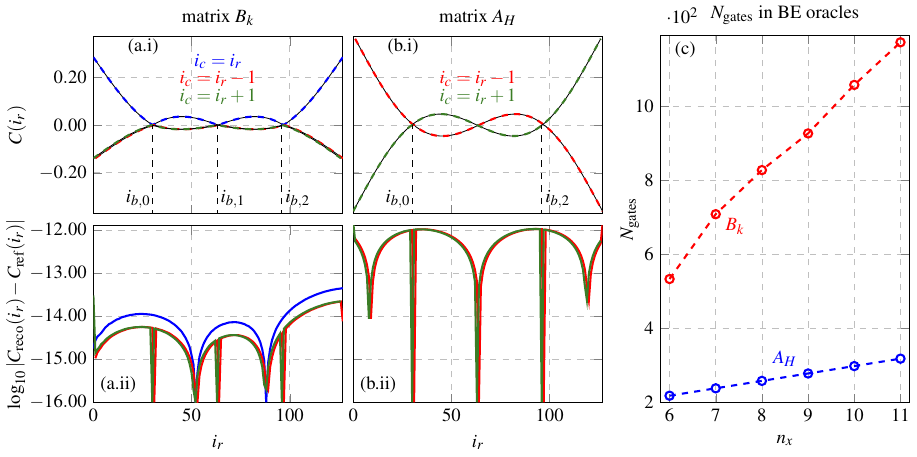}
\caption{
    \label{fig:B-profs} 
    (a.i) Plots showing the matrix elements on the main diagonal and sidebands in the matrix $B_k$ normalized according to Eq.~\eqref{eq:norm}.
    (b.i) Plots showing the matrix elements on the sidebands in the normalized matrix $A_H$.
    Here, the solid black lines indicate elements found using Eqs.~\eqref{eq:Ah-UW} and~\eqref{eq:Aa-UW}.
    The colored dashed lines indicate the matrix elements computed by the BE oracles encoding the matrices $B_k$ and $A_H$.\cite{code-KvN-LCHS}
    The vertical dashed lines indicate the row indices $i_{b,k}$ where the curves are not smooth. Here, $n_x = 7$ is used.
    (a.ii) and (b.ii): Errors in the matrix elements computed by the BE oracles. 
    (c): Number of STMC gates in the BE oracles encoding the matrices $A_H$ (blue markers) and $B_k$ (red markers).
}
\end{figure*}

The main diagonal of $A_L$ contains the smooth odd function $C^\prime(i_r) = 2 F_{i_r}$ [Eqs.~\eqref{eq:Ah-UW} and~\eqref{eq:KvN-LCHS-problem}] modified by the sign-function $\xi_{i_r}$ [Eq.~\eqref{eq:xi-sign}].
The function $C^\prime(i_r)$ can be easily encoded using QSVT circuit~\ref{circ:qsvt}.
Moreover, since we use $\sin(x)$ in Eq.~\eqref{eq:KvN-LCHS-problem}, one can avoid transformation~\eqref{eq:f-to-arcsin} and encode $C^\prime(i_r)$ by using just a few QSVT angles (in our case, two angles).
Otherwise, encoding the function $F(\arcsin(y))$ would significantly worsen the query complexity of the QSVT circuit.

The sidebands of $A_L$ contain the smooth function $C^\prime(i_r) = F_{i_r} + F_{i_c}$ modified by non-overlapping delta-functions.
Since $C^\prime(i_r)$ does not have definite parity, one needs to encode $F_{i_r}$ and $F_{i_c}$ by two separate QSVT circuits and then find their sum by using the LCU circuit (Fig.~\ref{circ:LCU-QSVT}).
Note that $F_{i_c}$ can be computed by applying the same QSVT angles used for $F_{i_r}$, but the state of the register $r_x$ should be shifted by $\pm 1$ using a quantum decrementor or incrementor, depending on which sideband is considered.

Due to the action of the sign and delta functions in Eq.~\eqref{eq:Ah-UW}, the main diagonals and sidebands contain non-smooth functions.
Let us denote these non-smooth functions by $C(i_r)$.
For instance, the circuit of $O_H$ for encoding one of the sidebands of the matrix $B_k$ is shown in Fig.~\ref{circ:KvN-BE-OH}.
Quantum arithmetic operators and controlled Pauli $X$ gates (Fig.~\ref{circ:arithm-vomp}) are used there to correctly treat the nodes $i_{b,0}$, $i_{b,1}$, and $i_{b,2}$ (Fig~\ref{fig:B-profs}). 
The exact values of the indices $i_{b,k}$ can be found numerically after the construction of the matrices $A_H$ and $A_L$.

To implement the sign-function $\xi_{i_r}$, we use the combination of quantum arithmetic comparators\cite{Suau21} shown in Fig.~\ref{circ:arithm-vomp} with the rotation gate $R_z(2\pi)$. 
In circuit~\ref{circ:KvN-BE-OH}, we use the fact that, in our problem~\eqref{eq:KvN-LCHS-problem}, the indices $i_{b,0}$ and $i_{b,2}$ are symmetric with respect to $i_{b,1} = 2^{n_x-1}-1$, and, therefore, two comparators $i_r < (i_{b,0}+1)$ and $i_r \geq i_{b,2}$ are replaced by a single comparator $i_r < (i_{b,0}+1)$ combined with the bit-inversion of the register $r_x$.
The same comparator at the end of the circuit is used to return the ancilla $a_c$ back to the zero state (because of this, one can use $O_H$ instead of $O_{\sqrt{H}}$ and $O_{\sqrt{H}}^\dagger$).
Note that the comparator is self-adjoint, and both comparators use the qubit $r_{x,n_x-1}$ for intermediate computations and do not change its state.
The comparator inverts the ancilla $a_c$ if the row index $i_r$ encoded in the register $r_x$ is smaller than $(i_{b,0}+1)$ and if the qubit $r_{x,n_x-1}$ is in the zero state.
If $\ket{1}_{r_{x,n_x-1}}$, then the comparator, in synergy with the controlled $X$ gates, inverts $a_c$ if $i_r \geq i_{b,2}$.
For its implementation, the comparator needs the phase gate,
\begin{equation}\label{eq:phase-gate}
    P(\theta) = 
    \begin{pmatrix}
        1 & 0 \\
        0 & e^{\yi\theta}
    \end{pmatrix}.
\end{equation}

To encode the delta-functions used in Eq.~\eqref{eq:Ah-UW}, we use controlled $X$ gates denoted in circuit~\ref{circ:KvN-BE-OH} as `$i_r = i_{b,k}$'.
An example of such a gate for $i_{b,k} = 12$ is shown in Fig.~\ref{circ:comp-equl}.
The delta-functions make the smooth function $F(x_{i_r}) + F(x_{i_c})$ fall to zero in the nodes $i_{b,k}$.
Since the purpose of the BE is to encode the matrix elements into the amplitudes of the states entangled with the zero state of all ancilla qubits used by $O_H$, one can introduce an ancilla $a_c$ initialized in the zero state and invert its state at $i_r = i_{b,k}$. 
In this case, the amplitude of $\ket{0}_{a_c}$ at $i_r = i_{b,k}$ becomes equal to zero.

The detailed circuits of the oracles $O_H$ for the matrices $A_H$, $B_k$, and $B_{\rm max}$ can be found in Ref.~\onlinecite{code-KvN-LCHS}.
In particular, the oracle $O_H$ for the matrix $B_{\rm max}$ is the same as the one shown in Fig.~\ref{circ:KvN-BE-OH}, but is modified by an extra $Z$ gate to correct the sign of the matrix elements (one can compare Eq.~\eqref{eq:Bmax} with Eq.~\eqref{eq:Bk}).
The oracle $O_H$ for $A_H$ is similar to the one shown in Fig.~\ref{circ:KvN-BE-OH} but does not include the comparators.

In Fig.~\ref{fig:B-profs}, one can see the scaling of the constructed BE oracles. 
The number of STMC gates in $A_H$ is strictly linear with $n_x$.
In particular, it is the incrementor and decrementor used in the oracle $O_M$, as well as the subcircuit~\eqref{circ:sin} used by the QSVT circuit~\ref{circ:LCU-QSVT}, that depends on $n_x$.
The number of the QSVT angles in the circuit~\ref{circ:LCU-QSVT} does not depend on $n_x$.
The number of gates in $B_k$ and $B_{\rm max}$ is also linear with $n_x$ but slightly varies for different $n_x$ because the number of phase gates in the comparator~\ref{circ:comp-non-equl} depends on the number of nonzero bits in the bitstring representation of the integers $i_{b,k}$.
The spectral norms (i.e. the maximum singular values) of the encoded matrices $A_H$ and $B_k$ are $0.60$ and $0.49$, respectively. 
(The spectral norms of the matrices $B_k$ and $B_{\rm max}$ are equal.)
Hence, the BE can be considered efficient because it scales logarithmically with $N_x$, the number of ancillae in the BE oracles does not depend on $N_x$, and the spectral norm of the encoded matrices is close to one.

\bibliography{main}

\begin{thebibliography}{10}
\expandafter\ifx\csname url\endcsname\relax
  \def\url#1{\texttt{#1}}\fi
\expandafter\ifx\csname urlprefix\endcsname\relax\def\urlprefix{URL }\fi
\expandafter\ifx\csname href\endcsname\relax
  \def\href#1#2{#2} \def\path#1{#1}\fi

\bibitem{Suzuki93}
M.~Suzuki, General decomposition theory of ordered exponentials, Proceedings of the Japan Academy, Series B 69~(7) (1993) 161--166.
\newblock \href {https://doi.org/10.2183/pjab.69.161} {\path{doi:10.2183/pjab.69.161}}.

\bibitem{Berry15}
D.~W. Berry, A.~M. Childs, R.~Cleve, R.~Kothari, R.~D. Somma, \href{https://link.aps.org/doi/10.1103/PhysRevLett.114.090502}{Simulating {H}amiltonian dynamics with a truncated {T}aylor series}, Phys. Rev. Lett. 114 (2015) 090502.
\newblock \href {https://doi.org/10.1103/PhysRevLett.114.090502} {\path{doi:10.1103/PhysRevLett.114.090502}}.
\newline\urlprefix\url{https://link.aps.org/doi/10.1103/PhysRevLett.114.090502}

\bibitem{Gilyen19}
A.~Gily\'{e}n, Y.~Su, G.~H. Low, N.~Wiebe, \href{https://doi.org/10.1145/3313276.3316366}{Quantum singular value transformation and beyond: Exponential improvements for quantum matrix arithmetics}, in: Proceedings of the 51st Annual ACM SIGACT Symposium on Theory of Computing, STOC 2019, Association for Computing Machinery, New York, NY, USA, 2019, p. 193–204.
\newblock \href {https://doi.org/10.1145/3313276.3316366} {\path{doi:10.1145/3313276.3316366}}.
\newline\urlprefix\url{https://doi.org/10.1145/3313276.3316366}

\bibitem{Martyn21}
J.~M. Martyn, Z.~M. Rossi, A.~K. Tan, I.~L. Chuang, \href{https://link.aps.org/doi/10.1103/PRXQuantum.2.040203}{Grand unification of quantum algorithms}, PRX Quantum 2 (2021) 040203.
\newblock \href {https://doi.org/10.1103/PRXQuantum.2.040203} {\path{doi:10.1103/PRXQuantum.2.040203}}.
\newline\urlprefix\url{https://link.aps.org/doi/10.1103/PRXQuantum.2.040203}

\bibitem{Martyn23}
J.~M. Martyn, Y.~Liu, Z.~E. Chin, I.~L. Chuang, \href{https://doi.org/10.1063/5.0124385}{{Efficient fully-coherent quantum signal processing algorithms for real-time dynamics simulation}}, The Journal of Chemical Physics 158~(2) (2023) 024106.
\newblock \href {http://arxiv.org/abs/https://pubs.aip.org/aip/jcp/article-pdf/doi/10.1063/5.0124385/16668116/024106\_1\_online.pdf} {\path{arXiv:https://pubs.aip.org/aip/jcp/article-pdf/doi/10.1063/5.0124385/16668116/024106\_1\_online.pdf}}, \href {https://doi.org/10.1063/5.0124385} {\path{doi:10.1063/5.0124385}}.
\newline\urlprefix\url{https://doi.org/10.1063/5.0124385}

\bibitem{Gaitan20}
F.~Gaitan, \href{https://doi.org/10.1038/s41534-020-00291-0}{Finding flows of a navier--stokes fluid through quantum computing}, npj Quantum Information 6~(1) (2020) 61.
\newblock \href {https://doi.org/10.1038/s41534-020-00291-0} {\path{doi:10.1038/s41534-020-00291-0}}.
\newline\urlprefix\url{https://doi.org/10.1038/s41534-020-00291-0}

\bibitem{Dodin20}
I.~Y. Dodin, E.~A. Startsev, \href{https://doi.org/10.1063/5.0056974}{On applications of quantum computing to plasma simulations}, Physics of Plasmas 28~(9) (2021) 092101.
\newblock \href {http://arxiv.org/abs/https://doi.org/10.1063/5.0056974} {\path{arXiv:https://doi.org/10.1063/5.0056974}}, \href {https://doi.org/10.1063/5.0056974} {\path{doi:10.1063/5.0056974}}.
\newline\urlprefix\url{https://doi.org/10.1063/5.0056974}

\bibitem{Joseph23}
I.~Joseph, Y.~Shi, M.~D. Porter, A.~R. Castelli, V.~I. Geyko, F.~R. Graziani, S.~B. Libby, J.~L. DuBois, \href{http://dx.doi.org/10.1063/5.0123765}{Quantum computing for fusion energy science applications}, Physics of Plasmas 30~(1) (Jan. 2023).
\newblock \href {https://doi.org/10.1063/5.0123765} {\path{doi:10.1063/5.0123765}}.
\newline\urlprefix\url{http://dx.doi.org/10.1063/5.0123765}

\bibitem{Akiba23}
T.~Akiba, Y.~Morii, K.~Maruta, \href{https://doi.org/10.1038/s41598-023-31009-9}{Carleman linearization approach for chemical kinetics integration toward quantum computation}, Scientific Reports 13~(1) (2023) 3935.
\newblock \href {https://doi.org/10.1038/s41598-023-31009-9} {\path{doi:10.1038/s41598-023-31009-9}}.
\newline\urlprefix\url{https://doi.org/10.1038/s41598-023-31009-9}

\bibitem{An23-diffusion}
D.~An, K.~Trivisa, Quantum algorithms for linear and non-linear fractional reaction-diffusion equations (2023).
\newblock \href {http://arxiv.org/abs/2310.18900} {\path{arXiv:2310.18900}}.

\bibitem{Liu23}
J.-P. Liu, D.~An, D.~Fang, J.~Wang, G.~H. Low, S.~Jordan, \href{https://doi.org/10.1007/s00220-023-04857-9}{Efficient quantum algorithm for nonlinear reaction--diffusion equations and energy estimation}, Communications in Mathematical Physics 404~(2) (2023) 963--1020.
\newblock \href {https://doi.org/10.1007/s00220-023-04857-9} {\path{doi:10.1007/s00220-023-04857-9}}.
\newline\urlprefix\url{https://doi.org/10.1007/s00220-023-04857-9}

\bibitem{Hsieh91}
D.~A. HSIEH, \href{https://onlinelibrary.wiley.com/doi/abs/10.1111/j.1540-6261.1991.tb04646.x}{Chaos and nonlinear dynamics: Application to financial markets}, The Journal of Finance 46~(5) (1991) 1839--1877.
\newblock \href {http://arxiv.org/abs/https://onlinelibrary.wiley.com/doi/pdf/10.1111/j.1540-6261.1991.tb04646.x} {\path{arXiv:https://onlinelibrary.wiley.com/doi/pdf/10.1111/j.1540-6261.1991.tb04646.x}}, \href {https://doi.org/https://doi.org/10.1111/j.1540-6261.1991.tb04646.x} {\path{doi:https://doi.org/10.1111/j.1540-6261.1991.tb04646.x}}.
\newline\urlprefix\url{https://onlinelibrary.wiley.com/doi/abs/10.1111/j.1540-6261.1991.tb04646.x}

\bibitem{Richards00}
D.~Richards, \href{http://www.jstor.org/stable/10.3998/mpub.16428}{Political Complexity: Nonlinear Models of Politics}, University of Michigan Press, 2000.
\newline\urlprefix\url{http://www.jstor.org/stable/10.3998/mpub.16428}

\bibitem{Joseph20}
I.~Joseph, \href{https://link.aps.org/doi/10.1103/PhysRevResearch.2.043102}{{Koopman--von Neumann} approach to quantum simulation of nonlinear classical dynamics}, Phys. Rev. Res. 2 (2020) 043102.
\newblock \href {https://doi.org/10.1103/PhysRevResearch.2.043102} {\path{doi:10.1103/PhysRevResearch.2.043102}}.
\newline\urlprefix\url{https://link.aps.org/doi/10.1103/PhysRevResearch.2.043102}

\bibitem{Liu21}
J.-P. Liu, H.~Kolden, H.~K. Krovi, N.~F. Loureiro, K.~Trivisa, A.~M. Childs, \href{http://dx.doi.org/10.1073/pnas.2026805118}{Efficient quantum algorithm for dissipative nonlinear differential equations}, Proceedings of the National Academy of Sciences 118~(35) (Aug. 2021).
\newblock \href {https://doi.org/10.1073/pnas.2026805118} {\path{doi:10.1073/pnas.2026805118}}.
\newline\urlprefix\url{http://dx.doi.org/10.1073/pnas.2026805118}

\bibitem{Conde23}
J.~Gonzalez-Conde, A.~T. Sornborger, \href{https://arxiv.org/abs/2308.16147}{Mixed quantum-semiclassical simulation} (2023).
\newblock \href {http://arxiv.org/abs/2308.16147} {\path{arXiv:2308.16147}}.
\newline\urlprefix\url{https://arxiv.org/abs/2308.16147}

\bibitem{Bondar19}
D.~I. Bondar, F.~Gay-Balmaz, C.~Tronci, \href{https://royalsocietypublishing.org/doi/abs/10.1098/rspa.2018.0879}{Koopman wavefunctions and classical–quantum correlation dynamics}, Proceedings of the Royal Society A: Mathematical, Physical and Engineering Sciences 475~(2229) (2019) 20180879.
\newblock \href {http://arxiv.org/abs/https://royalsocietypublishing.org/doi/pdf/10.1098/rspa.2018.0879} {\path{arXiv:https://royalsocietypublishing.org/doi/pdf/10.1098/rspa.2018.0879}}, \href {https://doi.org/10.1098/rspa.2018.0879} {\path{doi:10.1098/rspa.2018.0879}}.
\newline\urlprefix\url{https://royalsocietypublishing.org/doi/abs/10.1098/rspa.2018.0879}

\bibitem{Tronci21}
C.~Tronci, I.~Joseph, Koopman wavefunctions and clebsch variables in vlasov–maxwell kinetic theory, Journal of Plasma Physics 87~(4) (2021) 835870402.
\newblock \href {https://doi.org/10.1017/S0022377821000805} {\path{doi:10.1017/S0022377821000805}}.

\bibitem{Joseph23JPA}
I.~Joseph, \href{http://dx.doi.org/10.1088/1751-8121/ad0533}{Semiclassical theory and the {Koopman-van Hove} equation}, Journal of Physics A: Mathematical and Theoretical 56~(48) (2023) 484001.
\newblock \href {https://doi.org/10.1088/1751-8121/ad0533} {\path{doi:10.1088/1751-8121/ad0533}}.
\newline\urlprefix\url{http://dx.doi.org/10.1088/1751-8121/ad0533}

\bibitem{VanKampen1976}
N.~G. Van~Kampen, Stochastic differential equations, Physics reports 24~(3) (1976) 171--228.

\bibitem{Kleinert2006book}
H.~Kleinert, Path integrals in quantum mechanics, statistics, polymer physics, and financial markets, World Scientific Publishing Company, 2006.

\bibitem{Oksendal2013book}
B.~Oksendal, Stochastic differential equations: an introduction with applications, Springer Science \& Business Media, 2013.

\bibitem{Low17}
G.~H. Low, I.~L. Chuang, \href{https://link.aps.org/doi/10.1103/PhysRevLett.118.010501}{Optimal {Hamiltonian} simulation by quantum signal processing}, Physical Review Letters 118 (2017) 010501.
\newblock \href {https://doi.org/10.1103/PhysRevLett.118.010501} {\path{doi:10.1103/PhysRevLett.118.010501}}.
\newline\urlprefix\url{https://link.aps.org/doi/10.1103/PhysRevLett.118.010501}

\bibitem{Low19}
G.~H. Low, I.~L. Chuang, \href{https://doi.org/10.22331/q-2019-07-12-163}{{Hamiltonian} simulation by qubitization}, {Quantum} 3 (2019) 163.
\newblock \href {https://doi.org/10.22331/q-2019-07-12-163} {\path{doi:10.22331/q-2019-07-12-163}}.
\newline\urlprefix\url{https://doi.org/10.22331/q-2019-07-12-163}

\bibitem{Babbush23}
R.~Babbush, D.~W. Berry, R.~Kothari, R.~D. Somma, N.~Wiebe, \href{https://link.aps.org/doi/10.1103/PhysRevX.13.041041}{Exponential quantum speedup in simulating coupled classical oscillators}, Phys. Rev. X 13 (2023) 041041.
\newblock \href {https://doi.org/10.1103/PhysRevX.13.041041} {\path{doi:10.1103/PhysRevX.13.041041}}.
\newline\urlprefix\url{https://link.aps.org/doi/10.1103/PhysRevX.13.041041}

\bibitem{Barthe24}
A.~Barthe, M.~Cerezo, A.~T. Sornborger, M.~Larocca, D.~García-Martín, \href{https://arxiv.org/abs/2407.06290}{Gate-based quantum simulation of gaussian bosonic circuits on exponentially many modes} (2024).
\newblock \href {http://arxiv.org/abs/2407.06290} {\path{arXiv:2407.06290}}.
\newline\urlprefix\url{https://arxiv.org/abs/2407.06290}

\bibitem{Brassard02}
G.~Brassard, P.~H{\o}yer, M.~Mosca, A.~Tapp, \href{http://dx.doi.org/10.1090/conm/305/05215}{Quantum amplitude amplification and estimation}, Quantum Computation and Information 305 (2002) 53–74.
\newblock \href {https://doi.org/10.1090/conm/305/05215} {\path{doi:10.1090/conm/305/05215}}.
\newline\urlprefix\url{http://dx.doi.org/10.1090/conm/305/05215}

\bibitem{Wootters82}
W.~K. Wootters, W.~H. Zurek, \href{https://doi.org/10.1038/299802a0}{A single quantum cannot be cloned}, Nature 299~(5886) (1982) 802--803.
\newblock \href {https://doi.org/10.1038/299802a0} {\path{doi:10.1038/299802a0}}.
\newline\urlprefix\url{https://doi.org/10.1038/299802a0}

\bibitem{Leyton08}
S.~K. Leyton, T.~J. Osborne, \href{https://arxiv.org/abs/0812.4423}{A quantum algorithm to solve nonlinear differential equations} (2008).
\newblock \href {http://arxiv.org/abs/0812.4423} {\path{arXiv:0812.4423}}.
\newline\urlprefix\url{https://arxiv.org/abs/0812.4423}

\bibitem{Lloyd20}
S.~Lloyd, G.~D. Palma, C.~Gokler, B.~Kiani, Z.-W. Liu, M.~Marvian, F.~Tennie, T.~Palmer, Quantum algorithm for nonlinear differential equations (2020).
\newblock \href {http://arxiv.org/abs/2011.06571} {\path{arXiv:2011.06571}}.

\bibitem{Engel23}
A.~Engel, S.~E. Parker, \href{https://doi.org/10.1140/epjp/s13360-023-04205-9}{Correspondence between open bosonic systems and stochastic differential equations}, The European Physical Journal Plus 138~(6) (2023) 578.
\newblock \href {https://doi.org/10.1140/epjp/s13360-023-04205-9} {\path{doi:10.1140/epjp/s13360-023-04205-9}}.
\newline\urlprefix\url{https://doi.org/10.1140/epjp/s13360-023-04205-9}

\bibitem{Kyriienko21}
O.~Kyriienko, A.~E. Paine, V.~E. Elfving, \href{https://link.aps.org/doi/10.1103/PhysRevA.103.052416}{Solving nonlinear differential equations with differentiable quantum circuits}, Phys. Rev. A 103 (2021) 052416.
\newblock \href {https://doi.org/10.1103/PhysRevA.103.052416} {\path{doi:10.1103/PhysRevA.103.052416}}.
\newline\urlprefix\url{https://link.aps.org/doi/10.1103/PhysRevA.103.052416}

\bibitem{Demirdjian22}
R.~Demirdjian, D.~Gunlycke, C.~A. Reynolds, J.~D. Doyle, S.~Tafur, \href{https://doi.org/10.1007/s11128-022-03667-7}{Variational quantum solutions to the advection--diffusion equation for applications in fluid dynamics}, Quantum Information Processing 21~(9) (2022) 322.
\newblock \href {https://doi.org/10.1007/s11128-022-03667-7} {\path{doi:10.1007/s11128-022-03667-7}}.
\newline\urlprefix\url{https://doi.org/10.1007/s11128-022-03667-7}

\bibitem{Jaksch23}
D.~Jaksch, P.~Givi, A.~J. Daley, T.~Rung, \href{https://doi.org/10.2514/1.J062426}{Variational quantum algorithms for computational fluid dynamics}, AIAA Journal 61~(5) (2023) 1885--1894.
\newblock \href {http://arxiv.org/abs/https://doi.org/10.2514/1.J062426} {\path{arXiv:https://doi.org/10.2514/1.J062426}}, \href {https://doi.org/10.2514/1.J062426} {\path{doi:10.2514/1.J062426}}.
\newline\urlprefix\url{https://doi.org/10.2514/1.J062426}

\bibitem{Kacewicz06}
B.~Kacewicz, \href{https://www.sciencedirect.com/science/article/pii/S0885064X06000148}{Almost optimal solution of initial-value problems by randomized and quantum algorithms}, Journal of Complexity 22~(5) (2006) 676--690, special Issue: Information-Based Complexity Workshops FoCM Conference Santander, Spain, July 2005.
\newblock \href {https://doi.org/https://doi.org/10.1016/j.jco.2006.03.001} {\path{doi:https://doi.org/10.1016/j.jco.2006.03.001}}.
\newline\urlprefix\url{https://www.sciencedirect.com/science/article/pii/S0885064X06000148}

\bibitem{Gaitan24}
F.~Gaitan, \href{https://link.aps.org/doi/10.1103/PhysRevA.109.032604}{Circuit implementation of oracles used in a quantum algorithm for solving nonlinear partial differential equations}, Phys. Rev. A 109 (2024) 032604.
\newblock \href {https://doi.org/10.1103/PhysRevA.109.032604} {\path{doi:10.1103/PhysRevA.109.032604}}.
\newline\urlprefix\url{https://link.aps.org/doi/10.1103/PhysRevA.109.032604}

\bibitem{Oz23}
F.~Oz, O.~San, K.~Kara, \href{https://doi.org/10.1038/s41598-023-34966-3}{An efficient quantum partial differential equation solver with chebyshev points}, Scientific Reports 13~(1) (2023) 7767.
\newblock \href {https://doi.org/10.1038/s41598-023-34966-3} {\path{doi:10.1038/s41598-023-34966-3}}.
\newline\urlprefix\url{https://doi.org/10.1038/s41598-023-34966-3}

\bibitem{Criado23}
J.~C. Criado, M.~Spannowsky, \href{https://dx.doi.org/10.1088/2058-9565/acaa51}{Qade: solving differential equations on quantum annealers}, Quantum Science and Technology 8~(1) (2022) 015021.
\newblock \href {https://doi.org/10.1088/2058-9565/acaa51} {\path{doi:10.1088/2058-9565/acaa51}}.
\newline\urlprefix\url{https://dx.doi.org/10.1088/2058-9565/acaa51}

\bibitem{Nguyen24}
V.-D. Nguyen, L.~Wu, F.~Remacle, L.~Noels, \href{https://www.sciencedirect.com/science/article/pii/S0997753824000342}{A quantum annealing-sequential quadratic programming assisted finite element simulation for non-linear and history-dependent mechanical problems}, European Journal of Mechanics - A/Solids 105 (2024) 105254.
\newblock \href {https://doi.org/https://doi.org/10.1016/j.euromechsol.2024.105254} {\path{doi:https://doi.org/10.1016/j.euromechsol.2024.105254}}.
\newline\urlprefix\url{https://www.sciencedirect.com/science/article/pii/S0997753824000342}

\bibitem{Koopman31}
B.~O. Koopman, \href{https://www.pnas.org/doi/10.1073/pnas.17.5.315}{{Hamiltonian} systems and transformations in {Hilbert} space}, Proceedings of the National Academy of Sciences of the United States of America 17~(5) (1931) 315--318.
\newblock \href {https://doi.org/10.1073/pnas.17.5.315} {\path{doi:10.1073/pnas.17.5.315}}.
\newline\urlprefix\url{https://www.pnas.org/doi/10.1073/pnas.17.5.315}

\bibitem{Carleman32}
T.~Carleman, \href{https://doi.org/10.1007/BF02546499}{Application de la th\'eory des \'equations int\'egrales lin\'eaires aux syst\`emes d'\'equations diff\'erentielles non lin\'eaires}, Acta Mathematica 59 (1932) 63--87.
\newblock \href {https://doi.org/10.1007/BF02546499} {\path{doi:10.1007/BF02546499}}.
\newline\urlprefix\url{https://doi.org/10.1007/BF02546499}

\bibitem{Dodin21}
I.~Y. Dodin, E.~A. Startsev, \href{https://arxiv.org/abs/2105.07317}{Quantum computation of nonlinear maps} (2021).
\newblock \href {http://arxiv.org/abs/2105.07317} {\path{arXiv:2105.07317}}.
\newline\urlprefix\url{https://arxiv.org/abs/2105.07317}

\bibitem{Godunov59}
S.~K. Godunov, I.~Bohachevsky, \href{https://hal.science/hal-01620642}{{Finite difference method for numerical computation of discontinuous solutions of the equations of fluid dynamics}}, {Matemati{\v c}eskij sbornik} 47(89)~(3) (1959) 271--306.
\newline\urlprefix\url{https://hal.science/hal-01620642}

\bibitem{Harten83}
A.~Harten, \href{https://www.sciencedirect.com/science/article/pii/0021999183901365}{High resolution schemes for hyperbolic conservation laws}, Journal of Computational Physics 49~(3) (1983) 357--393.
\newblock \href {https://doi.org/https://doi.org/10.1016/0021-9991(83)90136-5} {\path{doi:https://doi.org/10.1016/0021-9991(83)90136-5}}.
\newline\urlprefix\url{https://www.sciencedirect.com/science/article/pii/0021999183901365}

\bibitem{Liu94WENO}
X.-D. Liu, S.~Osher, T.~Chan, \href{https://www.sciencedirect.com/science/article/pii/S0021999184711879}{Weighted essentially non-oscillatory schemes}, Journal of Computational Physics 115~(1) (1994) 200--212.
\newblock \href {https://doi.org/https://doi.org/10.1006/jcph.1994.1187} {\path{doi:https://doi.org/10.1006/jcph.1994.1187}}.
\newline\urlprefix\url{https://www.sciencedirect.com/science/article/pii/S0021999184711879}

\bibitem{Lin22Koopman}
Y.~T. Lin, R.~B. Lowrie, D.~Aslangil, Y.~Subaşı, A.~T. Sornborger, {Koopman--von Neumann} mechanics and the {Koopman} representation: A perspective on solving nonlinear dynamical systems with quantum computers (2022).
\newblock \href {http://arxiv.org/abs/2202.02188} {\path{arXiv:2202.02188}}.

\bibitem{Ambainis12}
A.~Ambainis, \href{http://drops.dagstuhl.de/opus/volltexte/2012/3426}{{Variable time amplitude amplification and quantum algorithms for linear algebra problems}}, in: C.~D{\"u}rr, T.~Wilke (Eds.), 29th International Symposium on Theoretical Aspects of Computer Science (STACS 2012), Vol.~14 of Leibniz International Proceedings in Informatics (LIPIcs), Schloss Dagstuhl--Leibniz-Zentrum fuer Informatik, Dagstuhl, Germany, 2012, pp. 636--647.
\newblock \href {https://doi.org/10.4230/LIPIcs.STACS.2012.636} {\path{doi:10.4230/LIPIcs.STACS.2012.636}}.
\newline\urlprefix\url{http://drops.dagstuhl.de/opus/volltexte/2012/3426}

\bibitem{Chakraborty19}
S.~Chakraborty, A.~Gilyén, S.~Jeffery, \href{https://drops.dagstuhl.de/entities/document/10.4230/LIPIcs.ICALP.2019.33}{The power of block-encoded matrix powers: Improved regression techniques via faster hamiltonian simulation}, Schloss Dagstuhl – Leibniz-Zentrum für Informatik, 2019.
\newblock \href {https://doi.org/10.4230/LIPICS.ICALP.2019.33} {\path{doi:10.4230/LIPICS.ICALP.2019.33}}.
\newline\urlprefix\url{https://drops.dagstuhl.de/entities/document/10.4230/LIPIcs.ICALP.2019.33}

\bibitem{Costa21}
P.~C. Costa, D.~An, Y.~R. Sanders, Y.~Su, R.~Babbush, D.~W. Berry, \href{https://link.aps.org/doi/10.1103/PRXQuantum.3.040303}{Optimal scaling quantum linear-systems solver via discrete adiabatic theorem}, PRX Quantum 3 (2022) 040303.
\newblock \href {https://doi.org/10.1103/PRXQuantum.3.040303} {\path{doi:10.1103/PRXQuantum.3.040303}}.
\newline\urlprefix\url{https://link.aps.org/doi/10.1103/PRXQuantum.3.040303}

\bibitem{Linden22}
N.~Linden, A.~Montanaro, C.~Shao, Quantum vs. classical algorithms for solving the heat equation, Communications in Mathematical Physics 395~(2) (2022) 601--641.

\bibitem{An24FF}
D.~An, A.~Onwunta, G.~Yang, \href{https://arxiv.org/abs/2410.13189}{Fast-forwarding quantum algorithms for linear dissipative differential equations} (2024).
\newblock \href {http://arxiv.org/abs/2410.13189} {\path{arXiv:2410.13189}}.
\newline\urlprefix\url{https://arxiv.org/abs/2410.13189}

\bibitem{Jennings24}
D.~Jennings, M.~Lostaglio, R.~B. Lowrie, S.~Pallister, A.~T. Sornborger, \href{https://arxiv.org/abs/2309.07881}{The cost of solving linear differential equations on a quantum computer: fast-forwarding to explicit resource counts} (2024).
\newblock \href {http://arxiv.org/abs/2309.07881} {\path{arXiv:2309.07881}}.
\newline\urlprefix\url{https://arxiv.org/abs/2309.07881}

\bibitem{Fang23}
D.~Fang, L.~Lin, Y.~Tong, \href{https://doi.org/10.22331/q-2023-03-20-955}{Time-marching based quantum solvers for time-dependent linear differential equations}, {Quantum} 7 (2023) 955.
\newblock \href {https://doi.org/10.22331/q-2023-03-20-955} {\path{doi:10.22331/q-2023-03-20-955}}.
\newline\urlprefix\url{https://doi.org/10.22331/q-2023-03-20-955}

\bibitem{Jin22Sch}
S.~Jin, N.~Liu, Y.~Yu, \href{https://arxiv.org/abs/2212.13969}{Quantum simulation of partial differential equations via {Schr\"odingerisation}} (2022).
\newblock \href {http://arxiv.org/abs/2212.13969} {\path{arXiv:2212.13969}}.
\newline\urlprefix\url{https://arxiv.org/abs/2212.13969}

\bibitem{Jin23Sch}
S.~Jin, N.~Liu, Y.~Yu, \href{https://link.aps.org/doi/10.1103/PhysRevA.108.032603}{Quantum simulation of partial differential equations: Applications and detailed analysis}, Phys. Rev. A 108 (2023) 032603.
\newblock \href {https://doi.org/10.1103/PhysRevA.108.032603} {\path{doi:10.1103/PhysRevA.108.032603}}.
\newline\urlprefix\url{https://link.aps.org/doi/10.1103/PhysRevA.108.032603}

\bibitem{Hu24Sch}
J.~Hu, S.~Jin, N.~Liu, L.~Zhang, \href{https://arxiv.org/abs/2403.10032}{Quantum circuits for partial differential equations via {Schr\"odingerisation}} (2024).
\newblock \href {http://arxiv.org/abs/2403.10032} {\path{arXiv:2403.10032}}.
\newline\urlprefix\url{https://arxiv.org/abs/2403.10032}

\bibitem{Lu24}
Z.~Lu, Y.~Yang, \href{https://www.sciencedirect.com/science/article/pii/S1540748924002487}{Quantum computing of reacting flows via {Hamiltonian} simulation}, Proceedings of the Combustion Institute 40~(1) (2024) 105440.
\newblock \href {https://doi.org/https://doi.org/10.1016/j.proci.2024.105440} {\path{doi:https://doi.org/10.1016/j.proci.2024.105440}}.
\newline\urlprefix\url{https://www.sciencedirect.com/science/article/pii/S1540748924002487}

\bibitem{An23}
D.~An, J.-P. Liu, L.~Lin, \href{https://link.aps.org/doi/10.1103/PhysRevLett.131.150603}{Linear combination of hamiltonian simulation for nonunitary dynamics with optimal state preparation cost}, Phys. Rev. Lett. 131 (2023) 150603.
\newblock \href {https://doi.org/10.1103/PhysRevLett.131.150603} {\path{doi:10.1103/PhysRevLett.131.150603}}.
\newline\urlprefix\url{https://link.aps.org/doi/10.1103/PhysRevLett.131.150603}

\bibitem{An23impr}
D.~An, A.~M. Childs, L.~Lin, Quantum algorithm for linear non-unitary dynamics with near-optimal dependence on all parameters (2023).
\newblock \href {http://arxiv.org/abs/2312.03916} {\path{arXiv:2312.03916}}.

\bibitem{Jin24}
S.~Jin, N.~Liu, C.~Ma, \href{https://arxiv.org/abs/2402.14696}{On schr\"odingerization based quantum algorithms for linear dynamical systems with inhomogeneous terms} (2024).
\newblock \href {http://arxiv.org/abs/2402.14696} {\path{arXiv:2402.14696}}.
\newline\urlprefix\url{https://arxiv.org/abs/2402.14696}

\bibitem{Clader22}
B.~D. Clader, A.~M. Dalzell, N.~Stamatopoulos, G.~Salton, M.~Berta, W.~J. Zeng, \href{http://dx.doi.org/10.1109/TQE.2022.3231194}{Quantum resources required to block-encode a matrix of classical data}, IEEE Transactions on Quantum Engineering 3 (2022) 1–23.
\newblock \href {https://doi.org/10.1109/tqe.2022.3231194} {\path{doi:10.1109/tqe.2022.3231194}}.
\newline\urlprefix\url{http://dx.doi.org/10.1109/TQE.2022.3231194}

\bibitem{Camps22}
D.~Camps, R.~Van~Beeumen, \href{http://dx.doi.org/10.1109/QCE53715.2022.00029}{Fable: Fast approximate quantum circuits for block-encodings}, in: 2022 IEEE International Conference on Quantum Computing and Engineering (QCE), IEEE, 2022.
\newblock \href {https://doi.org/10.1109/qce53715.2022.00029} {\path{doi:10.1109/qce53715.2022.00029}}.
\newline\urlprefix\url{http://dx.doi.org/10.1109/QCE53715.2022.00029}

\bibitem{Camps23}
D.~Camps, L.~Lin, R.~V. Beeumen, C.~Yang, \href{https://arxiv.org/abs/2203.10236}{Explicit quantum circuits for block encodings of certain sparse matrices} (2023).
\newblock \href {http://arxiv.org/abs/2203.10236} {\path{arXiv:2203.10236}}.
\newline\urlprefix\url{https://arxiv.org/abs/2203.10236}

\bibitem{Sunderhauf24}
C.~S{\"{u}}nderhauf, E.~Campbell, J.~Camps, \href{https://doi.org/10.22331/q-2024-01-11-1226}{Block-encoding structured matrices for data input in quantum computing}, {Quantum} 8 (2024) 1226.
\newblock \href {https://doi.org/10.22331/q-2024-01-11-1226} {\path{doi:10.22331/q-2024-01-11-1226}}.
\newline\urlprefix\url{https://doi.org/10.22331/q-2024-01-11-1226}

\bibitem{Novikau24-EVM}
I.~Novikau, I.~Dodin, E.~Startsev, Encoding of linear kinetic plasma problems in quantum circuits via data compression, Journal of Plasma Physics 90~(4) (2024) 805900401.
\newblock \href {https://doi.org/10.1017/S0022377824000795} {\path{doi:10.1017/S0022377824000795}}.

\bibitem{Novikau22}
I.~Novikau, E.~A. Startsev, I.~Y. Dodin, \href{https://link.aps.org/doi/10.1103/PhysRevA.105.062444}{Quantum signal processing for simulating cold plasma waves}, Phys. Rev. A 105 (2022) 062444.
\newblock \href {https://doi.org/10.1103/PhysRevA.105.062444} {\path{doi:10.1103/PhysRevA.105.062444}}.
\newline\urlprefix\url{https://link.aps.org/doi/10.1103/PhysRevA.105.062444}

\bibitem{Novikau23}
I.~Novikau, I.~Dodin, E.~Startsev, \href{https://link.aps.org/doi/10.1103/PhysRevApplied.19.054012}{Simulation of linear non-hermitian boundary-value problems with quantum singular-value transformation}, Phys. Rev. Appl. 19 (2023) 054012.
\newblock \href {https://doi.org/10.1103/PhysRevApplied.19.054012} {\path{doi:10.1103/PhysRevApplied.19.054012}}.
\newline\urlprefix\url{https://link.aps.org/doi/10.1103/PhysRevApplied.19.054012}

\bibitem{Liu24BE}
D.~Liu, W.~Du, L.~Lin, J.~P. Vary, C.~Yang, \href{https://arxiv.org/abs/2402.11205}{An efficient quantum circuit for block encoding a pairing {Hamiltonian}} (2024).
\newblock \href {http://arxiv.org/abs/2402.11205} {\path{arXiv:2402.11205}}.
\newline\urlprefix\url{https://arxiv.org/abs/2402.11205}

\bibitem{Dong22}
Y.~Dong, L.~Lin, Y.~Tong, \href{https://link.aps.org/doi/10.1103/PRXQuantum.3.040305}{Ground-state preparation and energy estimation on early fault-tolerant quantum computers via quantum eigenvalue transformation of unitary matrices}, PRX Quantum 3 (2022) 040305.
\newblock \href {https://doi.org/10.1103/PRXQuantum.3.040305} {\path{doi:10.1103/PRXQuantum.3.040305}}.
\newline\urlprefix\url{https://link.aps.org/doi/10.1103/PRXQuantum.3.040305}

\bibitem{QuCF}
{QuCF} framework, \url{https://github.com/QuCF/QuCF} (2024).

\bibitem{Jones19}
T.~Jones, A.~Brown, I.~Bush, S.~C. Benjamin, \href{https://doi.org/10.1038/s41598-019-47174-9}{Quest and high performance simulation of quantum computers}, Scientific Reports 9~(1) (2019) 10736.
\newblock \href {https://doi.org/10.1038/s41598-019-47174-9} {\path{doi:10.1038/s41598-019-47174-9}}.
\newline\urlprefix\url{https://doi.org/10.1038/s41598-019-47174-9}

\bibitem{HeinrichNovak01arxiv}
S.~Heinrich, E.~Novak, \href{https://doi.org/10.1007/978-3-642-56046-0_4}{Optimal summation and integration by deterministic, randomized, and quantum algorithms}, in: Monte Carlo and Quasi-Monte Carlo Methods 2000, Springer, 2002, pp. 50--62.
\newblock \href {http://arxiv.org/abs/arXiv:quant-ph/0105114} {\path{arXiv:arXiv:quant-ph/0105114}}, \href {https://doi.org/10.1007/978-3-642-56046-0_4} {\path{doi:10.1007/978-3-642-56046-0_4}}.
\newline\urlprefix\url{https://doi.org/10.1007/978-3-642-56046-0_4}

\bibitem{DongAn21MC}
D.~An, N.~Linden, J.-P. Liu, A.~Montanaro, C.~Shao, J.~Wang, \href{http://dx.doi.org/10.22331/q-2021-06-24-481}{Quantum-accelerated multilevel monte carlo methods for stochastic differential equations in mathematical finance}, Quantum 5 (2021) 481.
\newblock \href {https://doi.org/10.22331/q-2021-06-24-481} {\path{doi:10.22331/q-2021-06-24-481}}.
\newline\urlprefix\url{http://dx.doi.org/10.22331/q-2021-06-24-481}

\bibitem{Montanaro2015}
A.~Montanaro, Quantum speedup of monte carlo methods, Proceedings of the Royal Society A: Mathematical, Physical and Engineering Sciences 471~(2181) (2015) 20150301.

\bibitem{Lindblad76cmp}
G.~Lindblad, \href{https://doi.org/10.1007/BF01608499}{On the generators of quantum dynamical semigroups}, Commun. Math. Phys. 48 (1976) 119.
\newblock \href {https://doi.org/10.1007/BF01608499} {\path{doi:10.1007/BF01608499}}.
\newline\urlprefix\url{https://doi.org/10.1007/BF01608499}

\bibitem{Gorini76jmp}
V.~Gorini, A.~Kossakowski, E.~Sudarshan, \href{https://doi.org/10.1063/1.522979}{Completely positive dynamical semigroups of n-level systems}, J. Math. Phys. 17 (1976) 821.
\newblock \href {https://doi.org/10.1063/1.522979} {\path{doi:10.1063/1.522979}}.
\newline\urlprefix\url{https://doi.org/10.1063/1.522979}

\bibitem{KraussBook}
K.~Krauss, States, Effects and Operations: Fundamental Notions of Quantum Theory, Lecture Notes in Physics, Springer, New York, 1983.

\bibitem{Manzano20aipa}
D.~Manzano, \href{https://doi.org/10.1063/1.5115323}{A short introduction to the {L}indblad master equation}, AIP Advances 10 (2020) 025106.
\newblock \href {http://arxiv.org/abs/arXiv:1906.04478} {\path{arXiv:arXiv:1906.04478}}, \href {https://doi.org/10.1063/1.5115323} {\path{doi:10.1063/1.5115323}}.
\newline\urlprefix\url{https://doi.org/10.1063/1.5115323}

\bibitem{Arakawa1966jcp}
A.~Arakawa, Computational design for long-term numerical integration of the equations of fluid motion: {Two-dimensional} incompressible flow. {Part I}, J. Comp. Phys. 1~(1) (1966) 119--143.

\bibitem{Arakawa1977jcp}
A.~Arakawa, V.~Lamb, Computational design of the basic dynamical processes of the {UCLA} general circulation model, in: {J. Chang (Ed.), Methods in Computational Physics}, Vol.~17, Academic Press, 1977, pp. 173--265.

\bibitem{Childs12}
A.~M. Childs, N.~Wiebe, {Hamiltonian} simulation using linear combinations of unitary operations, Quantum Info. Comput. 12~(11–12) (2012) 901–924.

\bibitem{Barenco95}
A.~Barenco, C.~H. Bennett, R.~Cleve, D.~P. DiVincenzo, N.~Margolus, P.~Shor, T.~Sleator, J.~A. Smolin, H.~Weinfurter, \href{https://link.aps.org/doi/10.1103/PhysRevA.52.3457}{Elementary gates for quantum computation}, Phys. Rev. A 52 (1995) 3457--3467.
\newblock \href {https://doi.org/10.1103/PhysRevA.52.3457} {\path{doi:10.1103/PhysRevA.52.3457}}.
\newline\urlprefix\url{https://link.aps.org/doi/10.1103/PhysRevA.52.3457}

\bibitem{Claudon24}
B.~Claudon, J.~Zylberman, C.~Feniou, F.~Debbasch, A.~Peruzzo, J.-P. Piquemal, \href{https://doi.org/10.1038/s41467-024-50065-x}{Polylogarithmic-depth controlled-not gates without ancilla qubits}, Nature Communications 15~(1) (2024) 5886.
\newblock \href {https://doi.org/10.1038/s41467-024-50065-x} {\path{doi:10.1038/s41467-024-50065-x}}.
\newline\urlprefix\url{https://doi.org/10.1038/s41467-024-50065-x}

\bibitem{Dong21}
Y.~Dong, X.~Meng, K.~B. Whaley, L.~Lin, \href{http://dx.doi.org/10.1103/PhysRevA.103.042419}{Efficient phase-factor evaluation in quantum signal processing}, Physical Review A 103~(4) (Apr 2021).
\newblock \href {https://doi.org/10.1103/physreva.103.042419} {\path{doi:10.1103/physreva.103.042419}}.
\newline\urlprefix\url{http://dx.doi.org/10.1103/PhysRevA.103.042419}

\bibitem{Ying22}
L.~Ying, \href{https://arxiv.org/abs/2202.02671}{Stable factorization for phase factors of quantum signal processing} (2022).
\newblock \href {https://doi.org/10.48550/ARXIV.2202.02671} {\path{doi:10.48550/ARXIV.2202.02671}}.
\newline\urlprefix\url{https://arxiv.org/abs/2202.02671}

\bibitem{Dong23}
Y.~Dong, L.~Lin, H.~Ni, J.~Wang, \href{https://api.semanticscholar.org/CorpusID:260125356}{Robust iterative method for symmetric quantum signal processing in all parameter regimes}, ArXiv abs/2307.12468 (2023).
\newline\urlprefix\url{https://api.semanticscholar.org/CorpusID:260125356}

\bibitem{Haah20}
J.~Haah, \href{http://dx.doi.org/10.22331/q-2019-10-07-190}{Product decomposition of periodic functions in quantum signal processing}, Quantum 3 (2019) 190.
\newblock \href {https://doi.org/10.22331/q-2019-10-07-190} {\path{doi:10.22331/q-2019-10-07-190}}.
\newline\urlprefix\url{http://dx.doi.org/10.22331/q-2019-10-07-190}

\bibitem{Childs21}
A.~M. Childs, Y.~Su, M.~C. Tran, N.~Wiebe, S.~Zhu, \href{https://link.aps.org/doi/10.1103/PhysRevX.11.011020}{Theory of trotter error with commutator scaling}, Phys. Rev. X 11 (2021) 011020.
\newblock \href {https://doi.org/10.1103/PhysRevX.11.011020} {\path{doi:10.1103/PhysRevX.11.011020}}.
\newline\urlprefix\url{https://link.aps.org/doi/10.1103/PhysRevX.11.011020}

\bibitem{Haah20code}
Computation of angles for quantum signal processing in {F}\#, \url{https://github.com/microsoft/Quantum-NC/tree/main/src/simulation/qsp}, accessed: 04-2021 (2020).

\bibitem{code-KvN-LCHS}
{KvN-LCHS} quantum circuits, \url{https://github.com/QuCF/QuCF/wiki/LCHS} (2024).

\bibitem{Berry12}
D.~W. Berry, A.~M. Childs, Black-box {Hamiltonian} simulation and unitary implementation, Quantum Info. Comput. 12~(1–2) (2012) 29–62.

\bibitem{Suau21}
A.~Suau, G.~Staffelbach, H.~Calandra, \href{http://dx.doi.org/10.1145/3430030}{Practical quantum computing}, ACM Transactions on Quantum Computing 2~(1) (2021) 1–35.
\newblock \href {https://doi.org/10.1145/3430030} {\path{doi:10.1145/3430030}}.
\newline\urlprefix\url{http://dx.doi.org/10.1145/3430030}

\end{thebibliography}

\end{document}